\documentclass[a4paper,fleqn,usenatbib]{mnras5}

\usepackage{newtxtext,newtxmath}
\usepackage{xspace}

\usepackage[T1]{fontenc}
\usepackage{ae,aecompl}

\usepackage{tikz}
\usepackage{placeins}
\usepackage{graphicx}	
\usepackage{amsmath}	
\usepackage{lipsum}

\usepackage{array}
\newcolumntype{H}{>{\setbox0=\hbox\bgroup}c<{\egroup}@{}}

\newcommand{\eps}{\varepsilon}

\newcommand{\bs}[1]{\boldsymbol{#1}}

\newcommand{\AREPO}{\textsc{AREPO}\xspace}

\title[Interaction of a cold cloud with a hot wind]{Interaction of a cold cloud with a hot wind: the regimes of cloud growth and destruction and the impact of magnetic fields}

\author[Sparre et al.]{
Martin Sparre$^{1,2}$\thanks{E-mail: sparre@uni-potsdam.de}, Christoph Pfrommer$^{2,1}$ and Kristian Ehlert$^2$\\
\\
$^1$Institut f\"ur Physik und Astronomie, Universit\"at Potsdam, Karl-Liebknecht-Str.\,24/25, 14476 Golm, Germany\\
$^2$Leibniz-Institut f\"ur Astrophysik Potsdam (AIP), An der Sternwarte 16, 14482 Potsdam, Germany\\
}

\pubyear{2020}

\begin{document}
\label{firstpage}
\pagerange{\pageref{firstpage}--\pageref{lastpage}}
\maketitle

\begin{abstract}
  Multiphase galaxy winds, the accretion of cold gas through galaxy haloes, and
gas stripping from jellyfish galaxies are examples of interactions between cold
and hot gaseous phases. There are two important regimes in such systems. A
sufficiently \textit{small} cold cloud is destroyed by the hot wind as a result
of Kelvin-Helmholtz instabilities, which shatter the cloud into small pieces
that eventually mix and dissolve in the hot wind. On the contrary, stripped cold
gas from a \textit{large} cloud mixes with the hot wind to intermediate
temperatures, and then becomes thermally unstable and cools, causing a net accretion of
hot gas to the cold tail. Using the magneto-hydrodynamical code \AREPO, we
perform cloud crushing simulations and test analytical criteria for the
transition between the growth and destruction regimes to clarify a current
debate in the literature. We find that the hot-wind cooling time sets the
transition radius and not the cooling time of the mixed phase. Magnetic fields
modify the wind-cloud interaction. Draping of wind magnetic field enhances the
field upstream of the cloud and fluid instabilities are suppressed by a
turbulently magnetised wind beyond what is seen for a wind with a uniform
magnetic field.  We furthermore predict jellyfish galaxies to have ordered
magnetic fields aligned with their tails. We finally discuss how the results of
idealised simulations can be used to provide input to subgrid models in
cosmological \mbox{(magneto-)}hydrodynamical simulations, which cannot resolve
the detailed small-scale structure of cold gas clouds in the circum-galactic
medium.
\end{abstract}



\section{Introduction}\label{Intro}

In our galaxy formation paradigm, gas enters the interstellar medium (ISM), where stars are formed, through mergers and accretion of cold or hot gas. In the hot-accretion mode, gas shock is heated as it enters a galaxy halo and radiative processes cool it to lower temperatures \citep{1977MNRAS.179..541R,1978MNRAS.183..341W,1980MNRAS.193..189F,1991ApJ...379...52W}. In the cold-accretion regime, gas is accreted through the halo in cold filaments. The kinetic energy of these filaments is dissipated through a sequence of small shocklets and the associated temperature increase is quickly radiated away in these dense filaments. The existence of a cold accretion regime is supported by various hydrodynamical cosmological simulations \citep{2005MNRAS.363....2K,2006MNRAS.368....2D,2009Natur.457..451D}, even though modern hydrodynamical methods find this accretion regime to be less important than originally suggested \citep{2013MNRAS.429.3353N}, because hydrodynamical instabilities disrupt the streams. The urge to understand the stability of streams has led to several studies using analytical calculations and idealised high-resolution simulations \citep{2016MNRAS.463.3921M,2018MNRAS.477.3293P,2019MNRAS.484.1100M,2019MNRAS.490..181A,2019MNRAS.489.3368B}. 

The circumgalactic medium (CGM) -- the spatial region outside the galaxy disc but still inside the virial radius -- likely exhibits more complex physics than what can be described by simple cold and hot accretion regimes. Simulations for example show that ISM winds are continuously adding gas to the CGM, from which gas is recycled into the ISM as well \citep{2017MNRAS.470.4698A,2018MNRAS.481..835O,2020arXiv200616316F}. Idealised simulations furthermore suggest that physics on small scales comparable to the cooling length or Field length may be important for the evolution of the gas \citep{2018MNRAS.473.5407M,2019MNRAS.482.5401S,2018arXiv180610688L}, indicating that cosmological simulations may lack the necessary resolution to resolve the CGM gas. Several groups have performed cosmological simulations specifically targeting an extra high resolution in the CGM, and they usually find denser and smaller structure at higher resolution, even though the relevance of small-scale structure (on parsec scales) is debated \citep{2018arXiv181006566P,2018arXiv181105060C,2019ApJ...882..156H,2019MNRAS.482L..85V}. From a theoretical point of view, understanding the physical processes in the CGM is therefore an important challenge for future simulations.

\begin{table*}
\centering
\begin{tabular}{ll|ccrl}
\hline\hline
\# (1) &Name (2) & $\bs{B}$ in wind (3) & $\bs{B}$ in cloud (4) &  $R_{\text{cloud}}/\Delta x$ (5) & Note\\
\hline
{\tt 1} &{\tt 1-NoMF} & None, $\beta=\infty$ & None, $\beta=\infty$ & 64\\
{\tt 2} &{\tt 2-WindTurb-CloudNone} & Turbulent, $\beta=10$ & None, $\beta=250$ & 64 & For numerical reasons we set $\beta=250$ in \\
 & &  &  & & the wind rather than $\infty$ (see text for details). \\
{\tt 3} &{\tt 3-WindNone-CloudTangled} & None, $\beta=\infty$ & Tangled, $\beta=10$ & 64 \\
{\tt 4} &{\tt 4-WindTurb-CloudTangled} & Turbulent, $\beta=10$ & Tangled, $\beta=10$ & 64 \\
{\tt 4-HR} &{\tt 4-HR-WindTurb-CloudTangled} & Turbulent, $\beta=10$ & Tangled, $\beta=10$ & 128 & High-resolution simulation \\
{\tt 5} &{\tt 5-WindUniform-CloudTangled} & Uniform, $\beta=10$ & Tangled, $\beta=10$ & 64 \\
\hline\hline
\end{tabular}
\caption{The simulations presented in this paper. We will refer to a simulation either by its number (column 1) or name (column 2). In column 3 and 4 we describe how the magnetic fields in the wind and cloud are initialised -- we quote the value of $\beta \equiv P_\text{th}/P_{B}$. Column 5 states the resolution in terms of number of cells per cloud radius (in one dimension).}
\label{Table:SimulationOverview}
\end{table*}

\begin{figure}
\centering
\includegraphics[width=0.7\linewidth]{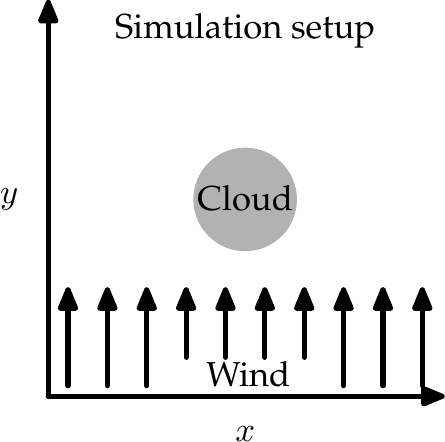}
\caption{A sketch of a cloud crushing simulation setup.}
\label{Sketch-1}
\end{figure}

A large fraction of the CGM is warm and diffuse, and it is hence hard to observe in emission, but recently remarkable progress has been made by observing a distribution of Ly$\alpha$ haloes around $z\sim 2$ galaxies \citep{2018Natur.562..229W}. The strongest constraints on the CGM of low-redshift galaxies derive from spectroscopy of galaxies with haloes along the sight-lines of distant quasars. Studying samples of nearby galaxies with the \emph{Cosmic Origins Spectrograph} (COS, mounted on the \emph{Hubble Space Telescope}) has made it possible to constrain the baryon budget in various gas phases of the CGM \citep{2014ApJ...786...54P,2014ApJ...792....8W,2014ApJ...796..136B,2017A&A...607A..48R,2017ARA&A..55..389T}.

\citet{2014ApJ...792....8W} identified a multiphase distribution of CGM gas with a cool gas phase (with a temperature of $T\sim 10^4$ K) in the CGM, co-existing with a warm-hot phase (traced by the \ion{O}{vi} absorption line) and a hot phase with $T>10^7$ K (see their fig. 11). Such detection's call for a better theoretical understanding of multiphase gas. Remarkably dense and cold gas clouds are also observed in the multiphase outflows in the CGM of interacting starburst galaxies \citep{2009ApJS..181..272G,2013ApJ...768...75R}; the most visually appealing example being the Messier 82 galaxy \citep{2009ApJ...697.2030S,2015ApJ...814...83L,2020A&ARv..28....2V}. It is a theoretical puzzle how dense and cold gas survives in the CGM, since simulations show that ISM clouds influenced by a starburst wind \citep[as described in][]{1985Natur.317...44C,2018ApJ...860..135S,2020arXiv200104384Y} are dissolved before they become co-moving with the wind \citep{2015ApJ...805..158S,2016ApJ...822...31B,2017ApJ...834..144S,2020MNRAS.tmp.2098H}.

A unique window to study physical processes associated with the interaction of a hot wind with a cold gaseous phase is provided by \emph{jellyfish}-galaxies. They reside in the outskirts of galaxy clusters, and show ram-pressure stripping of dense gas of the ISM that is exposed to a hot, magnetised wind that the  galaxy feels as it moves through the intra-cluster medium (ICM). The presence of such galaxies can be explained by cosmological magnetohydrodynamical simulations \citep{2019MNRAS.483.1042Y}, which reliably model ram pressure stripping in clusters. Recently, \citet{2019ApJ...870...63C} observed a long star-forming tail of dense star-forming gas (length of 60 kpc and width of 1.5 kpc) in the jellyfish galaxy, D100, which is in the vicinity of the Coma cluster. The presence of such a long tail of dense gas, implies that the gas in this case survives the transport from the ISM far into the ICM. 

Such a regime, where cold gas can survive being transported to large distances, is expected when the radiative cooling time-scale is sufficiently short in comparison to the time-scale of hydrodynamical instabilities \citep{2017MNRAS.470..114A,2018MNRAS.480L.111G,2019arXiv190902632L,2019MNRAS.tmp.2995G}. If this is the case clouds will indeed experience growth rather than destruction. This does not only apply to jellyfish galaxies, but potentially also to cold accretion filaments as they fragment into large clouds in galaxy haloes \citep{2019AJ....158..124F,2019arXiv191005344M}. Currently, there is a debate about the exact criteria for the transition between the growth and destruction regimes. The criteria from \citet{2018MNRAS.480L.111G} and \citet{2019arXiv190902632L} for example differ since they rely on the radiative cooling time-scale of the mixed gas (from the hot wind and cold cloud) and the hot wind, respectively. A different suggestion comes from \citet{2015MNRAS.449....2M}, which concludes that magnetic draping, occurring when magnetic field lines in the wind are wrapped up upstream from the cold cloud \citep{2008ApJ...677..993D}, is the key for causing cloud survival.

In this paper we investigate various theoretical criteria for the division between the cloud growth and destruction regime in idealised simulations of cold clouds interacting with a hot wind. We include a range of configurations of the magnetic field. This enables us to study how multiphase gas may arise in the winds of starbursts and normal galaxies. We asses how a medium strength magnetic field affects this criterion, and we furthermore estimate how magnetic fields affect the gas structure in our simulations. In Sect.~\ref{SimulationOverview} and \ref{Sec:Results} we describe our initial conditions and simulations in the cloud destruction regime. In Sect.~\ref{GrowthRegime} we present simulations investing the criterion for the transition between the destruction and growth regime. In Sect.~\ref{Sec:Discussion} we discuss implications of our work related to jellyfish galaxies, cold accretion flows and the development of subgrid models for numerical galaxy formation simulations.

\section{Simulation overview}\label{SimulationOverview}

In this paper we perform three--dimensional cloud crushing simulations, where a cold and dense spherical cloud is influenced by a hot and diffuse supersonic wind. Such a setup is sketched in Fig.~\ref{Sketch-1}.

We here introduce the setup used for simulating clouds in the destruction regime, and in Sect.~\ref{GrowthRegime} we focus on establishing a criterion that describes the transition to the growth regime. A key goal is to estimate the role of the configuration of the magnetic field in the wind and in the cloud, since we present for the first time cloud crushing simulations with a turbulent magnetic wind.

\subsection{Cloud and wind properties}

The cold cloud is initialised with a temperature of $T_\text{cloud}=10^4$~K and a density of $n_\text{cloud}=0.1$~cm$^{-3}$, and the hot wind has $T_\text{wind}=5\times 10^6$~K and $n_\text{wind}=2\times 10^{-4}$~cm$^{-3}$, such that the cloud is in pressure equilibrium with the wind. The simulations are carried out in the $t=0$ rest-frame of the cloud, and the hot wind moves in the positive $y$-direction with a sonic Mach number of 2. The cloud is initialised with a radius of 25 pc. All gas cells are initialised to have a solar metallicity.

At our fiducial resolution level we use 64 cells per cloud radius and in our high-resolution simulation we use 128 cells per radius. We hence use an identical resolution as e.g., by \citet{2015ApJ...805..158S}. The box size is $(L_x,L_y, L_z) = (400,1600,400)$ pc, which is sufficient to simulate the shock front upstream from the cloud and also to avoid dense gas leaving the simulation domain before it is mixed with the hot wind.

\begin{figure*}
\centering
\includegraphics[width=0.77\linewidth]{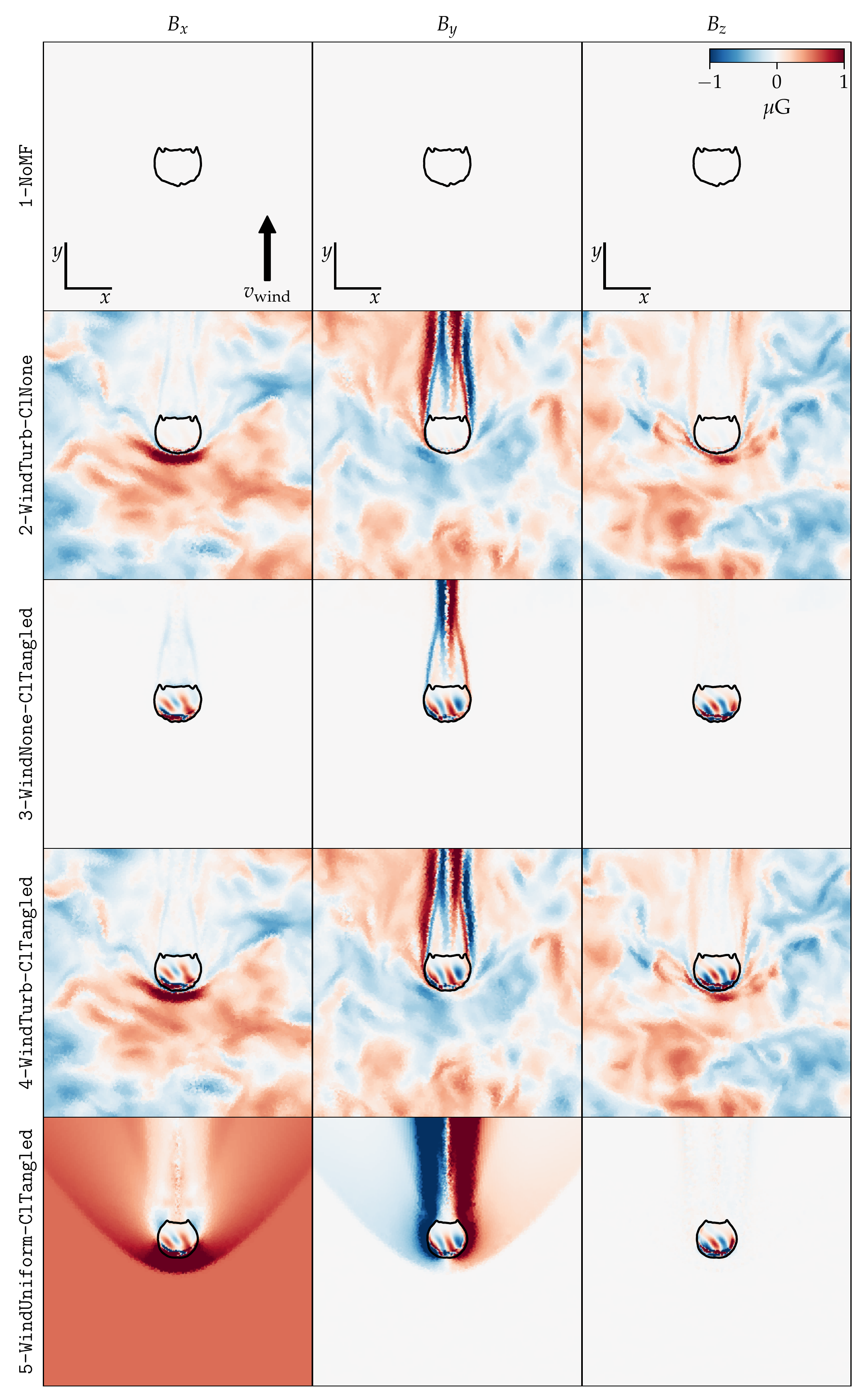}
\caption{A slice in the $z=0$ plane at one cloud crushing time ($t_\text{cc}$). We show the five different magnetic field configurations in our simulations. In the simulations with a turbulent magnetic wind (simulation {\tt 2} and {\tt 4}) field lines are wound up upstream from the cloud, causing an increased magnetic field in the $\hat{\mathbfit{x}}$- and $\hat{\mathbfit{z}}$-directions. In the simulation with a uniform magnetic wind, perpendicular to the wind velocity (simulation {\tt 5}), a draping layer exists for the $B_x$-component, but is absent for the two other magnetic field components. A draped magnetic field in front of the cloud protects against disruption \citep{2008ApJ...677..993D} -- an effect we will quantify in the remaining parts of this paper.}
\label{Fig801_Bx}
\end{figure*}

\subsection{Magnetic field configurations}

The magnetic field configurations in the simulations are summarised in Table~\ref{Table:SimulationOverview} (see also Fig.~\ref{Fig801_Bx}). The first simulation, named {\tt 1-NoMF}, is a purely hydrodynamical simulation with magnetohydrodynamics (MHD) disabled. This simulation serves the purpose of comparing the MHD simulations to a non-MHD analogue.

In our simulations with a magnetic wind we either inject a constant magnetic field along the $x$-axis, which is perpendicular to the wind velocity, or we inject a turbulent magnetic field. In either case we scale the average magnitude of the $\bs{B}$-field to match the $\beta$-value set in Table~\ref{Table:SimulationOverview}. For the simulations with a turbulent magnetic wind, we generate a Gaussian field with a power spectrum of the form, $P_i(k) \propto k^2 \left|\tilde{B}_i(k)\right|^2$, where the absolute square of the Fourier transformation of each of the magnetic field component $B_i$ is (we follow App. A of \citealt{2018MNRAS.481.2878E}, but see also \citealt{2007MNRAS.378..662R})
\begin{align}
\left|\tilde{B}_i(k)\right|^2 = \begin{cases}
A,  & \text{ if } k< k_\text{inj}.\\
A \left( \frac{k}{k_\text{inj}} \right)^{-11/3}, & \text{ if } k\geq k_\text{inj}.\\
\end{cases}
\end{align}
Here $k$ is the wavenumber, $k=1/\sqrt{x^2+y^2+z^2}$. The injection scale is set to $k_\text{inj}=1/(\sqrt{3}L_x)$. We hence have white noise on scales larger than the side length of the box, and on smaller scales, $k\geq k_\text{inj}$, we have Kolmogorov turbulence. The normalisation of the power spectrum is chosen such that the magnetic field strength, $\sqrt{\langle \bs{B}\rangle^2}$, is as specified by the $\beta$-value in Table~\ref{Table:SimulationOverview}, where $\beta=P_\rmn{th}/P_B$ is the thermal-to-magnetic pressure ratio.

The turbulent cubic box has a side length of $L_x$ and periodic boundary conditions. This allows us to read in multiple instances of the box along the $y$-axis in the injection region (our simulation boxes are always rectangular such that $L_x=L_z<L_y$), so the magnetic field is set throughout the simulation domain. Examples of simulations with magnetic turbulent fields in the wind can be seen in simulation {\tt 2-WindTurb-CloudNone} and {\tt 4-WindTurb-CloudTangled} in Fig.~\ref{Fig801_Bx}.

For the simulations with a magnetic field inside the cloud, we follow \citet{2015MNRAS.449....2M} and generate a tangled divergence-free magnetic field as a superposition of 11 fields generated according to,
\begin{align}
\bs{B}=\cos (\alpha a)\hat{\bs{c}} + \sin (\alpha a)\hat{\bs{b}} ,
\end{align}
where $\hat{\bs{a}}$, $\hat{\bs{b}}$ and $\hat{\bs{c}}$ are unit vectors constituting a right-handed coordinate system (randomly drawn from a spherically symmetric distribution), $a=|\bs{a}|$, and the coherence length $\alpha$ is $0.1R_\text{cloud}$. A superposition of such fields is per construction force free and divergence free.

At the interface between the cloud and the wind we carry out a procedure to isolate the cloud's magnetic field from the wind (such that the radial component vanishes at the cloud's surface), while also keeping the condition, $\text{div}\,\bs{B}=0$. This is done by recursively removing $\text{div}\,\bs{B}$ and isolating the $\bs{B}$-field, until a solution is obtained, where the divergence measured across single cells at the cloud boundary is less than 5 per cent of the magnetic field strength. For a detailed description of the recursive algorithm, the reader is referred to Appendix A of \citet{2018MNRAS.481.2878E}.

For the simulation {\tt 2-WindTurb-CloudNone} we wish to have a $\beta=10$ turbulent field in the wind and a vanishing magnetic field in the cloud. We require a vanishing magnetic field divergence near the cloud boundary. Hence, we initiate the cloud with $\beta=250$, which gives a much more stable cloud than with $\beta=\infty$.

\subsection{Cooling function}

We use the implementation of the cooling function as described in the galaxy formation model of \citet{2013MNRAS.436.3031V}. This is the same implementation used in the large-scale cosmological simulations, Illustris-TNG \citep{2018MNRAS.475..676S,2018MNRAS.475..648P,2018MNRAS.480.5113M,2018MNRAS.477.1206N} and Illustris \citep{2014Natur.509..177V,2014MNRAS.445..175G}. The cooling rate of the gas is decomposed into 1) cooling from primordial species (hydrogen and helium), which is following \citealt{1996ApJS..105...19K}, 2) Compton cooling off of the cosmic microwave background (see \citealt{2013MNRAS.436.3031V} for details), and 3) metal-line cooling based on tables of CLOUDY models \citep{1998PASP..110..761F,2013RMxAA..49..137F} assuming a spatially uniform UV background (from \citealt{2009ApJ...703.1416F}) and ionisation equilibrium. We use a temperature floor of $5\times 10^3$ K, which is the lowest temperature a gas cell can have in our simulations.

\subsection{Simulation code, refinement and boundary conditions}

The simulations are carried out with the moving mesh code \AREPO \citep{2010MNRAS.401..791S,2016MNRAS.455.1134P}. We use a refinement scheme such that all cells are within a factor of two of the \emph{target mass}, which is the simulation input parameter determining the mass resolution of a simulation. Voronoi cells are derefined (refined) if the mass per cell falls below (rises above) half (twice) that parameter. Due to our almost equal-mass-refinement criterion the spatial resolution of the hot wind is lower than for the cold cloud. At our default resolution level we set the target mass, such that we have a linear cell size for the hot wind of $\Delta x = L_x/128$ in the initial conditions. To gain a higher resolution in the domains between the dense gas and the diffuse wind, we use a neighbour refinement scheme enforcing the linear size of each gas cells to be at maximum eight times larger than any of its neighbours; gas cells exceeding this size are refined.

The simulation box is periodic in the $x$ and $z$ directions. We have not enabled outflow conditions at the upper $y$-boundary, but instead we use a periodic $y$-boundary and place an \emph{injection region} at the lower $y$-boundary. Here new cells are created as the wind moves in the positive $y$-direction, and the density, temperature, metallicity, magnetic field and cell volume are fixed to the prescribed properties of the hot wind. This yields a classical windtunnel setup. For details on the setup, see \citet{2019MNRAS.482.5401S}.

\section{The cloud destruction regime}\label{Sec:Results}

To visualise the magnetic field at the initial stages of our simulations, the field components are plotted at $t=t_\text{cc}$ in Fig.~\ref{Fig801_Bx}, where $t_\text{cc}$ is the time-scale for the initial shock crushing the cold cloud,
\begin{align}
t_\text{cc} \equiv \frac{R_\text{cloud}}{\varv_\text{wind}} \sqrt{\frac{\rho_\text{cloud}}{\rho_\text{wind}}}. \label{tcc}
\end{align}
Here $R_\text{cloud}$ is the cloud radius, $\varv_\text{wind}$ is the wind velocity, and $\rho_\text{cloud}$ and $\rho_\text{wind}$ are the mass densities of the cloud and wind, respectively. In simulation {\tt 1-NoMF} the magnetic field is strictly 0, because MHD is disabled in the simulation.

In simulation {\tt 5-WindUniform-CloudTangled}, where the wind is magnetised in the $\hat{\bs{x}}$-direction, which is perpendicular to the wind velocity and the cloud's symmetry axis, we see an enhanced magnetic field in-between the bow shock and the cloud. This is caused by two effects. First, the magnetic field is adiabatically compressed across the shock, which causes the field component perpendicular to the shock normal to increase as the density, $B_\perp\propto n$, along the stagnation line. Second, as the cloud moves through the magnetised plasma, it sweeps up magnetic field to build up a dynamically important sheath around the object -- this is the effect of \emph{magnetic draping} \citep{2008ApJ...677..993D}.

After an initial ramp-up phase, the layer's strength in steady-state is set by a competition between assembling new magnetic field in the layer and field lines slipping around the cloud. In the draping layer, the magnetic energy density $\eps_{B,\,\rmn{drape}} \simeq \alpha\rho_\text{wind}^{} \varv_\text{wind}^2$ with $\alpha\simeq2$ \citep{2008ApJ...677..993D}, is solely given by the ram pressure $\rho_\text{wind}^{} \varv_\text{wind}^2$ and {\em completely} independent of the magnetic energy density in the wind, $\eps_{B,\,\rmn{wind}}$. Assuming that the sphere with radius $R$ and volume $V$ is wrapped into a draping layer of constant thickness $l_\rmn{drape}=R/(6\alpha M_\rmn{A}^2)$ over an area $A=2\pi\,R^2$ of the half-sphere (where $M_\rmn{A}=\varv_\text{wind}/\varv_\text{A}$ is the Alfv\'enic Mach number, $\varv_\text{A}$ is the Alfv{\'e}n speed, and $R$ is the curvature radius at the stagnation point), we estimate the magnetic energy of the draping layer: 
\begin{align}
  E_{B,\,\rmn{drape}} 
  = \frac{B_\rmn{drape}^2}{8\pi}\, A\,l_\rmn{drape} = 
  \frac{B_\rmn{drape}^2}{8\pi}\,\frac{A\,R}{6\alpha\,M_A^2} 
  =\eps_{B,\,\rmn{wind}}\,\frac{V}{2}.
\end{align}
This  {\it Archimedes principle of magnetic draping}\footnote{See also the lectures notes by Pfrommer on \emph{A Pedagogical Introduction to Magnetic Draping} (2011, Kavli Institute for Theoretical Physics workshop): \url{http://online.itp.ucsb.edu/online/gclusters11/}} states that the ramp-up phase lasts for a crossing time of the half-sphere before we enter steady state, independent of the magnetic field strength of the ICM. Once the system has reached steady state, modes with wave length $\lambda\lesssim 10 l_\rmn{drape}= R/(\beta M_\rmn{s}^2)$ are stabilised against the Kelvin-Helmholtz instability \citep{2007ApJ...670..221D}, where $M_\rmn{s}$ is the sonic Mach number and we assume the adiabatic index $\gamma=5/3$. Hence, a smaller wind magnetic field has no consequence for the time-scale to reach a steady state but implies a narrower thickness of the draping layer that stabilises only small-scale modes while modes with $\lambda \gtrsim R/(\beta M_\rmn{s}^2)$ can still get unstable. Such a layer with an enhanced magnetic field is only present for the $B_x$-component for this simulation as the $B_y$ and $B_z$ components are vanishing upstream from the cloud. Inside the cloud the magnetic field is tangled as in the initial conditions.

For the simulations with a turbulent magnetic wind, {\tt 2-WindTurb-CloudNone} and {\tt 4-WindTurb-CloudTangled}, there is no preferred direction of the prescribed magnetic field in the wind. Layers of enhanced magnetic field strength can therefore develop for both the $B_x$- and $B_z$-component. The presence of an enhanced magnetic field for both the $x$- and $z$-components is potentially important, because draping can increase the drag force on clouds as well as suppress fluid instabilities. Such effects have previously only been established using uniform magnetic fields at dense objects \citep{2008ApJ...677..993D,2015MNRAS.449....2M}, but not with a turbulent magnetic field.

\begin{figure*}
\centering
\includegraphics[width=\linewidth]{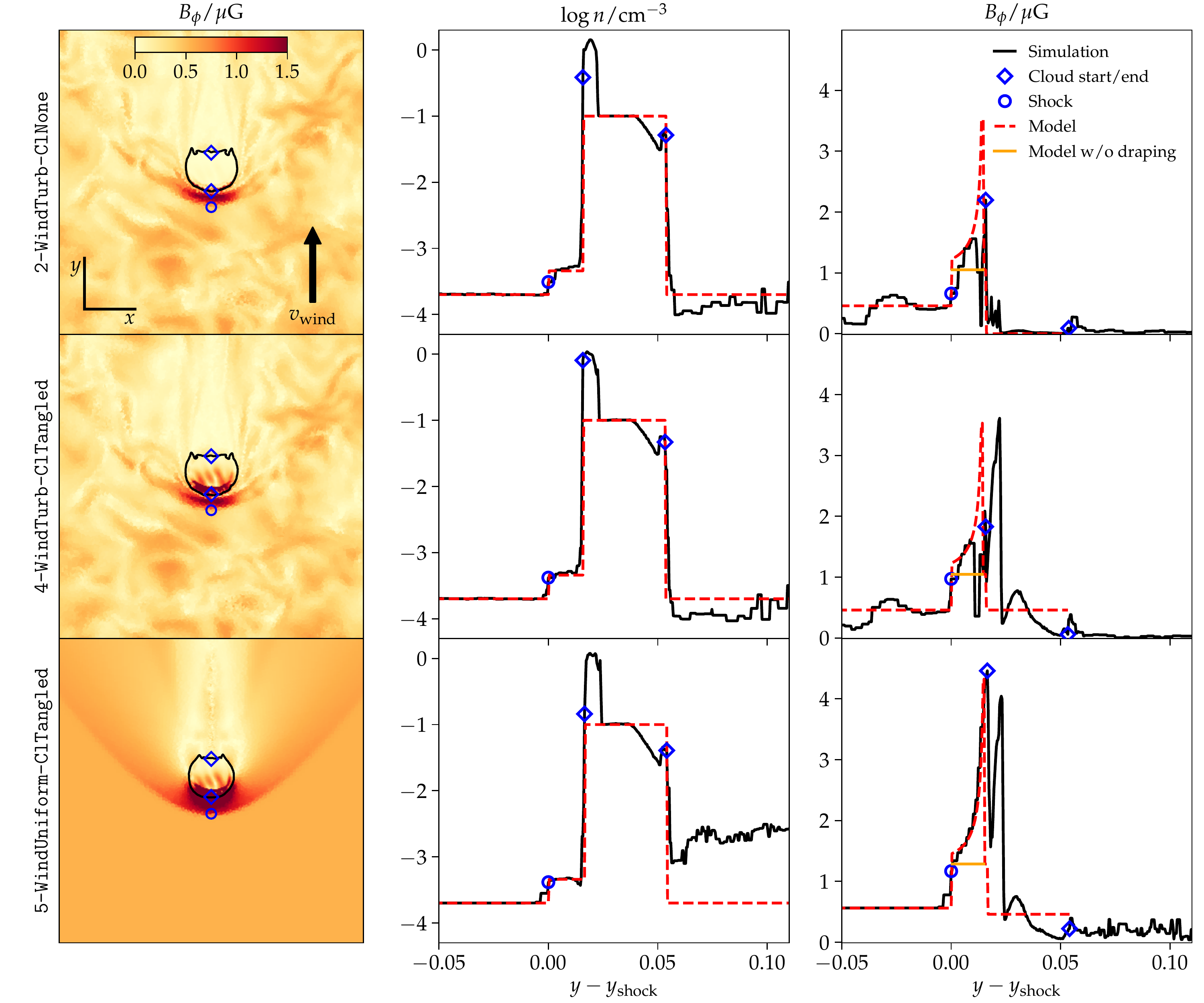}
\caption{Magnetic field amplification due to shock compression and magnetic draping. The left panel shows the $B_\phi$ component at time $t_\text{cc}$. The central panel shows number density along the stagnation line (the axis along the wind direction through the symmetry axis of the cloud). The location of the bow shock is marked with a circle, and the cloud's head and tail are marked with $\Diamond$-symbols. The dashed red line shows the theoretical expectation with $n=n_\text{wind}$ upstream of the bow shock, $n=2.286\, n_\text{wind}$ in the post-shock region according to the Rankine-Hugoniot jump conditions, and $n=n_\text{cloud}$ inside the cloud, not modelling adiabatic compression of the gas at the cloud head. The right panel shows $B_\phi$ along the stagnation line. The red dashed line shows a model with an average value $\langle B_\phi \rangle$ in the wind and in the cloud, and in the post-shock region we take into account magnetic draping as well as adiabatic compression of $B_\phi$ at the shock. The horizontal orange line shows a model of the post-shock region, where the increase in $B_\phi$ is entirely caused by adiabatic shock compression (draping is omitted). The uniform-wind simulation shows excellent agreement with our model (dashed line), whereas the turbulent-wind simulations fluctuate because the magnetic energy is also shared with the $B_z$ component.}
\label{Fig807_Draping}
\end{figure*}

\subsection{Shock compression and magnetic draping}

Pioneering simulations has revealed how magnetic fields may evolve during the interaction of a cloud in a diffuse wind \citep{1994ApJ...433..757M,1996ApJ...473..365J,1999ApJ...510..726M,1999ApJ...527L.113G,2008ApJ...677..993D,2010NatPh...6..520P}. Radiative cooling does, however, significantly alter the rate at which instabilities develop \citep{2015MNRAS.449....2M}. We therefore revisit the basic problem of magnetic draping (as laid out in detail by \citealt{2008ApJ...677..993D}) in our MHD simulations, which include a state-of-the-art treatment of radiative cooling.

We specifically quantify how the magnetic field is amplified due to shock compression and magnetic draping. We therefore study the toroidal magnetic field component, $B_\phi \equiv \sqrt{ B_x^2 + B_z^2}$, in the simulations with a magnetised wind in Fig.~\ref{Fig807_Draping}. As we see, $B_\phi$ is enhanced in the shocked region, in between the cloud and the bow shock. To quantify the behaviour of $B_\phi$ we plot the density and the magnetic field along the stagnation line (the line along the $y$-axis passing through the bow shock's head and the head of the cloud). We compare the density profile to a simple analytical model, which assumes 1) a constant density of the wind $n_\text{wind}$ upstream from the shock, 2) a shock compressed density of $2.286\,n_\text{wind}$ according to the Rankine Hugoniot jump conditions for a shock with Mach number $M=2$ and adiabatic index of $5/3$ in the post-shock region, 3) a cloud density of $n_\text{cloud}$, and 4) an initial density of the wind downstream from the cloud.

\begin{figure*}
\centering
\begin{minipage}{.46\linewidth}
\includegraphics[width=1.0\linewidth]{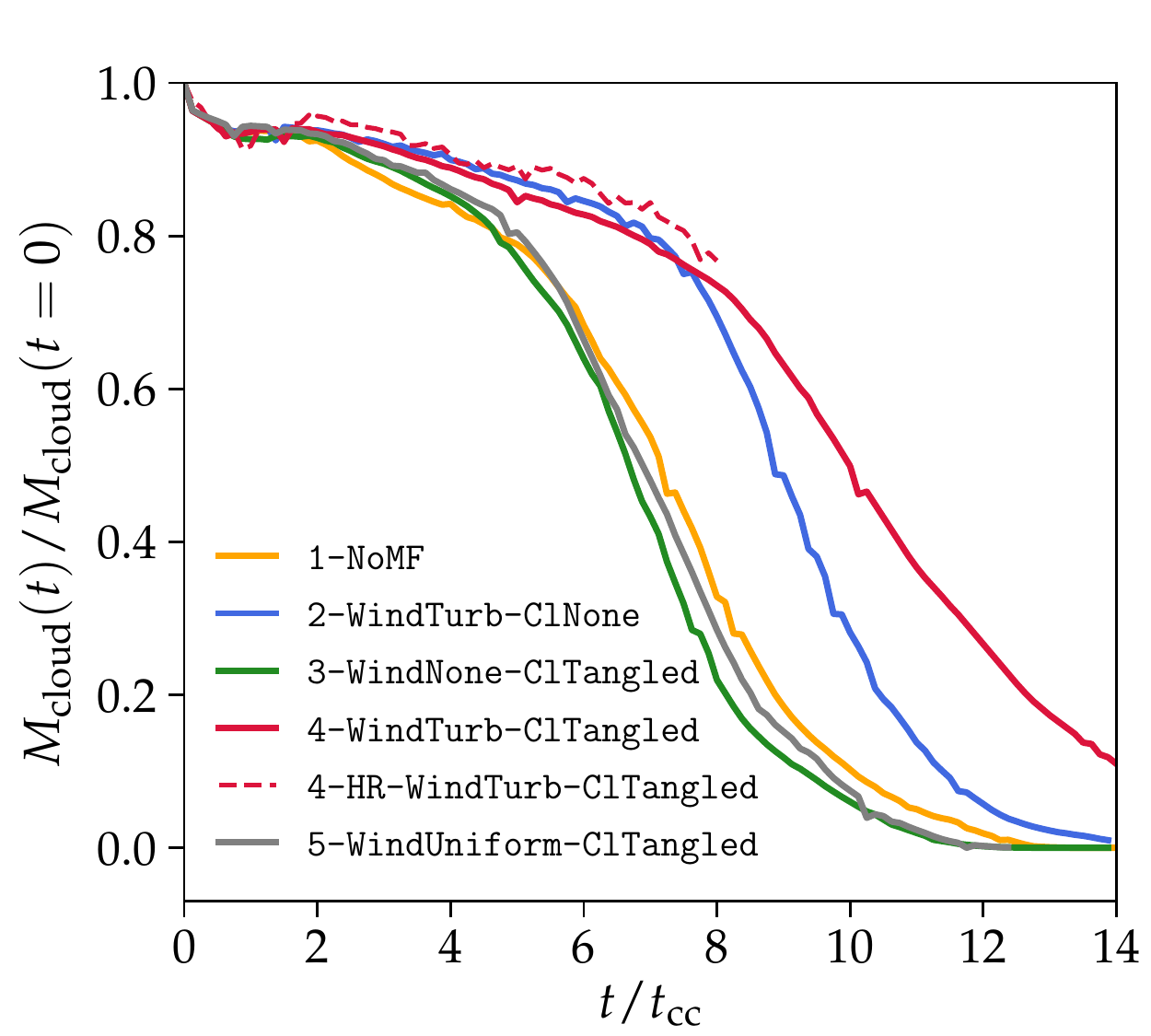}
\end{minipage}
\begin{minipage}{.46\linewidth}
\includegraphics[width=1.0\linewidth]{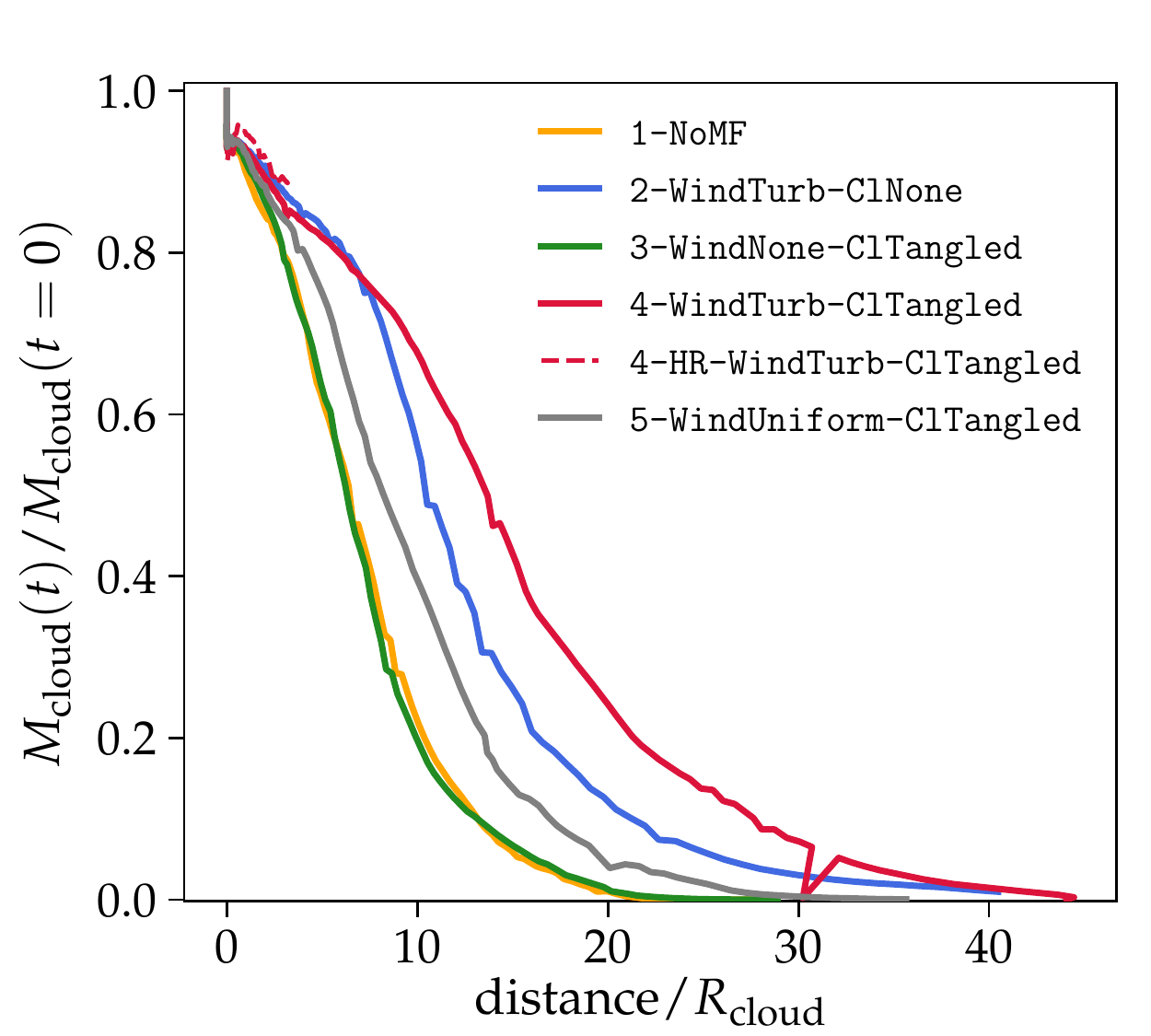}
\end{minipage}
\caption{\emph{Left panel}: the survival fraction on dense gas with $n \geq n_\text{cloud}/3$ as a function of time. A turbulent magnetic wind (simulation {\tt 2} and {\tt 4}) enhances the presence of dense gas at late times $t\gtrsim 6t_\text{cc}$. \emph{Right panel}: The dense cloud-gas survives travelling to larger distances in the presence of a turbulent magnetic wind, as probed by the survival fraction versus the median distance travelled. In the simulation with a uniform magnetic wind (simulation {\tt 5}) the material at fixed cloud mass fraction survives to a larger travelling distance in comparison to the simulations without a magnetic wind ({\tt 1} and {\tt 3}), but still falls short to the simulations with a turbulent wind.}
\label{Fig000_CloudSurvival_3D}
\end{figure*}

We see a perfect agreement of model and our simulations except for the head of the cloud and the wake. The wind ram pressure causes the head to be adiabatically compressed by a similar amount for the different simulations. Downstream from the cloud, the density is increased above the initial wind density in the simulation with a uniform magnetic wind, whereas it is decreased in the simulations with a turbulent magnetic wind. In all the simulations the thermal pressure dominates and is roughly constant (within a factor of two) downstream from the cloud along the plotted axis, but the actual density distribution is different such that we have lower density and higher temperature in the simulations with a turbulent magnetic wind in comparison to the uniform wind simulation. The density is higher in the wake of the cloud (along the axis studied in Fig.~\ref{Fig807_Draping}) in the simulation with a uniform magnetic wind because magnetic draping here only protects against Kelvin-Helmholtz instabilities along the orientation of the magnetic field (in the $x$-direction), so instabilities can act in the $z$-direction and cause ablation of gas, which is advected downstream by the wind \citep{2007ApJ...670..221D,2019MNRAS.489.3368B}.

Having established that our simple analytical model describes the density well upstream of the cloud, we proceed and investigate the magnitude of the $B_\phi$-component in Fig.~\ref{Fig807_Draping} (right panel). We compare our simulations to a model with an average value $\langle B_\phi \rangle$ in the wind and in the cloud. Our theoretical model takes into account the adiabatic compression of $B_\phi$ at the shock as well as the magnetic draping at the cloud according to \citet{2008ApJ...677..993D},
\begin{align}
\frac{B_\phi}{n} = \frac{B_\phi^\text{wind}}{n^\text{wind}} \times \frac{1}{\sqrt{  1-\left[R/(y-y_\text{cloud})\right]^3 } },
\end{align}
where $R$ is the curvature radius of the cloud at the stagnation point and $y_\text{cloud}$ is the cloud center. These parameters are fitted to match the curvature near the head of the cloud. For simulation {\texttt 2} and {\texttt 4} we fit a curvature radius of $R=26.6$ pc, and for simulation {\texttt 5} we obtain $R=23.0$ pc. The uniform-wind simulation shows excellent agreement with our model, whereas in the turbulent-wind simulations the model provides an upper envelope of the simulated values because the magnetic energy is also shared with the $B_z$ component. This demonstrates that magnetic draping occurs in all simulations because the post-shock regions have a $B_\phi$-value that is enhanced above a model which does not include draping and only takes into account the adiabatic compression of $B_\phi$ (labelled as \emph{Model without draping} in the figure).

The simulations with a turbulent magnetic wind have a dip in the $B_\phi$-value in between the draping layer and the head of the cloud. This pattern arises frequently in these simulations, as the magnetic polarity in the draping layer changes its orientation. Such a dip is absent in the simulation with a uniform magnetic field, since the draping layer's orientation is fixed throughout the simulation. Additionally, some small degree of numerical resistivity arises as draped field lines of different magnetic polarities are moved together at the grid resolution. This numerical reconnection is strongest immediately upstream the cloud. Overall, we conclude that draping of field lines as the wind sweeps up the cloud increases the magnetic field strength in front of the cloud in all of the simulations with a magnetic wind.

In all simulations with magnetised clouds, we observe a second peak of $B_\phi$ at the head of the cloud in the right-hand panels of Fig.~\ref{Fig807_Draping}. This is due to adiabatic compression of the density at the head of the cloud. However, the $B_\phi$ enhancement is much narrower in comparison to the density enhancement. The reason for this is again the changing polarity of a tangled field. Because of our magnetic isolation procedure, we produced a mostly tangential magnetic field at the cloud boundary, which is naturally enhanced upon adiabatic compression. Once the ram-pressure enhanced density encounters radial magnetic fields in the inner regions, the toroidal field plummets and recovers with a much smaller toroidal field enhancement due to the tangled nature of the cloud field. This smaller enhancement can also be appreciated as a striped feature of $B_\phi$ in the bottom two panels in the left column of Fig.~\ref{Fig807_Draping}.

\begin{figure*}
\centering
\includegraphics[width=\linewidth]{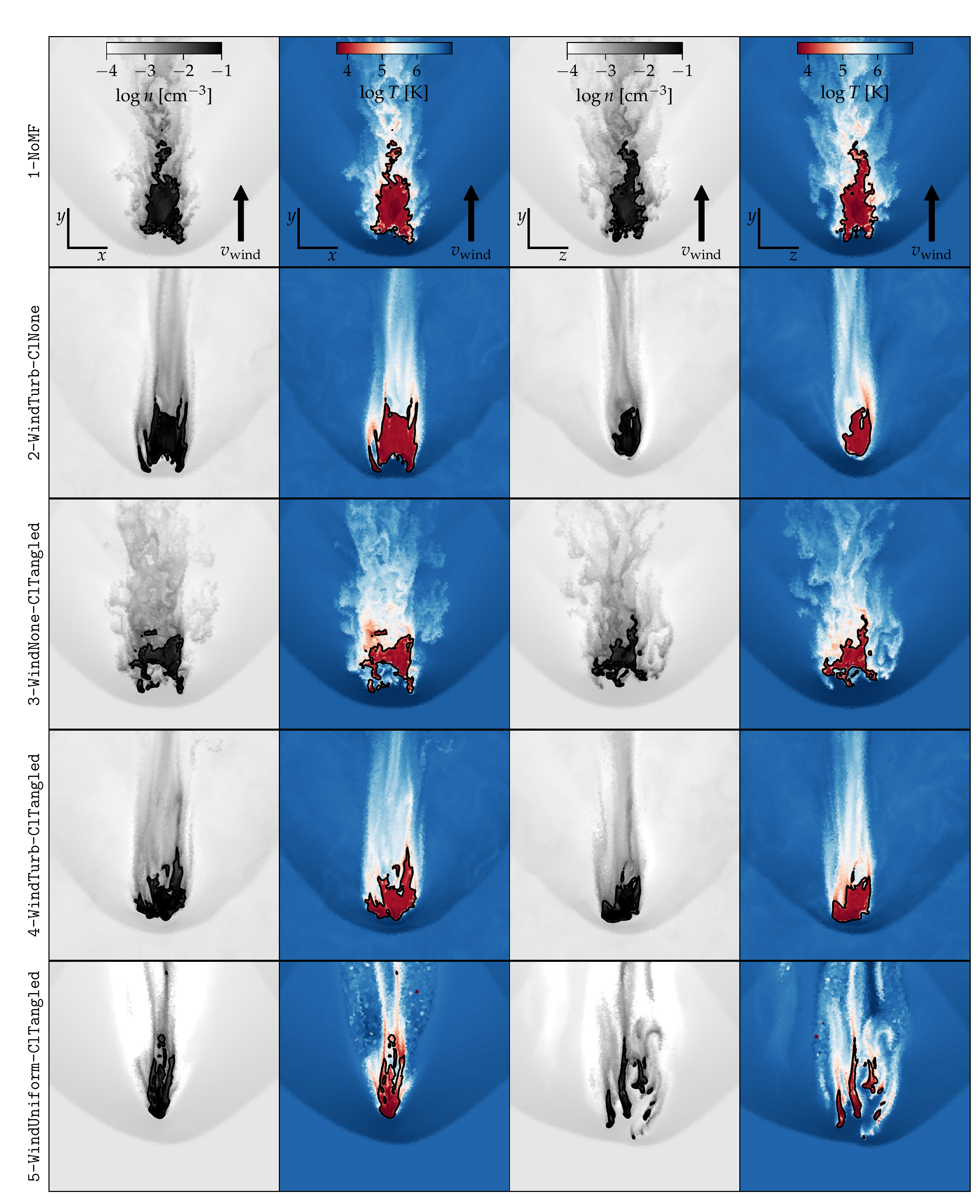}
\caption{Density and temperature slices in the $z=0$ plane (column 1 and 2) and $x=0$ plane (column 3 and 4) at $t=6t_\text{cc}$ for our simulations. In the absence of a magnetic wind (simulation {\tt 1} and {\tt 3}) Kelvin--Helmholtz--billows form downstream from the dense cloud; this can be seen both in the panels showing temperature and density. In simulation {\tt 5} draping of a uniform wind magnetic field suppresses cloud destruction in the $x$-direction, but not in the $z$-direction as seen in the $(z,y)$-projection (bottom right panel). In the presence of a turbulent magnetic wind (simulation {\tt 2} and {\tt 4}), instabilities are suppressed both in the $x$- and $z$-directions implying an extended cloud lifetime.}
\label{Fig802_Density}
\end{figure*}

\begin{figure}
\centering
\includegraphics[width=\linewidth]{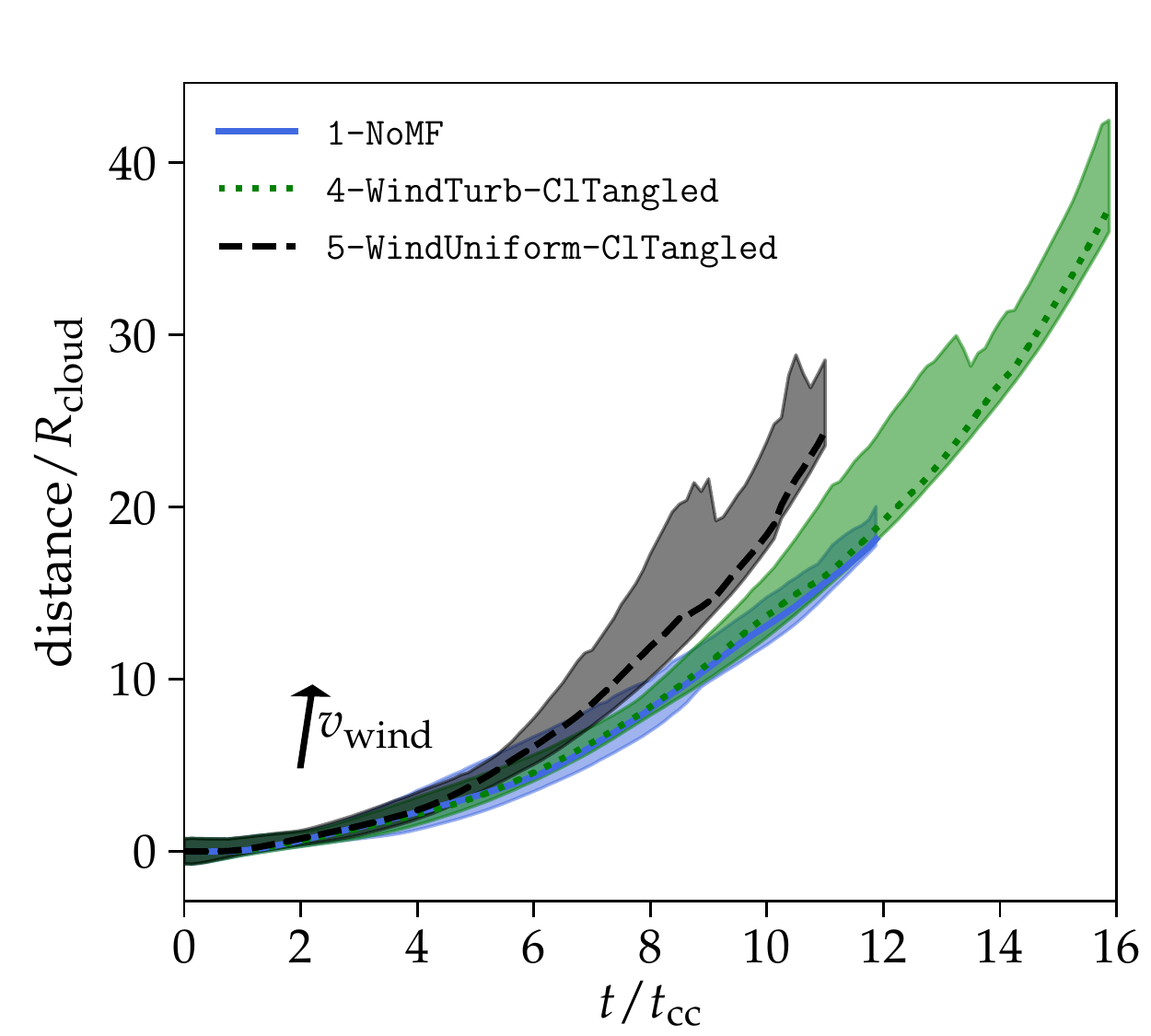}
\caption{For the dense gas, with a density above a third of the initial cloud density, we compute the median distance travelled and the 5--95 percentiles as a function of time. We only show times where the mass of dense gas is more than 2 per cent of the initial cloud mass. Until $11t_\text{cc}$ the dense material is accelerated at a similar rate in the simulation with a turbulent wind ({\tt 4}) and the simulation without magnetic field ({\tt 1}). After this time all gas is, however, evaporated in the latter simulation, whereas dense gas survives in the MHD simulation and it is continuously accelerated. At around $7 t_\text{cc}$ the simulation with a uniform magnetic wind ({\tt 5}) is accelerated more efficiently in comparison to the other two simulations shown. This occurs at the same time as an instability evolves in the $y$--$z$-plane. While experiencing this excessive acceleration, dense gas is evaporated at a fast rate.}
\label{Fig001_CloudSurvival_Distance_3D}
\end{figure}

\subsection{Magnetic field in the downstream gas}

In the wake of the cloud, the $B_y$-value, which is in the direction of the wind velocity, is enhanced in comparison to the wind's toroidal field component (see simulation {\tt 2}--{\tt 5} in Fig.~\ref{Fig801_Bx}). \citet{2016MNRAS.455.1309B} found a similar enhancement of the magnetic field in the wake of a simulated cloud interacting with a uniform magnetic wind (see their fig. 9). Such a field is present even, when the magnetic field in the cloud or wind is negligible (simulation {\tt 2} and {\tt 3}, respectively). This shows that the magnetic field in the wake of the cloud is seeded both by gas stripping from the cloud and draping of the wind. Adiabatic compression and/or shear amplification furthermore amplify the $B_y$ component of the magnetic field. Since both effects increase the $B_y$-value compared to $B_x$ and $B_z$, it is not surprising that we see a magnetic field aligned with the tail of the cloud in all simulations (except for simulation {\tt 1}, where magnetic fields are absent). Aligned magnetic fields in the tail of clouds are hence theoretically expected, when there is either a magnetic field in the cloud or wind in a cloud--wind interaction.

\subsection{Survival of dense gas}

A frequently used characteristic of a cloud's survival, is the time evolution of mass in dense gas with $n\geq n_\text{cloud}/3$. This is shown in the left panel of Fig.~\ref{Fig000_CloudSurvival_3D}. A clear trend is that the two simulations with a turbulent magnetic wind (simulation {\tt 2} and {\tt 4}) have an extended lifetime (for example measured by the time, where 50 or 75 per cent of the dense gas mass is evaporated) in comparison to the other simulations. This is a direct consequence of magnetic draping, which suppresses the fast growth of the Kelvin-Helmholtz instability \citep{2007ApJ...670..221D,2008ApJ...677..993D}.

Interestingly, the simulation with a uniform magnetic wind (simulation {\tt 5}) does not have an extended lifetime compared to the simulations without a magnetised wind. Intuitively, this would be expected, because draping of the $B_x$-component occurs. To shed more light on how gas is evaporated and accelerated we show the dense gas mass as a function of the median distance travelled by the same dense gas reservoir (Fig.~\ref{Fig000_CloudSurvival_3D}, right panel). Here we clearly see the effect of draping; a larger amount of dense gas survives being moved downstream in simulation {\tt 5} in comparison to the simulations without a magnetised wind ({\tt 1}, {\tt 3}). The simulations with a turbulent magnetic wind ({\tt 2} and {\tt 4}) maintain even larger survival mass fractions downstream in comparison to the case with a uniform magnetic wind. This is expected because of draping occurring in two dimensions.

To visualise how instabilities occur in the various simulations we plot density and temperature slices in Fig.~\ref{Fig802_Density}. For simulation {\tt 5} an instability shatters the cloud to smaller fragments in the $y$--$z$ plane, but the cloud appears stable in the $x$--$y$ plane. This is because draping only protects the cloud against instabilities in the plane of the magnetic field \citep{2007ApJ...670..221D}. Neither of the other simulations show a qualitative difference between the cloud's behaviour in the two planes. We note that turbulent signatures from the Kelvin--Helmholtz instability are visible downstream from the cloud in the simulations without a magnetic wind. In the simulations with a turbulent magnetic wind this instability is suppressed.

We note that a magnetic field also suppresses the Kelvin--Helmholtz instability (in comparison to purely hydrodynamical simulations) in the context of cylindrical cold streams \citep{2019MNRAS.489.3368B}. Our result is also consistent with \citet{2020arXiv200807537V}, who found that a magnetic field reduces gas mixing in the circumgalactic medium of Milky Way mass galaxies.

The presence of a magnetic field within the cloud plays a minor role in extending a cloud's lifetime. This can be seen by comparing {\tt 1-NoMF} and {\tt 3-WindNone-CloudTangled}, which have an almost identical behaviour in both of the panels in Fig.~\ref{Fig000_CloudSurvival_3D}. This result is consistent with \citet{2015MNRAS.449....2M}, where the wind's magnetic field was found to be more important than that of the cloud.

In Fig.~\ref{Fig000_CloudSurvival_3D} we check for convergence by comparing simulation {\tt 4} and {\tt 4-HR}. They show a similar evolution with the deviation being less than five per cent so that our fiducial resolution of 64 cells per cloud radius in the dense gas is sufficient to obtain convergence. We ran simulation {\tt 4-HR} until $8 t_\text{cc}$, which is sufficient to establish convergence.

\subsection{Acceleration of dense gas}

\begin{figure*}
\centering
\includegraphics[width=\linewidth]{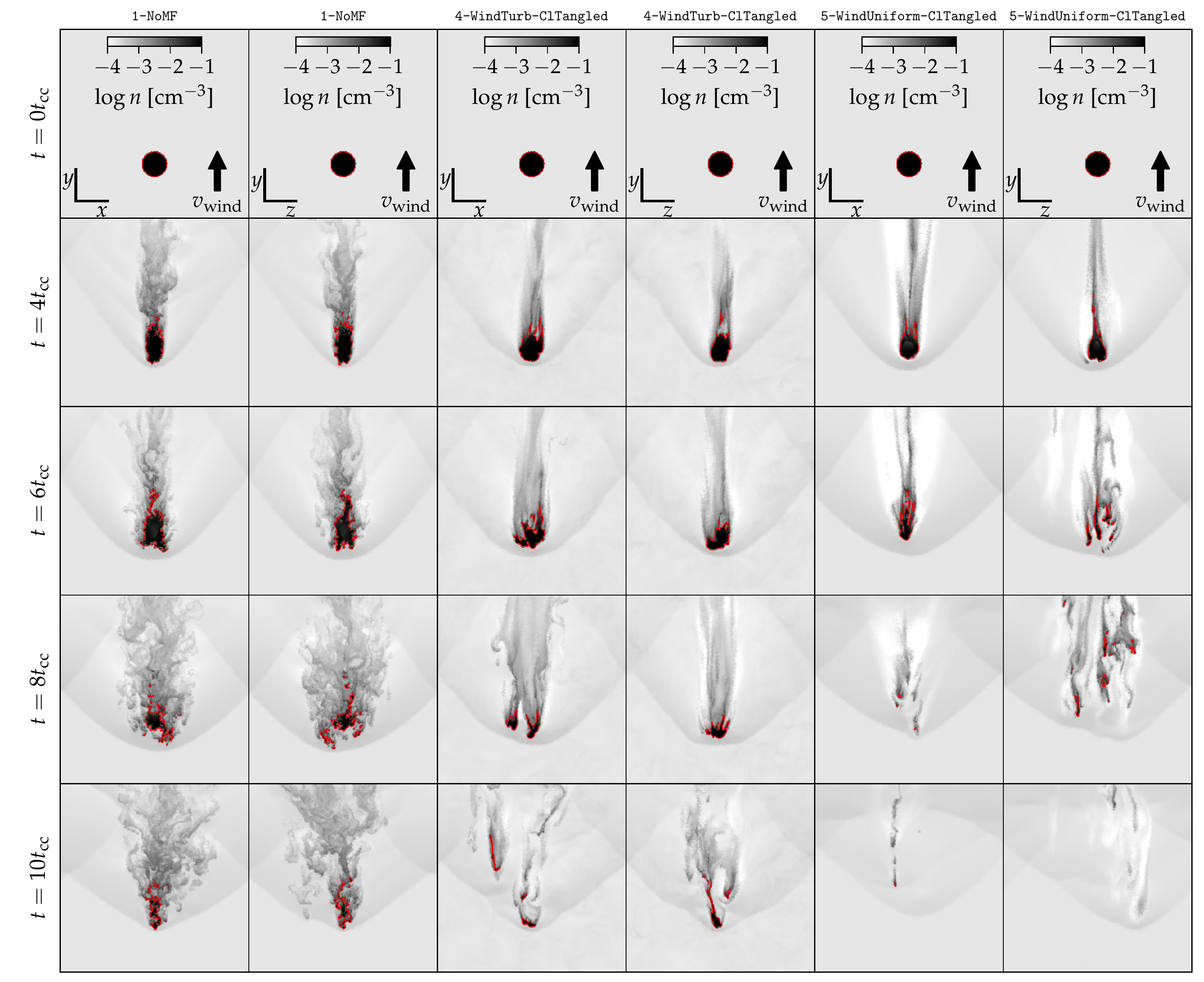}
\caption{Density slices for simulation {\tt 1}, {\tt 4} and {\tt 5} visualising the destruction and acceleration of clouds in Fig.~\ref{Fig001_CloudSurvival_Distance_3D}. Simulation {\tt 5} shatters in the $y$--$z$-projection at $t\gtrsim 6t_\text{cc}$ and is subsequently evaporated, but at the same time the dense fragments are accelerated efficiently (see Fig.~\ref{Fig001_CloudSurvival_Distance_3D}). The initial fragmentation occurs in the $y$--$z$-plane, because instabilities in the $x$--$y$-plane are suppressed by magnetic draping. For simulation {\tt 1} and {\tt 4} fragmentation occurs at later times.
}
\label{Fig804_Density}
\end{figure*}

To study the cloud's velocity evolution we plot the median and 5--95 percentile of the distance travelled for the dense gas as a function of time in Fig.~\ref{Fig001_CloudSurvival_Distance_3D}. We focus on simulation {\tt 1}, {\tt 4} and {\tt 5} to avoid too many lines in the plot. For each simulation the evolution is shown until 98 per cent of the dense gas is evaporated. The simulations with a turbulent wind and no magnetic field ({\tt 4} and {\tt 1}, respectively) follow a similar evolution until the cloud is evaporated in the simulation without magnetic fields. From $7 t_\text{cc}$ to $11t_\text{cc}$ the simulation with a uniform wind ({\tt 5}) experiences excessive acceleration compared to the two other simulations. This is because the dense cloud fragments into smaller sub-clouds, which are efficiently accelerated, but also subsequently destroyed. This process is visualised in the sliced density plots in Fig.~\ref{Fig804_Density}, where it can be seen that simulation {\tt 5} fragments in the $y$--$z$-projection, where draping is not able to protect the cloud against instabilities. The two other simulations are able to resist fragmentation beyond $7t_\text{cc}$.

In \citet{2019MNRAS.482.5401S} we studied 2D and 3D simulations with radiative cooling (but without a magnetic field). By comparing 2D and 3D simulations we found that some instabilities are suppressed in 2D, because instabilities are unable to develop along the $z$-axis (our simulations were carried out in the $x$--$y$ plane). This is comparable to what we see in simulation {\tt 5}, where draping protects against instabilities along the direction of the wind's magnetic field. In \citet{2019MNRAS.482.5401S} we did, however, also find that instabilities in the $x$--$y$-plane can grow faster in 2D compared to 3D, because the wind cannot move around the gas along the $z$-direction. This may explain why the simulation with a uniform wind ({\tt 5}) is both destroyed and accelerated faster than the simulations without a magnetic field included (simulation {\tt 1}).

In simulations with a uniformly magnetised wind \citet{2015MNRAS.449....2M} also identified such a fragmentation, and an efficient acceleration of the fragments. Our simulation {\tt 5} is qualitatively consistent with theirs, but note that we use a different density contrast and Mach number, so quantitative agreement is not expected.

Going back to Fig.~\ref{Fig001_CloudSurvival_Distance_3D}, we see a larger scatter in distance (measured by the plotted 5--95 percentile) near the end in simulations with a magnetic wind, in comparison to the hydrodynamical simulation. The reason is that a magnetic field in the wind protects those dense gas fragments against fast fragmentation that meet the conditions for draping a (small-scale) turbulent magnetic field and accelerates them in the downstream by the magnetic tension force while others encounter a (larger-scale) mostly homogeneous magnetic field so that the Kelvin-Helmholtz instability can act perpendicular to the direction of this magnetic field. For this particular setup, the gas is nevertheless evaporated before being accelerated to the hot wind's velocity (the arrow in the figure marks the slope corresponding to the hot wind's injection velocity). The following section (Sect.~\ref{GrowthRegime}) explores a different regime, where clouds grow rather than being destroyed, such that the gas survive being accelerated to the wind velocity.

\begin{table}
\centering
\begin{tabular}{HclcH}
\hline\hline
\# (0) & $\chi\equiv \rho_\text{cloud}/\rho_\text{wind}$  &  $R_\text{cloud}/$pc & Regime \\
\hline\hline
&&$ \bs{M = 0.5}$&\\
\hline\hline
{\tt $\chi$100-$\zeta$0.01-R1500} &100  & $6.41 \times 10^{-1}$& destruction\\
{\tt $\chi$100-$\zeta$0.03-R470} &100  & $1.92\times 10^{0}$ & growth \\
{\tt $\chi$100-$\zeta$0.1-R150} &100  & $5.77\times 10^{0}$ & growth \\
{\tt $\chi$100-$\zeta$0.3-R47} &100  & $1.73 \times 10^{1}$ & growth \\
{\tt $\chi$100-$\zeta$1-R15} &100  & $5.19\times 10^{1}$ & growth \\
\hline
{\tt $\chi$1000-$\zeta$0.01-R19000} &1000  & $2.37\times 10^{2}$ & destruction\\
{\tt $\chi$1000-$\zeta$0.03-R6000} &1000  & $7.11\times 10^{2}$ & growth\\
{\tt $\chi$1000-$\zeta$0.1-R1900} &1000  & $2.13\times 10^{3}$  & growth\\
{\tt $\chi$1000-$\zeta$1-R190} &1000  & $6.40\times 10^{3}$ & growth \\
{\tt $\chi$1000-$\zeta$10-R19} &1000  & $1.92\times 10^{4}$ &  growth \\
\hline\hline
&&$ \bs{M = 1.5}$&\\
\hline\hline
{\tt $\chi$100-$\zeta$10-R1.5} &100  & $1.50\times 10^{0}$ & destruction \\
{\tt $\chi$100-$\zeta$1-R15} &100  & $1.50\times 10^{1}$ & destruction \\
{\tt $\chi$100-$\zeta$0.3-R47} &100  & $4.74\times 10^{1}$ & destruction \\
{\tt $\chi$100-$\zeta$0.1-R150} &100  & $1.50\times 10^{2}$ & growth \\
{\tt $\chi$100-$\zeta$0.03-R470} &100  & $4.74\times 10^{2}$ & runaway cool.\\
{\tt $\chi$100-$\zeta$0.01-R1500} &100  & $1.50\times 10^{3}$& runaway cool.\\
\hline
{\tt $\chi$333-R5} &333  & $5.00\times 10^{0}$ & destruction \\
{\tt $\chi$333-R19} &333  & $5.00\times 10^{1}$ & destruction \\
{\tt $\chi$333-R190} &333  & $1.58\times 10^{2}$ & destruction \\
{\tt $\chi$333-R1900} &333  & $5.00\times 10^{2}$ & destruction \\
{\tt $\chi$333-R6000} &333  & $1.58\times 10^{3}$ & growth\\
{\tt $\chi$333-R19000} &333  & $5.00\times 10^{3}$ & growth\\
\hline
{\tt $\chi$1000-$\zeta$10-R19} &1000  & $1.90\times 10^{1}$ & destruction \\
{\tt $\chi$1000-$\zeta$1-R190} &1000  & $1.90\times 10^{2}$ & destruction \\
{\tt $\chi$1000-$\zeta$0.1-R1900} &1000  & $1.90\times 10^{3}$ & destruction \\
{\tt $\chi$1000-$\zeta$0.03-R6000} &1000  & $6.00\times 10^{3}$ & destruction\\
{\tt $\chi$1000-$\zeta$0.01-R19000} &1000  & $1.90\times 10^{4}$ & growth\\

\hline\hline
&& $\bs{ M = 4.5}$&\\
\hline\hline
{\tt $\chi$100-$\zeta$0.01-R1500} &100  & $1.13\times 10^{1}$& destruction \\
{\tt $\chi$100-$\zeta$0.03-R470} &100  & $3.40\times 10^{1}$ & destruction \\
{\tt $\chi$100-$\zeta$0.1-R150} &100  & $1.02\times 10^{2}$ & destruction \\
{\tt $\chi$100-$\zeta$0.3-R47} &100  & $3.06\times 10^{2}$ & destruction \\
{\tt $\chi$100-$\zeta$1-R15} &100  & $9.19 \times 10^{2}$ & growth \\
\hline
{\tt $\chi$1000-$\zeta$0.01-R19000} &1000  & $4.19 \times 10^{3}$ & destruction\\
{\tt $\chi$1000-$\zeta$0.03-R6000} &1000  & $1.26\times 10^{4}$ & destruction\\
{\tt $\chi$1000-$\zeta$0.1-R1900} &1000  & $3.77\times 10^{4}$ & destruction \\
{\tt $\chi$1000-$\zeta$1-R190} &1000  & $1.13\times 10^{5}$ & growth \\
\hline\hline
\end{tabular}
\caption{An overview of the simulations analysed in Sect.~\ref{GrowthRegime}. These simulations have a cloud temperature and density of $T_\text{cloud}=10^4$ K and $n_\text{cloud}=0.1$ cm$^{-3}$, respectively. The hot wind has a density and temperature of $n_\text{cloud}/\chi$ and $\chi T_\text{cloud}$, respectively. The cloud radius is shown in column 2, and column 3 shows whether a simulation reveals the cloud to be in the destruction, growth or runaway cooling regime (see text for details).}
\label{Table:SimulationOverviewB}
\end{table}

\section{The cloud growth regime}\label{GrowthRegime}

\subsection{The criterion for cloud growth}\label{GrowthRegimeCriterion}

For a sufficiently large cloud radius, we encounter a different regime in which the cloud mass increases with time, instead of experiencing destruction. This was e.g. demonstrated in 2D simulations of \citet{2017MNRAS.470..114A}. The instabilities disrupting a cloud do, however, work differently in 2D and 3D \citep{2019MNRAS.482.5401S}, since some instability modes are by construction suppressed in 2D simulations. Recently the criterion for whether clouds grow or dissolve in 3D simulations has been studied by \citet{2018MNRAS.480L.111G} and \citet{2019arXiv190902632L}. Essential for either criterion is the cooling time-scale,
\begin{align}
t_\text{cool}\equiv \frac{3 n k_\text{B} T  }{2n_\text{H}^2 \Lambda},\label{tcooleq}
\end{align}
where $\Lambda=\Lambda(n_\text{H},T,Z)$ is the cooling function usually measured in units of erg cm$^3$ s$^{-1}$. We here summarise the two criteria for cloud growth.

\subsubsection{Criterion based on the wind cooling time}

\citet{2019arXiv190902632L} show that a cloud grows mass from the hot wind via mixing provided the cooling time of the hot gas is smaller than the predicted cloud survival time-scale:
\begin{align}
t_\text{cool,wind} < 10 t_\text{cc} \tilde{f}, \label{LiCritEq}
\end{align}
where the scale dependent behaviour, mainly caused by cooling and conduction, is parametrised as,
\begin{align}
\tilde{f}\equiv (0.9 \pm 0.1) \left(\frac{2R_\text{cloud}}{\text{1 pc}}\right)^{0.3} \left( \frac{n_\text{wind}}{0.01 \text{ cm}^{-3}}\right)^{0.3} \left( \frac{\varv_\text{wind}}{100 \text{ km s}^{-1}}\right)^{0.6}.\label{fudge}
\end{align}
The right hand side of Eq.~\eqref{LiCritEq} is determined by simulations of clouds in the destruction regime.

\subsubsection{Criterion based on the mixed gas cooling time}

Instead, \citet{2018MNRAS.480L.111G} have proposed a different criterion that involves the cooling time-scale of the mixed gas, rather than the hot gas. They estimate the temperature of the stripped cold gas that mixes with the surrounding hot gas, as
\begin{align}
T_\text{mix} \equiv \sqrt{T_\text{wind}T_\text{cloud}}.
\end{align}
Under the assumption that the cold, mixed and hot phases are in pressure equilibrium, we can associate a density of $n_\text{wind}\times T_\text{wind}/ T_\text{mix}$ with the mixed phase. This makes it possible to determine the cooling time-scale of the mixed gas ($t_\text{cool,mix}$) based on Eq.~\eqref{tcooleq}. 

\citet{2018MNRAS.480L.111G} derive a criterion for cloud growth by requiring the mixed gas to be able to cool faster than the time-scale of the hydrodynamical destruction of the cloud,
\begin{align}
t_\text{cool,mix} < t_\text{cc}. \label{GronkeCritEq}
\end{align}
This criterion has been obtained by hydrodynamical and MHD simulations exploring an extensive set of cloud and wind properties \citep{2019MNRAS.tmp.2995G}. The most fundamental difference between Eqs.~\eqref{LiCritEq} and \eqref{GronkeCritEq} is whether it is the cooling time of the hot or of the mixed gas that is relevant for cloud survival.

\subsection{Testing the cloud growth criterion in simulations}

To test the two criteria for cloud growth presented above we create a set of simulations with Mach numbers, cloud radii and $\chi$-values, as shown in Table~\ref{Table:SimulationOverviewB}. We first ran a set of simulations with a Mach number of $M=1.5$ designed to test whether Eq.~\eqref{LiCritEq} or Eq.~\eqref{GronkeCritEq} best describe the transition between the growth and the destruction regime. We subsequently ran additional simulations with $M=0.5$ and $M=4.5$ to test the Mach number dependence of the results.

\subsubsection{Simulation details}

As in previous sections we use $T_\text{cloud}=10^4$ K, $n_\text{cloud}=0.1$ cm$^{-3}$, and a solar metallicity for the gas cells. We continue to use a temperature floor of $5\times 10^3$ K. We use a turbulent magnetic field in the wind and a tangled magnetic field in the clouds (we use setup \texttt{4} from Table~\ref{Table:SimulationOverview}), both initialised with $\beta = 10$.

The simulations of this section are performed at a lower resolution than in the previous sections. Such a trade-off is necessary, because we need extremely large box sizes to resolve the mixed downstream gas. For $M=0.5$ and $1.5$ we use a box size of $L_x,L_y,L_z = 16 R_\text{cloud},384 R_\text{cloud},16 R_\text{cloud}$, which is sufficient for capturing the growth of the mixed gas. For $M=4.5$ we increase $L_y$ by a factor of 3, which is necessary to avoid dense gas leaving the simulation box during a simulation. Our resolution is 7 and 15 cells per cloud radius for the simulations with $\chi=100$ and $\chi=1000$, respectively. In Sect.~\ref{subsubconvergence} and Appendix \ref{AppendixConvergence} we discuss convergence of our results, and conclude that our resolution is high enough to resolve whether clouds are in the growth or destruction regime.

\begin{figure*}
\centering
\begin{minipage}{.32\linewidth}
\includegraphics[width=1.0\linewidth]{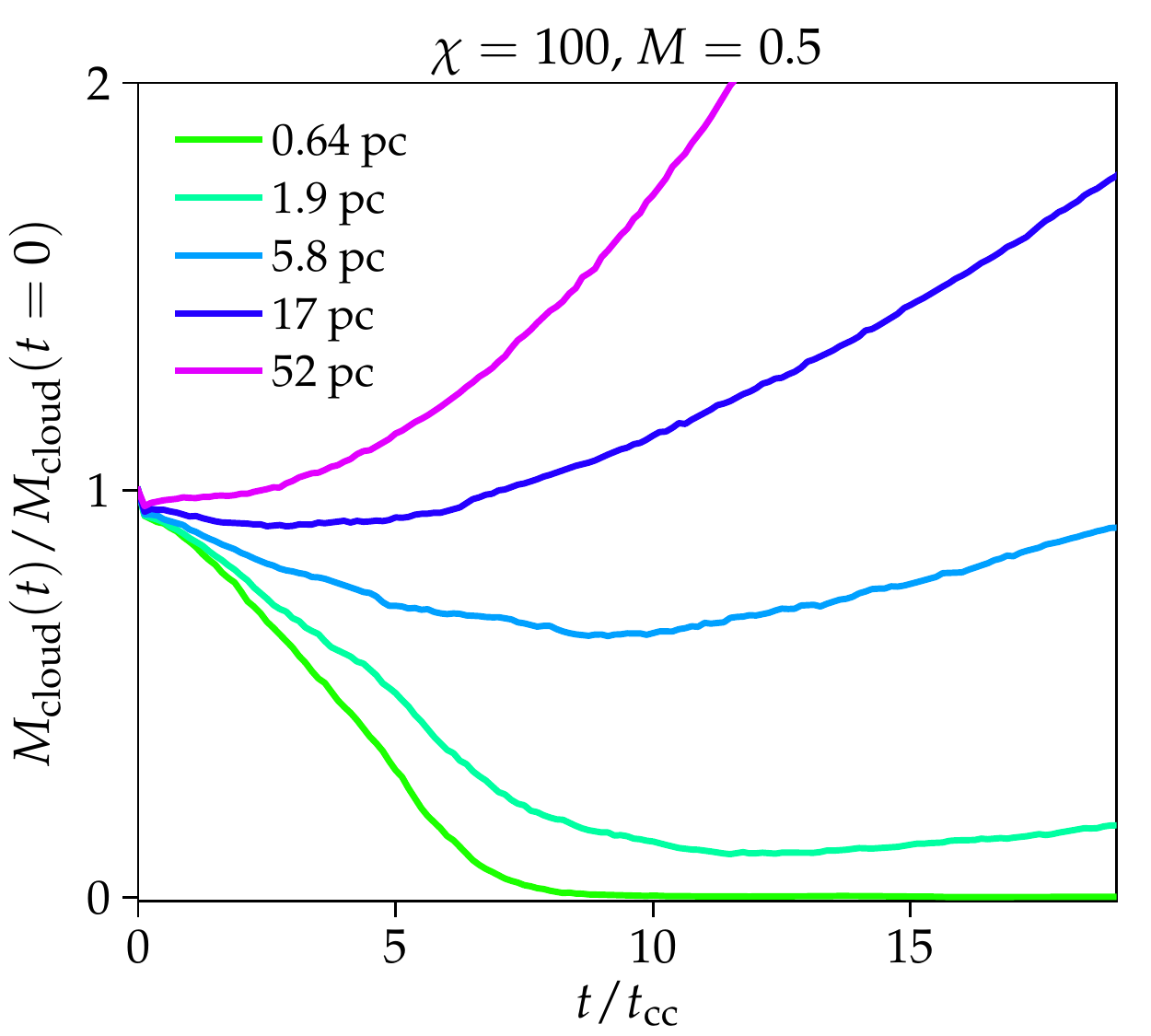}
\end{minipage}
\begin{minipage}{.32\linewidth}
\includegraphics[width=1.0\linewidth]{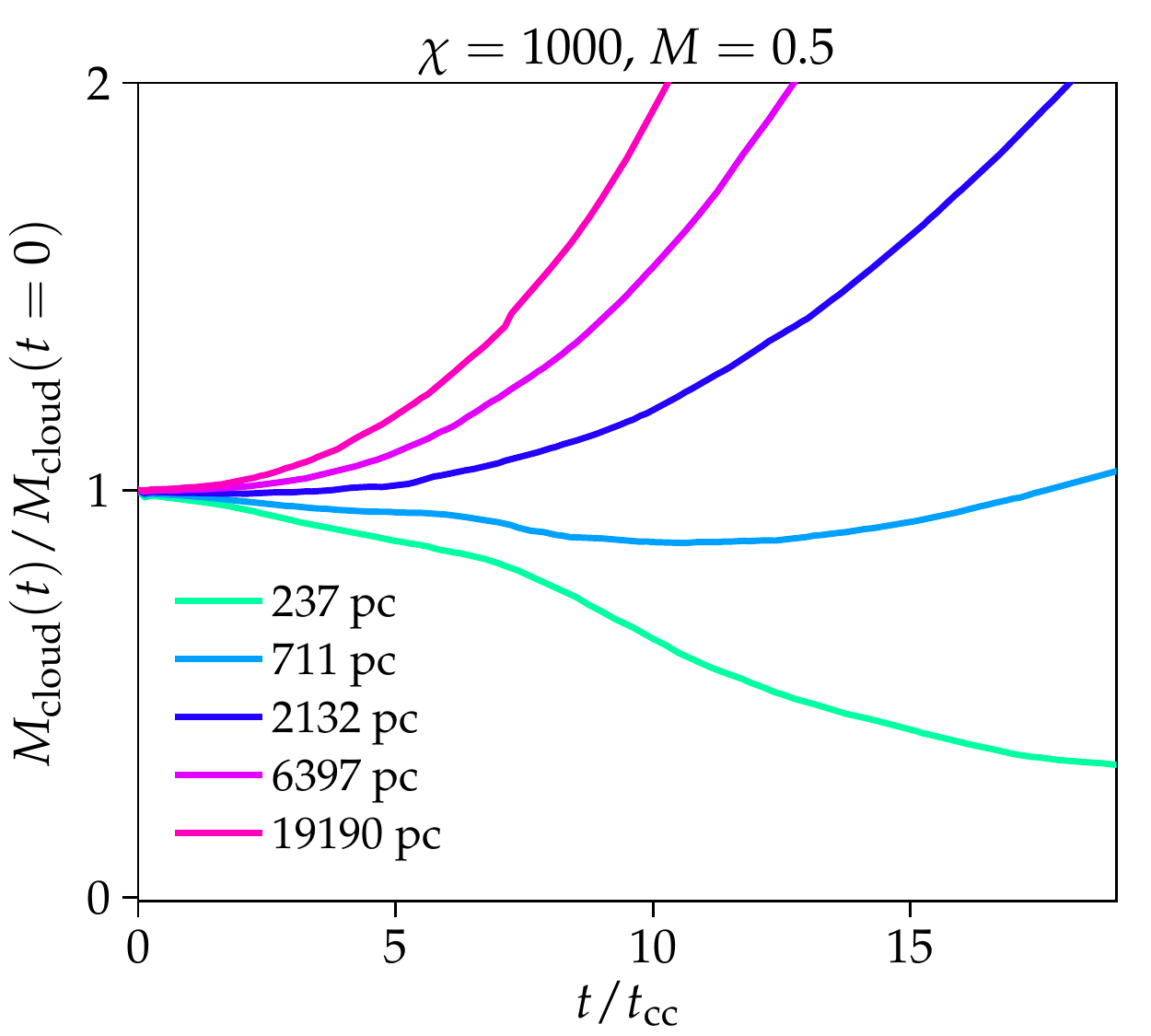}
\end{minipage}
\begin{minipage}{.32\linewidth}
\includegraphics[width=1.0\linewidth]{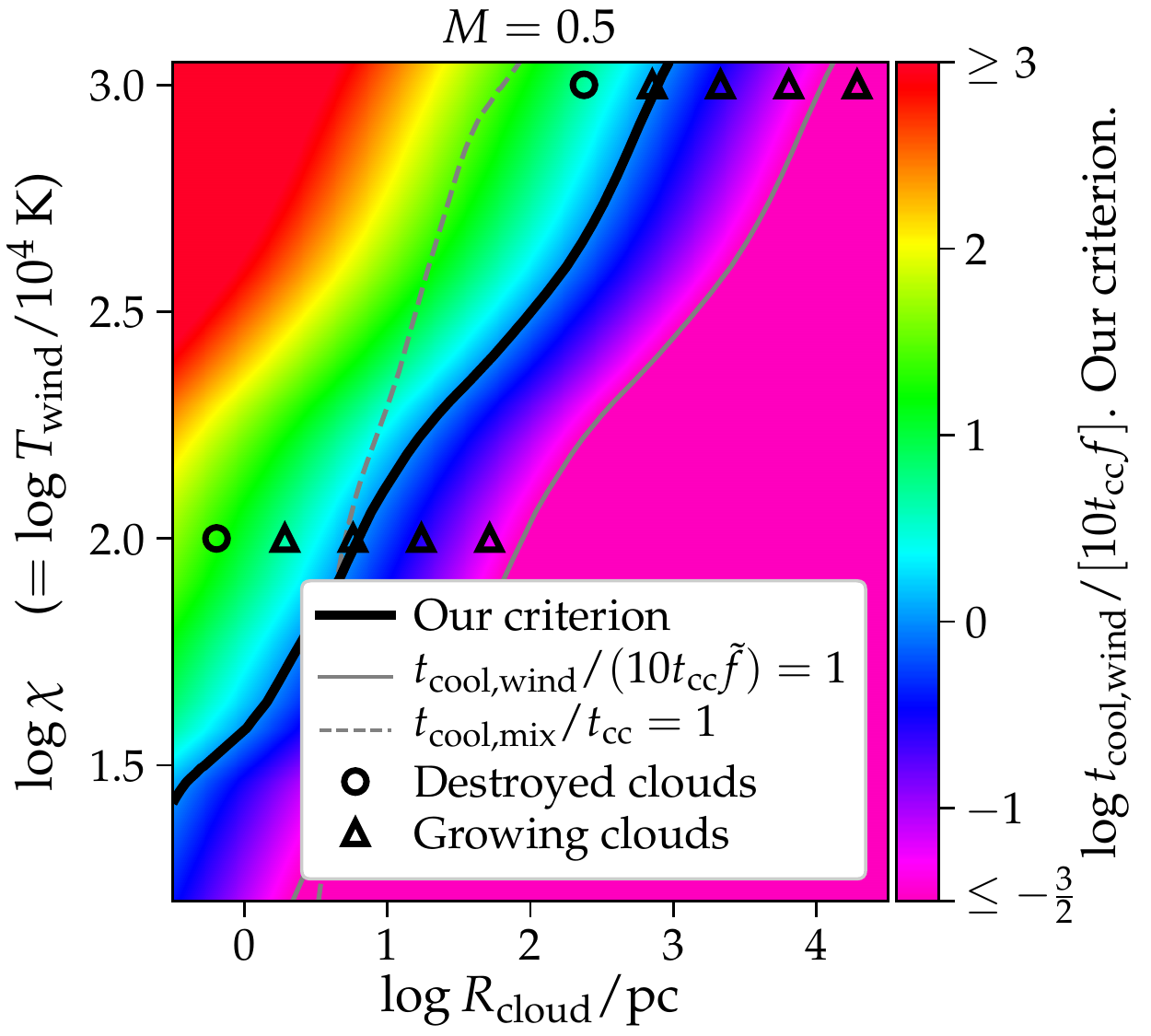}
\end{minipage}
\begin{minipage}{.32\linewidth}
\includegraphics[width=1.0\linewidth]{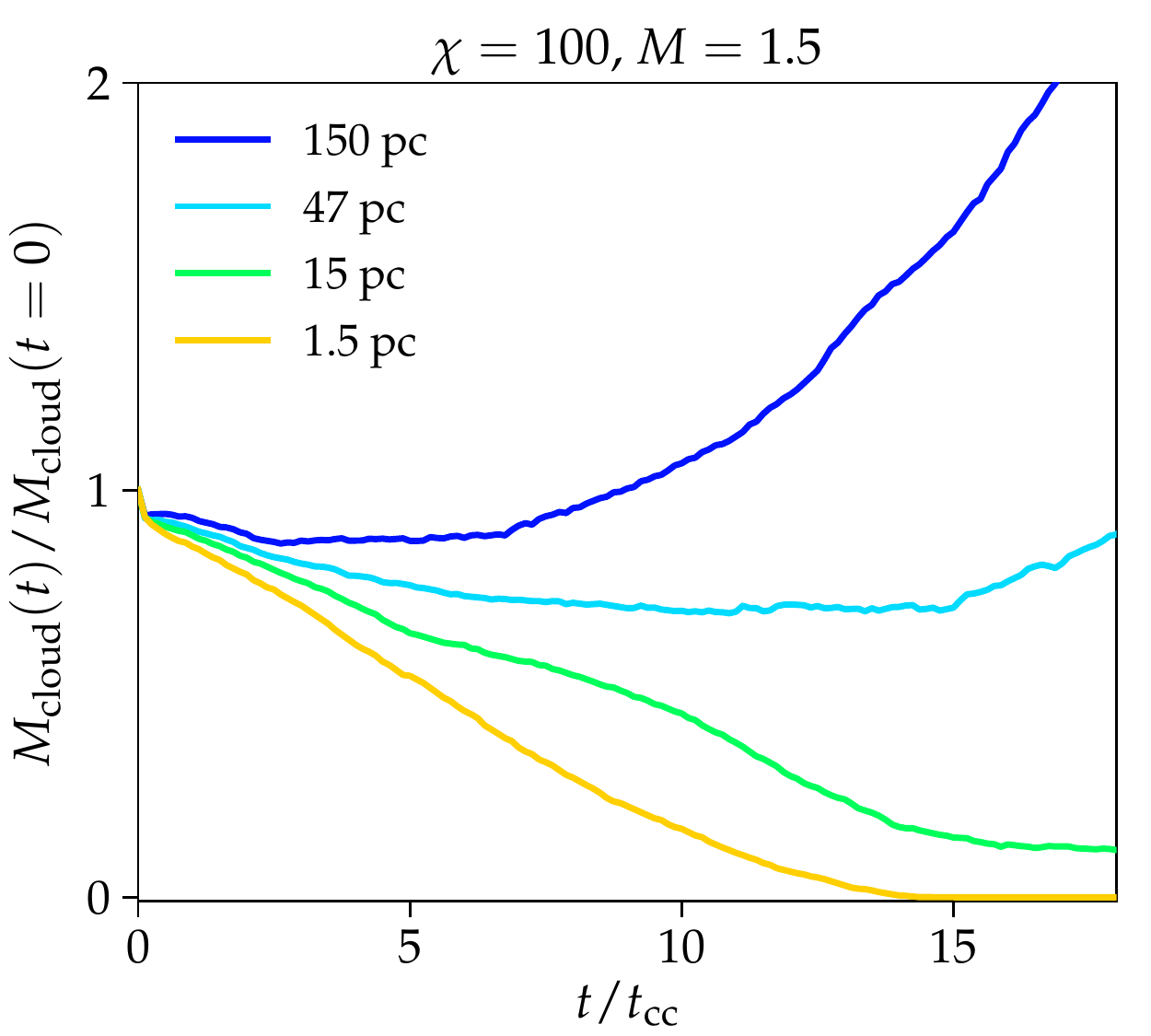}
\end{minipage}
\begin{minipage}{.32\linewidth}
\includegraphics[width=1.0\linewidth]{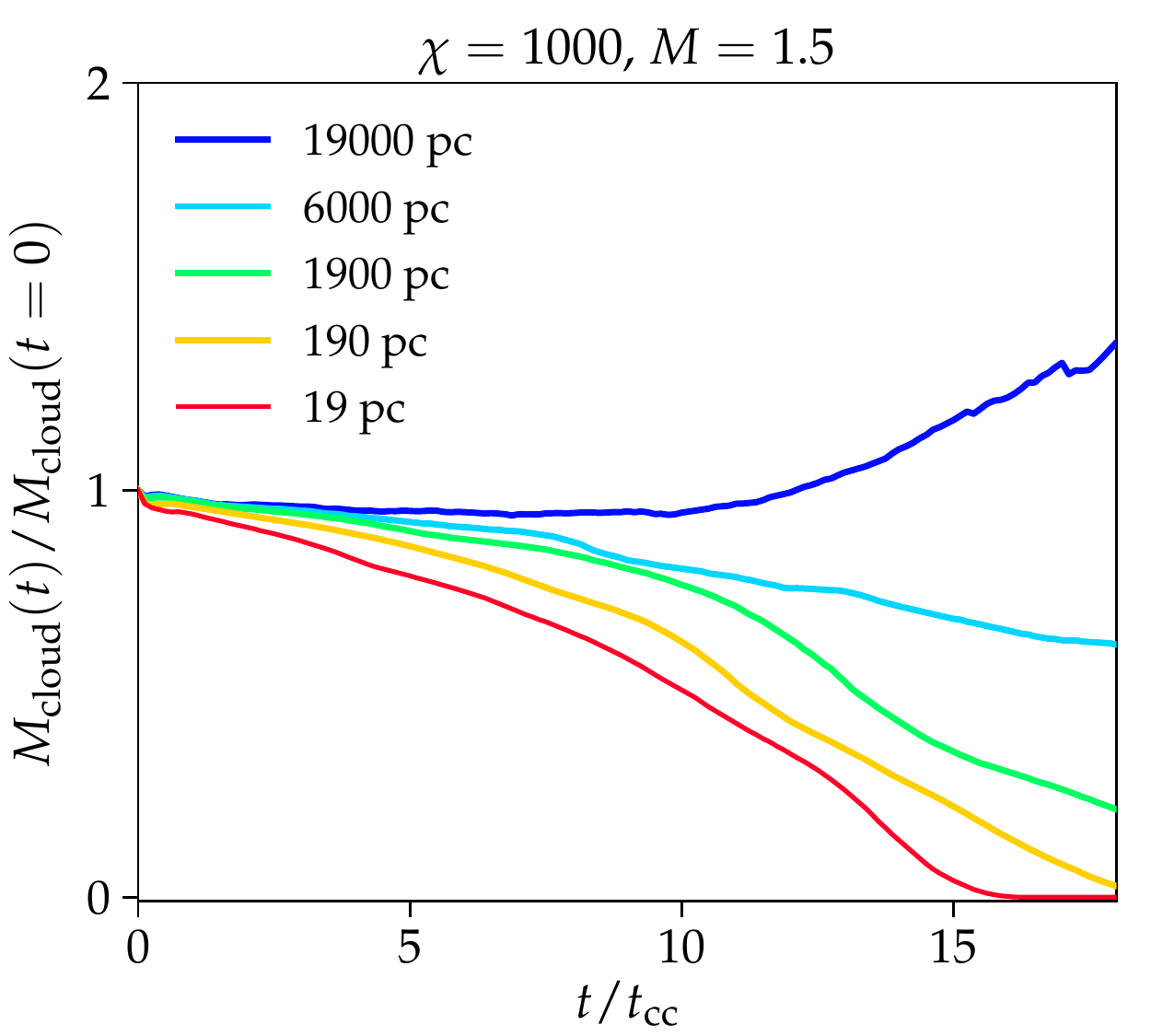}
\end{minipage}
\begin{minipage}{.32\linewidth}
\includegraphics[width=1.0\linewidth]{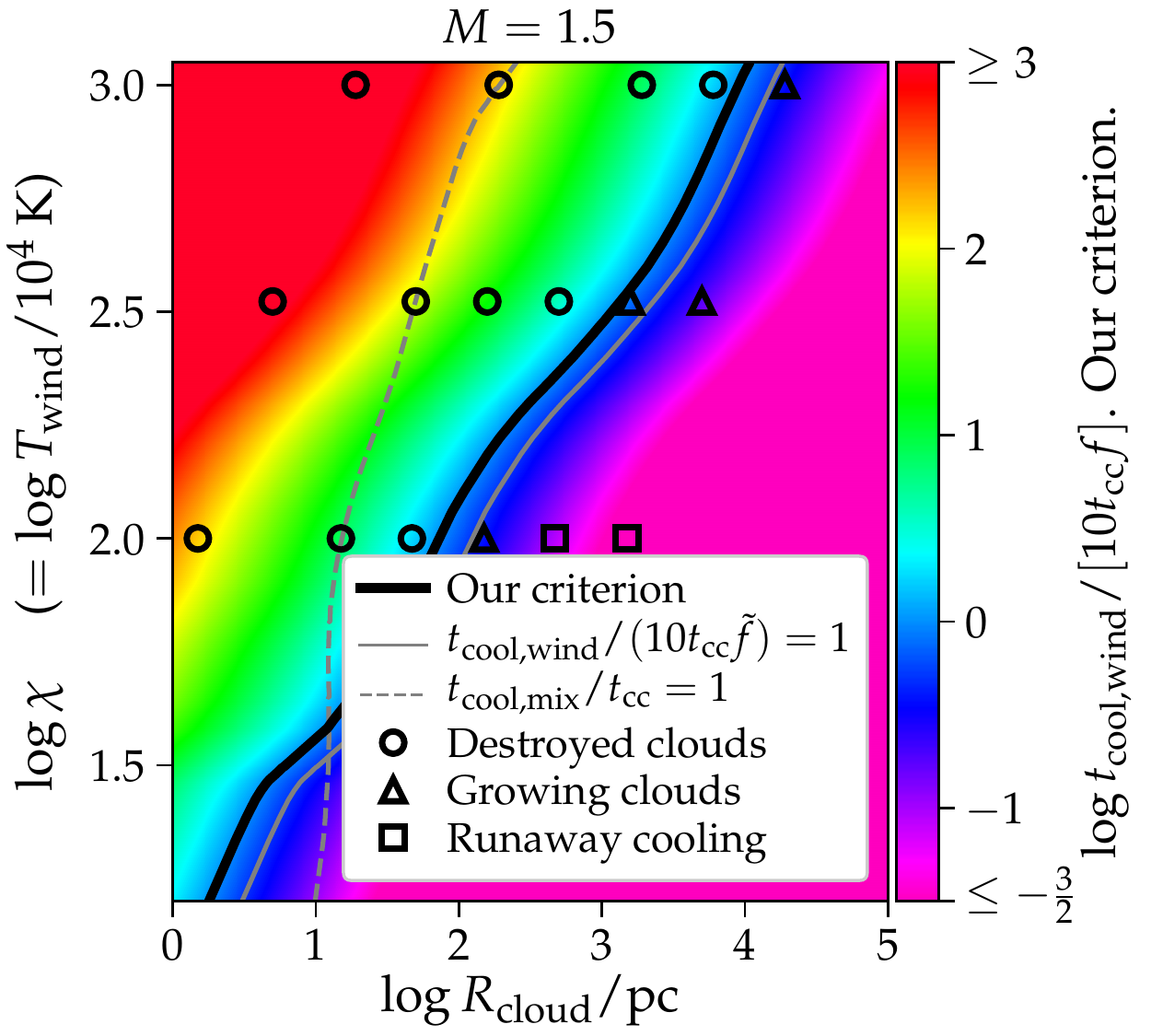}
\end{minipage}
\begin{minipage}{.32\linewidth}
\includegraphics[width=1.0\linewidth]{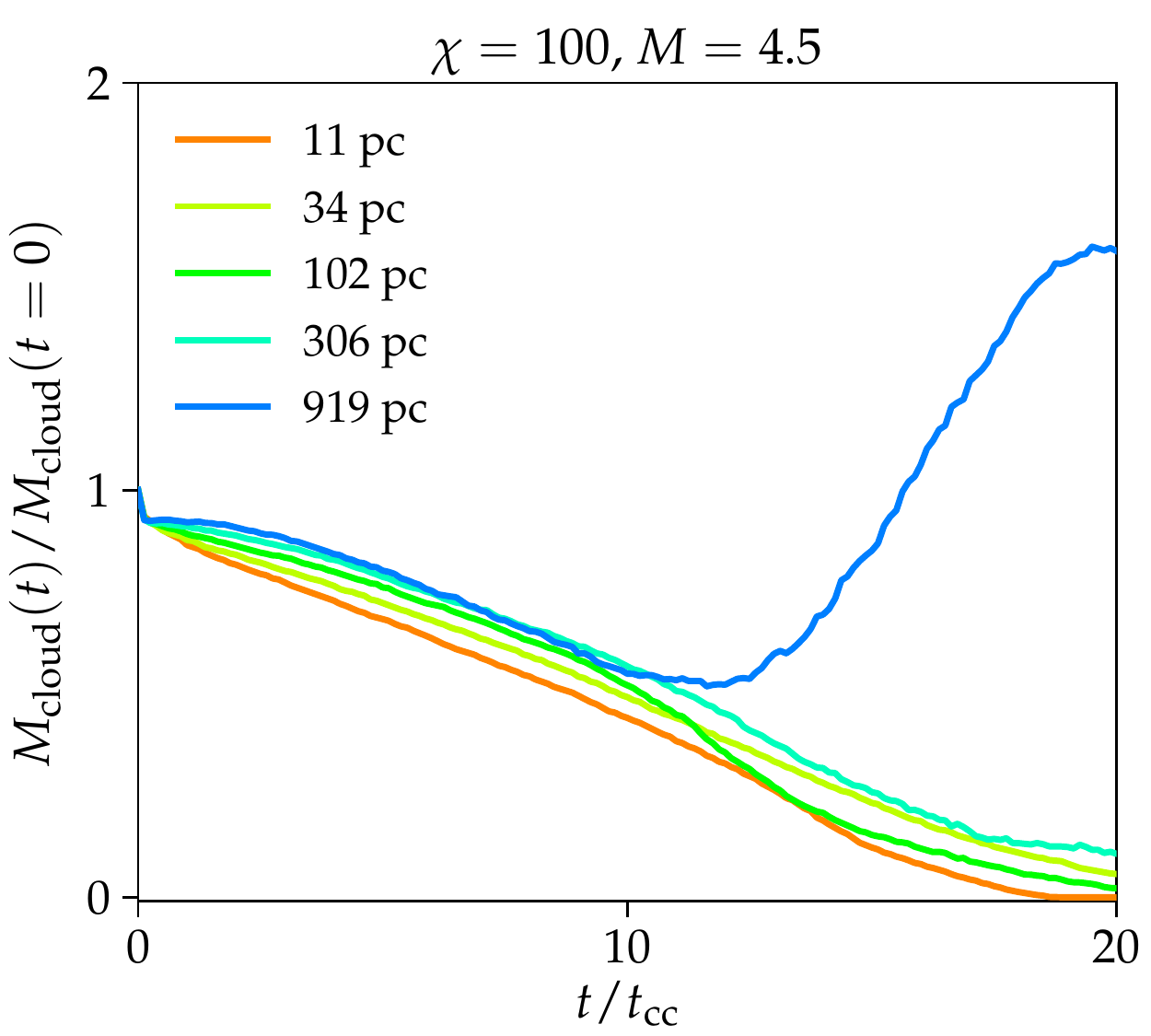}
\end{minipage}
\begin{minipage}{.32\linewidth}
\includegraphics[width=1.0\linewidth]{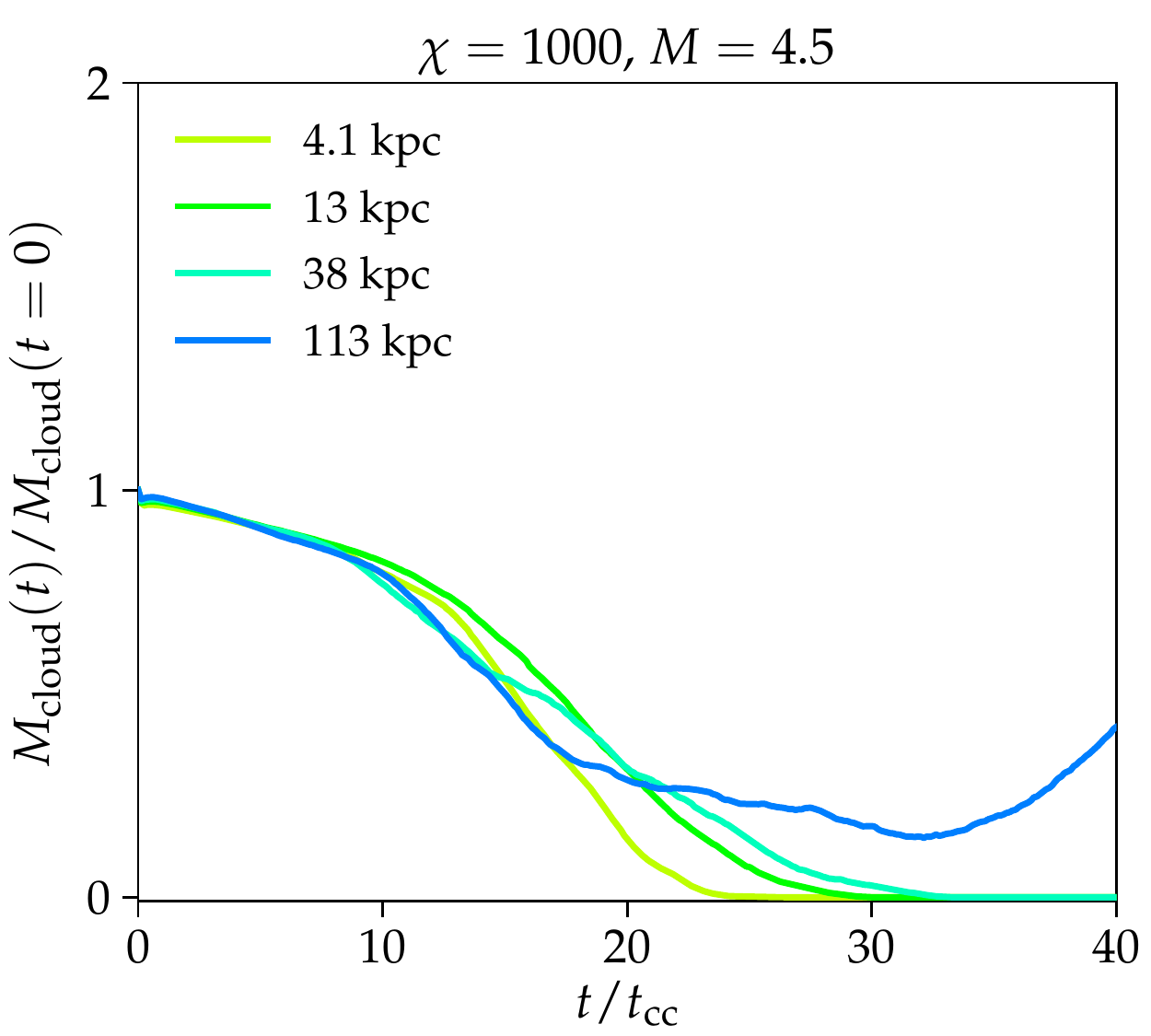}
\end{minipage}
\begin{minipage}{.32\linewidth}
\includegraphics[width=1.0\linewidth]{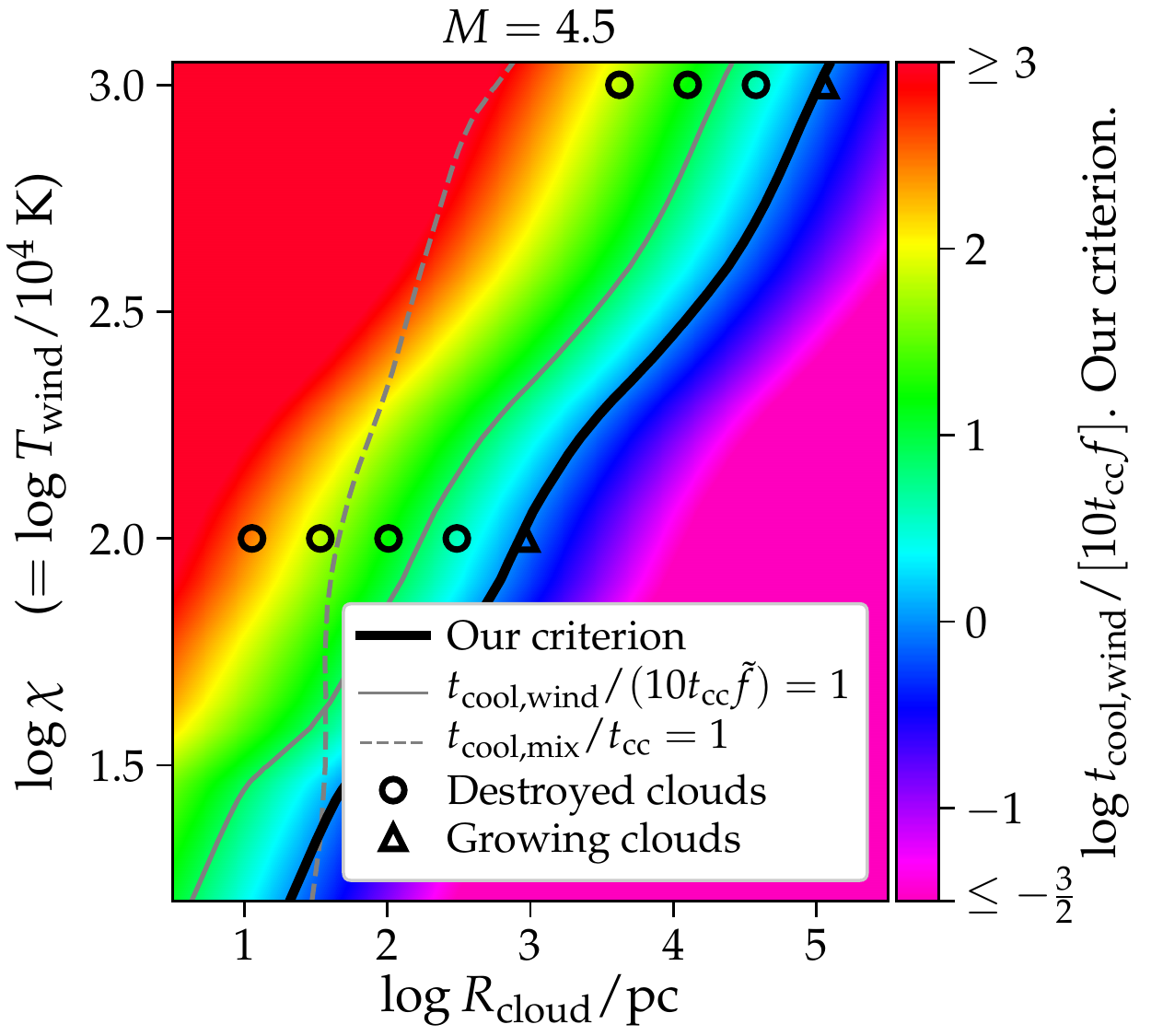}
\end{minipage}
\caption{With our simulations we test three criteria for the transition between the destruction and growth regime. We perform simulations with $M=0.5$, $1.5$ and $4.5$ (upper, middle and lower panels, respectively).  We show simulations with $\chi = 100$ (left panels) and $\chi=1000$ (central panels), and in the right panels we summarise whether a simulation is in the growth or destruction regime. Our favoured criterion (thick solid lines in the right panels) nicely separates destroyed from growing clouds. The only remarkable exception is the simulation with $M=0.5$, $\chi=100$ and $R_\text{cloud}=1.9$ pc, which is growing despite of showing a radius three times smaller than our predicted transition. We do, however, expect some scatter around the transition, so we do not regard this as a problem for our criterion. We conclude that our criterion well captures whether a simulation is in the destruction or growth regime.}
\label{MachNumber0.5}
\label{MachNumber1.5}
\label{MachNumber4.5}
\end{figure*}

\subsubsection{Defining the dense gas}

In the literature of cloud crushing simulations different criteria are used to define the dense gas. In the previous sections we defined dense gas with $n\geq n_\text{cloud}/3$. This definition is for example also used by \citet{2018MNRAS.480L.111G}. \citet{2019arXiv190902632L} favour a criterion, where the dense gas consists of the phase denser than the geometric mean of the wind and the cloud density, $n\geq \sqrt{n_\text{cloud}n_\text{wind}}$. For the density contrasts studied throughout this paper, $100\leq\chi\leq 1000$, the latter definition includes gas of lower densities in comparison to the former definition.

We compare the two definitions in Appendix~\ref{MassDefinition}. We see a quantitatively different time evolution of the gas mass associated with the two criteria. If a cloud has a radius close to the critical transition for cloud growth the dense gas definition can change the regime of a cloud. But for clouds well in the destruction or growth regime it plays no role. A remarkable difference is that the evolution of the gas with $n\geq \sqrt{n_\text{cloud}n_\text{wind}}$ yields a smoother, more monotonic increase in gas mass in the simulations, where we see a growth near the end. The growth is also present when using the criterion $n\geq n_\text{cloud}/3$, but it is less monotonic.

In the remaining parts of this paper, we use $n\geq \sqrt{n_\text{cloud}n_\text{wind}}$ to define the dense gas phase. This is the most robust criterion, due to the monotonic increase (decrease) of simulations in the growth (destruction) regime. By using this criterion we are also consistent with \citet{2019arXiv190902632L}, which uses a criterion for cloud growth, which shares many similarities with our favoured criterion (see below).

\subsubsection{Testing transition criteria for cloud growth}\label{GrowthCriterion}

The mass evolution of our simulations are shown in the left and central panels of Fig.~\ref{MachNumber1.5}. To assess the criterion for cloud growth the outcome of the simulations is summarised in the right panels. It is marked whether a simulation is in the regime of cloud growth, cloud destruction or runaway cooling. As a numerical criterion for a cloud to be in the growth regime in a simulation with $M=0.5$ or $M=1.5$ we require an increase in the dense gas mass (i.e. $\dot{M}_\text{cloud}>0$) measured at $12.5 t_\text{cc}$. For $M=4.5$ the growth starts occurring at later times, so here we define a cloud to be growing based on the last two $t_\text{cc}$ shown in the panels. Looking at the figure, these criteria well match our intuition of significant growth.

The simulations with $M=1.5$ have the most complete sampling of the different regimes, so we start by characterising these. For $M=1.5$ there exists a radius, where clouds transition between a destruction and a growth regime. For $\chi=100$ and $\chi=1000$ we find clouds to be in the growth regime for $R_\text{cloud}\gtrsim 150$ pc and $R_\text{cloud}\gtrsim 19000$ pc, respectively. The mass evolution of the simulations with $M=1.5$ and $\chi=333$ is shown in detail in Appendix~\ref{AppendixChi333}, and again we find a transition radius below which clouds are in the destruction regime.

\begin{figure}
\centering
\includegraphics[width=\linewidth]{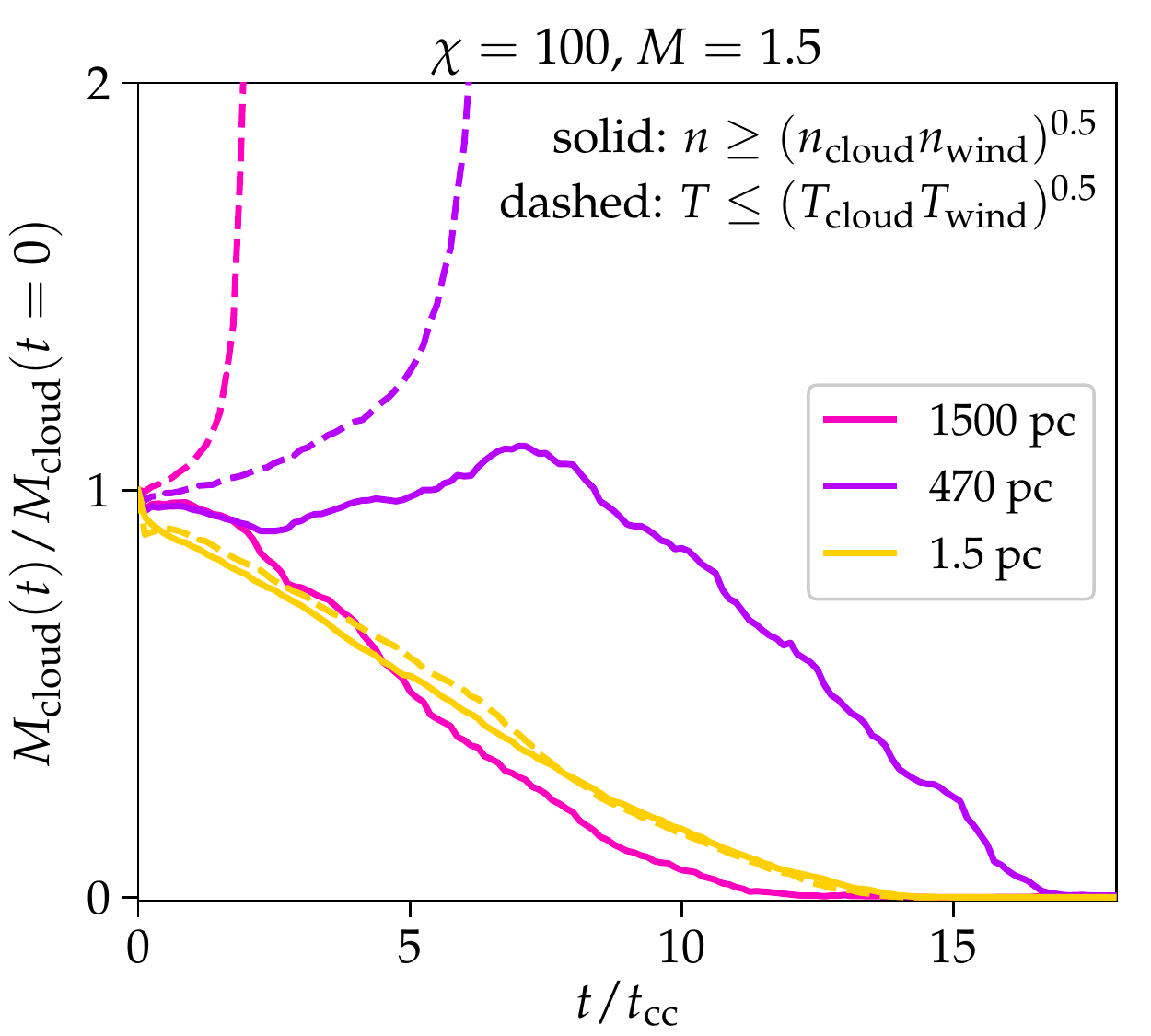}
\caption{For a wind temperature of $T_\text{wind}=10^6$ K ($\chi =100$) the simulations with $R_\text{cloud}=470$ pc and $1500$ pc undergo \emph{runaway cooling}, meaning that the wind radiates away its thermal energy on a shorter time-scale than it takes for the wind to reach the cloud from the injection region of a simulation. The gas mass with $T\leq \sqrt{T_\text{cloud}T_\text{wind}}$ increases in time (dashed lines) because the wind cools, but the mass of dense gas with $n\geq \sqrt{n_\text{cloud}n_\text{wind}}$ decreases (solid lines), because the cloud expands into a low-pressure medium. For a cloud in the destruction regime (we show the simulation with $R_\text{cloud}= 1.5$ pc) the evolution of the survival fraction is almost independent of whether a density or temperature threshold is used to define the cloud's mass.}
\label{FigX321_RunawayCooling_Chi100_3D}
\end{figure}

As summarised by \citet{2019arXiv190902632L} there also exists a regime, where the hot wind radiates away its thermal energy on a shorter time-scale than it takes for the hot wind to travel a cloud radius. Our two largest simulations with $M=1.5$ and $\chi=100$ are in this regime, as we see from the time evolution of the cloud mass in Fig.~\ref{FigX321_RunawayCooling_Chi100_3D}. The mass in dense gas decreases as a function of time, because the cloud is expanding in a low-pressure medium, whereas the mass in cold gas increases, because the wind cools to low temperatures. We simply refer to this as \emph{runaway cooling}. In our simulations this happens when the distance from the injection region to the initial coordinate of the cloud (which is $12R_\text{cloud}$) is larger than the cooling radius, $t_\text{cool,wind}/\varv_\text{wind}$, of the hot wind. 

In the right panels of Fig.~\ref{MachNumber1.5} we compare the regime of our simulations to three different curves describing different cloud growth criteria: $t_\text{cool,mix}/t_\text{cc}=1$ (from Eq.~\eqref{GronkeCritEq}, dashed grey), $t_\text{cool,wind} /[ 10 t_\text{cc} \tilde{f}]=1$ (from Eq.~\eqref{LiCritEq}, solid grey), and our own criterion (from Eq.~\eqref{OurFinalCrit}, see below). The latter criterion best describes our simulations, and it can be written as $t_\text{cool,wind} <  10 t_\text{cc} f$, where
\begin{align}
f &= 2 \left( \frac{M}{1.5}\right)^{-2.5} \tilde{f}\label{DefinitionF}\\
&= 1.8 \times \left(\frac{2R_\text{cloud}}{\text{1 pc}}\right)^{0.3} \left( \frac{M}{1.5}\right)^{-2.5}\left( \frac{n_\text{wind}}{0.01 \text{ cm}^{-3}}\right)^{0.3} \left( \frac{\varv_\text{wind}}{100 \text{ km s}^{-1}}\right)^{0.6}.
\end{align}
As can be seen from Eq.~\eqref{DefinitionF} our criterion differs from that of \citet{2019arXiv190902632L} by a factor of two, which indicates that the magnetic field extends the cloud life-time, and an additional Mach number dependence\footnote{We denote the factor entering our growth criterion by $f$, and the factor in the criterion of \citet{2019arXiv190902632L} by $\tilde{f}$.}. \citet{2019arXiv190902632L} mostly studied simulations with $M\leq 1$ (because this is the most relevant value for the circumgalactic medium of galaxies), so this is why their criterion does not reveal an explicit Mach number dependence. As a result, our criterion can be re-written as,
\begin{align}
\frac{t_\text{cool,wind}}{t_\text{cc}} = \frac{3}{2X^2\mu^2}
\frac{ k_\text{B} T_\text{wind}\varv_\text{wind}  }{R_\text{cloud}\Lambda_\text{wind}\sqrt{n_\text{wind}n_\text{cloud}}} < 10 f,\label{OurFinalCrit}
\end{align}
where $X$ is the Hydrogen mass fraction in the wind, and $\mu$ is the mean molecular weight of the gas in the wind. $t_\text{cool,wind}$ enters the criterion, because radiative cooling at a temperature slightly lower than $T_\text{wind}$ is the most time-consuming, rate-limiting step for the hot gas to cool to the cold cloud temperature. We further assess the physics of the growth criterion in Sec.~\ref{MassGrowthOrigin}.

This criterion does not only describe the critical radius for cloud growth for the simulations with $M=1.5$ well, but it also holds for $M=0.5$ and $4.5$. The exact form of $f$ as shown in Eq.~\eqref{DefinitionF} has been determined \emph{by eye} rather than by a formal fit. This approach is sufficient to divide the destroyed from the growing clouds.

We note, that our simulations confirm the result from \citet{2019arXiv190902632L} that the hot wind's cooling time-scale enters the growth criterion. This is reassuring because we use a very similar physical setup with a similar initial cloud temperature and cooling function. Our main difference from \citet{2019arXiv190902632L} is hence, that we introduce a Mach number dependence, which is required to explain our simulations with $M=0.5$ and $M=4.5$.

Our simulations poorly match the criterion involving the cooling time-scale of the mixed gas (from Eq.~\ref{GronkeCritEq}). For $M=1.5$ and $\chi=1000$ the transition radius occurs at a 100 times larger radius than predicted by that criterion. One potential reason for the discrepancy with the results of \citet{2018MNRAS.480L.111G} is the different set-up, because they switch off cooling of the hot wind (i.e. for temperatures above $0.6 T_\text{wind}$), implying that cooling is expected to be relatively more important for intermediate temperatures in comparison to our simulations and those by \citet{2019arXiv190902632L}. \citet{2018MNRAS.480L.111G} also use a different initial cloud temperature and temperature floor of $4\times 10^4$ K, which is larger than ours. The cooling function increases drastically from our cloud temperature of $10^4$ K to $4\times 10^4$ K, so that cooling of dense gas is slower in our simulations.

There are also numerical differences between our simulations and that by \citet{2018MNRAS.480L.111G}. They use a cloud tracking algorithm to maintain the head of the cloud near the lower boundary of their simulation box. This makes sure that no dense gas leaves their box. Instead of using a cloud tracking algorithm we use a sufficiently large simulation box, to avoid dense gas flowing out at the upper boundary. We have confirmed that no significant amount of dense gas leaves our simulations at the times analysed in our figures. The difference between our favoured growth criterion and that by \citet{2018MNRAS.480L.111G} is hence not caused by whether or not a cloud tracking algorithm is used.

\citet{2020arXiv200900525K} recently analysed simulations revealing a smaller transition radius compared to our simulations, e.g. for simulations with $M=1.5$ and $\chi=100$. The differences are, however, mostly caused by the two studies using different definitions of cloud growth. If we analyse our simulations with a similar growth definition as \citet{2020arXiv200900525K} we obtain a transition radius similar to their simulations. We present a quantitative analysis revealing the role of the growth definition in Appendix~\ref{SharmaGrowthCriterion}. We note, that our growth criterion is quite conservative and has the tendency to select the radius, where dense gas starts growing downstream in the vicinity of the main cloud. This criterion is for example relevant for studying the star forming tails in jellyfish galaxies or for the mass loading of galactic winds. A less stringent definition as presented in \citet{2020arXiv200900525K} has the tendency to pick a transition radius that only allows dense gas to grow far downstream from the main cloud (and in some cases even after the dense cloud core has resolved). Another disadvantage of such a criterion is that it is more sensitive to numerical parameters, such as boundary conditions, box size and cooling of the hot wind. There is no per se correct definition because each choice implies advantages and disadvantages, depending on the exact application at hand.

\begin{figure}
\centering
\includegraphics[width=\linewidth]{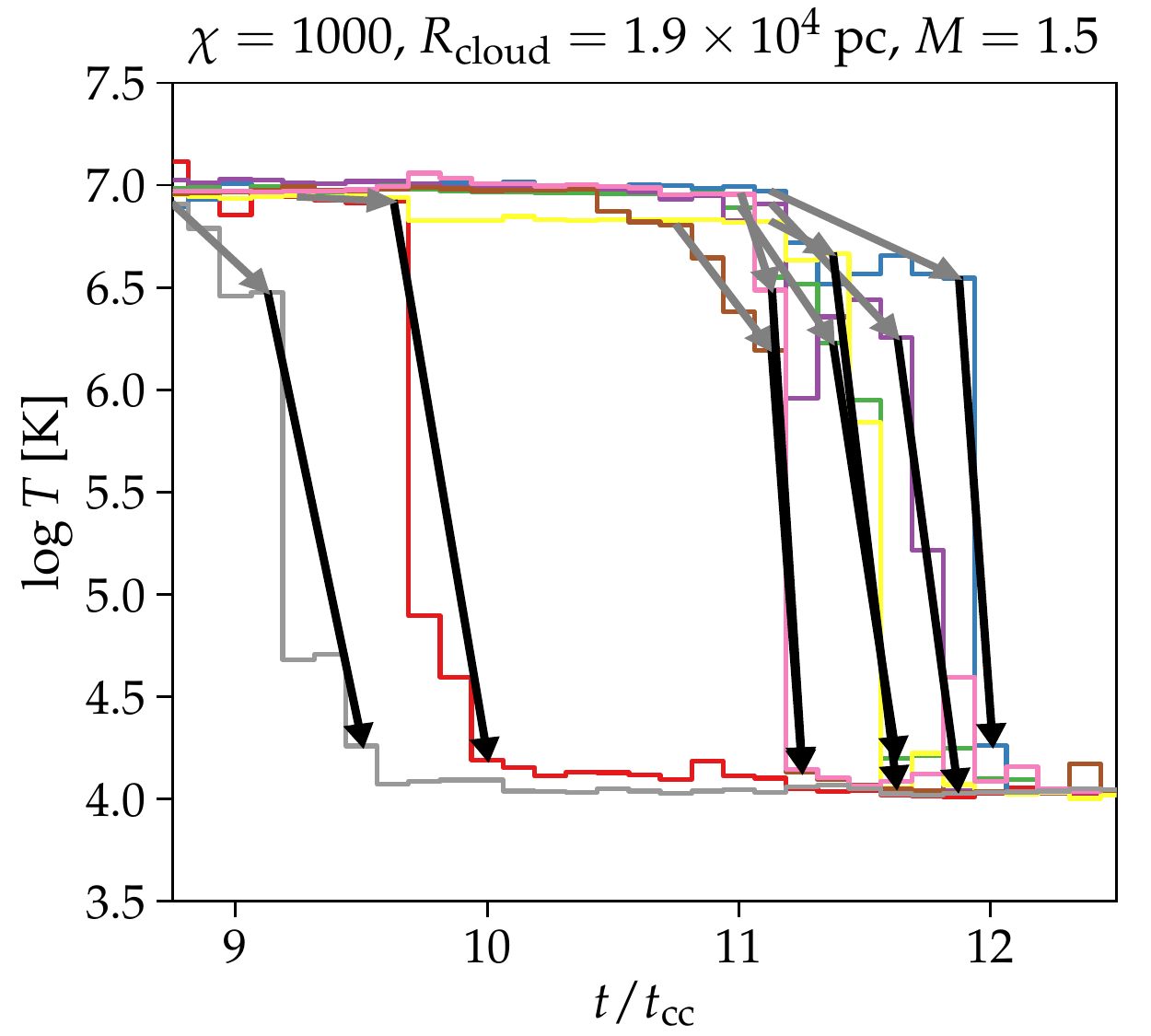}
\caption{To study how the cold cloud grows we show eight gas tracers belonging to the cold phase at $t=12.5t_\text{cc}$, and the hot phase at $8.75t_\text{cc}$. Each gas tracer goes through an epoch of a slow decrease in temperature from $10^7$~K to $10^{6.5}$~K (see grey arrows). After reaching $10^{6.5}$~K the gas cools very rapidly to $\lesssim 10^{4.3}$ K (see black arrows). The bottleneck in cooling from the wind to the cloud temperature is the initial decline from $10^7$~K to $10^{6.5}$~K.}
\label{FigX2000_Tracers_chi1000}
\includegraphics[width=\linewidth]{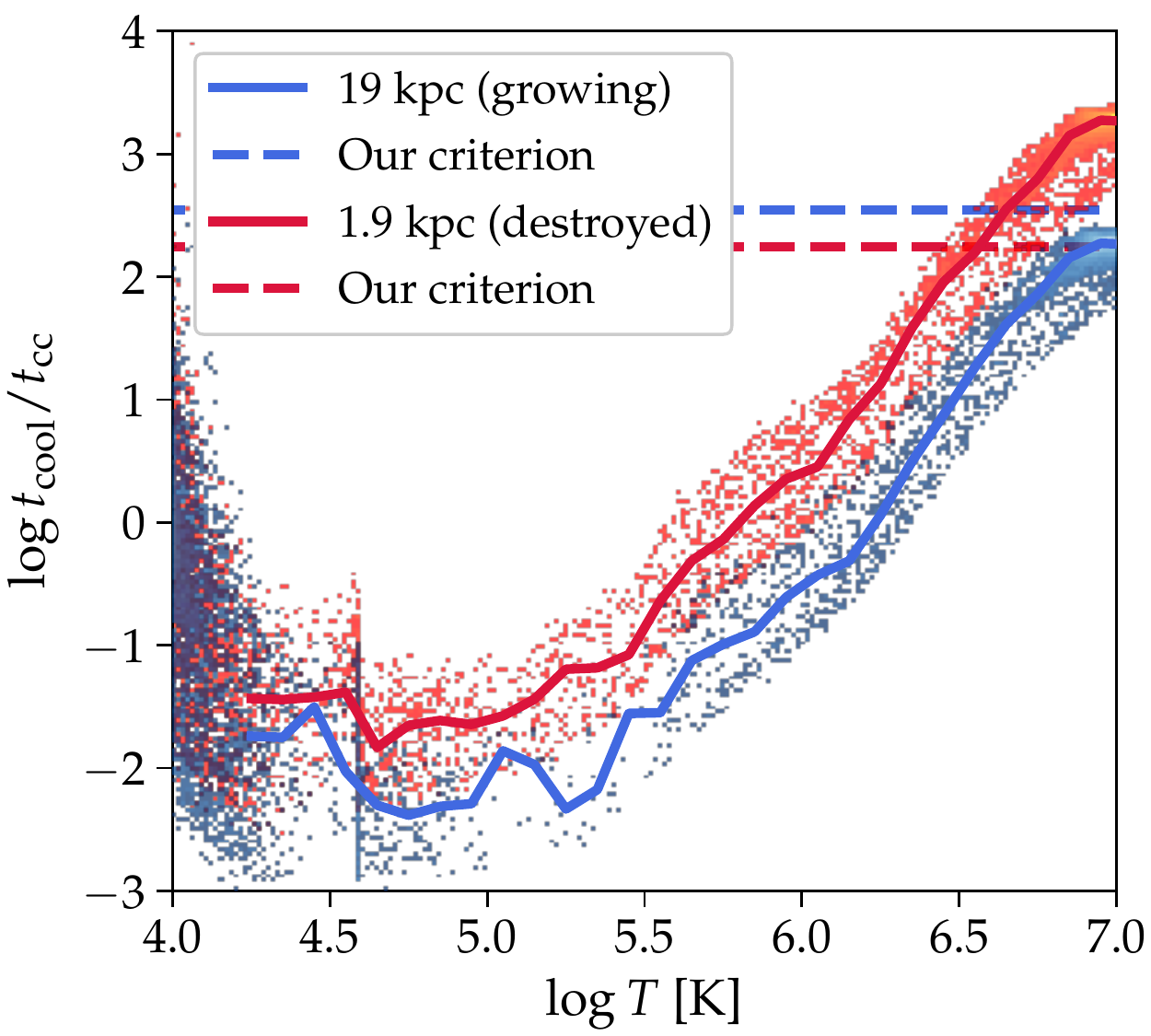}
\caption{We show the values of $t_{\rmn{cool}}/t_{\rmn{cc}}$ versus temperature of all Voronoi cells (dots) for a simulation in the growth regime ($R_\text{cloud}=19$ kpc, blue) and in the destruction regime ($R_\text{cloud}=1.9$ kpc, red) at $t_\text{cc}$. Both simulations have $\chi=1000$ and $M=1.5$. The vertical offset of the median values of the distribution (solid lines) by one order of magnitude is due to the linear dependence of $t_{\rmn{cc}}$ on the cloud radius. We show our criterion for the transition scale from the destruction to the growing regime, $t_{\rmn{cool,wind}}/t_{\rmn{cc}}= 10 f$ with dashed lines. For the cloud in the destruction regime the cooling time of the wind is significantly larger than the transition cooling time so that even mixing or density fluctuations are not sufficient to trigger a fast enough cooling of the wind so that the cloud experiences shattering and will eventually be destroyed. By contrast, the scaled wind cooling time of the large cloud is below the critical threshold, so that mixing of the hot wind facilitates the onset of cooling and causes cloud growth.}
\label{Fig7777_tcool_vs_T}
\end{figure}

\subsubsection{Origin of the mass growth}\label{MassGrowthOrigin}

To study how gas from the hot phase loses its thermal energy and increases the density, such that it becomes part of the cold and dense phase, we analyse homogeneously-distributed Lagrangian tracers \citep[see][]{2013MNRAS.435.1426G} in our simulations. In the simulation with $M=1.5$, $T_\text{wind}=10^7$~K and $R_\text{cloud}=1.9\times 10^4$~pc, which is in the cloud growth regime, we select tracers belonging simultaneously to the hot gas (i.e. $T\geq 0.5 T_\text{wind}$) at $t=8.75t_\text{cc}$ and to the cold cloud (i.e. with $T\leq 3 T_\text{cloud}$ and $n\geq n_\text{cloud}/3$) at $t=12.5 t_\text{cc}$. These tracers have hence joined the cold cloud from the hot wind in between these two times. The temperature evolution for eight (randomly selected) tracers is shown in Fig.~\ref{FigX2000_Tracers_chi1000}. Each gas tracer shows two cooling phases. A slow mixing and cooling phase, where the temperature decreases from $T\simeq 10^7$ K to $T\simeq 10^{6.5}$ K, and a subsequent rapid cooling phase where the temperature cools to $\lesssim 10^{4.5}$ K. In the figure these phases are marked with grey and black arrows, respectively, for each tracer particle. The time it takes to reach $T\simeq 10^{6.5}$ K from $T\simeq 10^7$ K is typically between $0.5t_\text{cc}$ and $1.0t_\text{cc}$, and the following cooling is more rapid.

In Fig.~\ref{Fig7777_tcool_vs_T} we assess the criterion responsible for the initial decline in temperature before the fast cooling sets in. We show $t_{\rmn{cool}}/t_{\rmn{cc}}$ for a simulation of a growing cloud ($R_\text{cloud}=19$ kpc) and of a cloud in the destruction regime ($R_\text{cloud}=1.9$ kpc). In both simulations, we adopt a Mach number $M=1.5$ and $\chi=1000$. Because $t_{\rmn{cc}}\propto R_{\rmn{cloud}}$, the scaled cooling time of the large cloud is on average smaller by an order of magnitude than that of the small cloud. We have over-plotted our growth criterion, $t_{\rmn{cool,wind}}/t_{\rmn{cc}}= 10 f$, with dashed lines for the two clouds sizes ($R_\text{cloud}$ enters our growth criterion through $f$ in Eq.~\ref{DefinitionF}). Clearly, the scaled cooling time of the hot wind is shorter than the critical cooling time for the large cloud, which enables cooling of the wind upon interacting with the stripped cold gas. Contrarily, the cooling time of the small cloud exceeds the threshold by more than an order of magnitude, precluding cooling to play an important role so that non-radiative gas dynamics dominates the shattering process and eventually causes the small cloud to be dissolved in the wind.

We can get a first glimpse on the mechanism by looking at the distribution of cooling times. Figure~\ref{Fig7777_tcool_vs_T} reveals that the individual gas cells (see background points in the plot) have a significant scatter around the median distribution (solid lines). The distribution shows two possible paths for the hot wind to cool to lower temperatures. The hot wind can mix with the cold gas, which slowly reduces its temperature to one half or a third of its original value, $T_{\rmn{mix}}\simeq 10^{6.5}-10^{6.7}$ K, where $t_\text{cool}$ is short enough for the gas to rapidly cool to a much lower temperature; we refer to this as a \emph{mixing mechanism}. It is also possible that the turbulent wake of the ram-pressure stripped cloud excites compressible fluctuations that interact with the hot wind. These fluctuations may cause fluctuations in the cooling time (which depends on density as $1/n$) of the hot wind. The regions of increased density show a 3 to 5 times shorter radiative cooling time in comparison to the median cooling time of the hot wind at $10^7$ K. We refer to this path as the \emph{fluctuation mechanism}.

We defer a detailed analysis of the initial cooling mechanism (which also needs to address the role of magnetic fields) to future work. However, we note that the cooling time-scale of the hot gas enters our criterion because rapid cooling has to start quite close to the hot wind temperature (usually 2-3 times lower than $T_\text{wind}$). We caution that because the cooling time-scale of the hot wind enters our growth criterion, this does not imply that runaway cooling occurs in the wind---to initiate rapid cooling and equivalently cloud growth we need initial mixing of the wind and stripped cold material or compressible fluctuations to enhance the cooling rate of the wind and to slightly decrease its temperature.

\subsubsection{Numerical convergence}\label{subsubconvergence}

As mentioned above it is computationally demanding to run simulations of clouds in the growth regime, since very large simulation boxes are needed to ensure that no dense gas leaves the simulated domain. To test for convergence, a subset of the simulations with $M=1.5$ from Table~\ref{Table:SimulationOverviewB} are run at an eight times finer mass resolution in comparison to what we have presented so far in this section. The evolution of these simulations is shown in Appendix~\ref{AppendixConvergence}. In summary, the radius, at which clouds transition from the growth to the destruction regime, is independent of resolution, so the classification of cloud regimes is well converged.

\begin{figure}
\centering
\includegraphics[width=\linewidth]{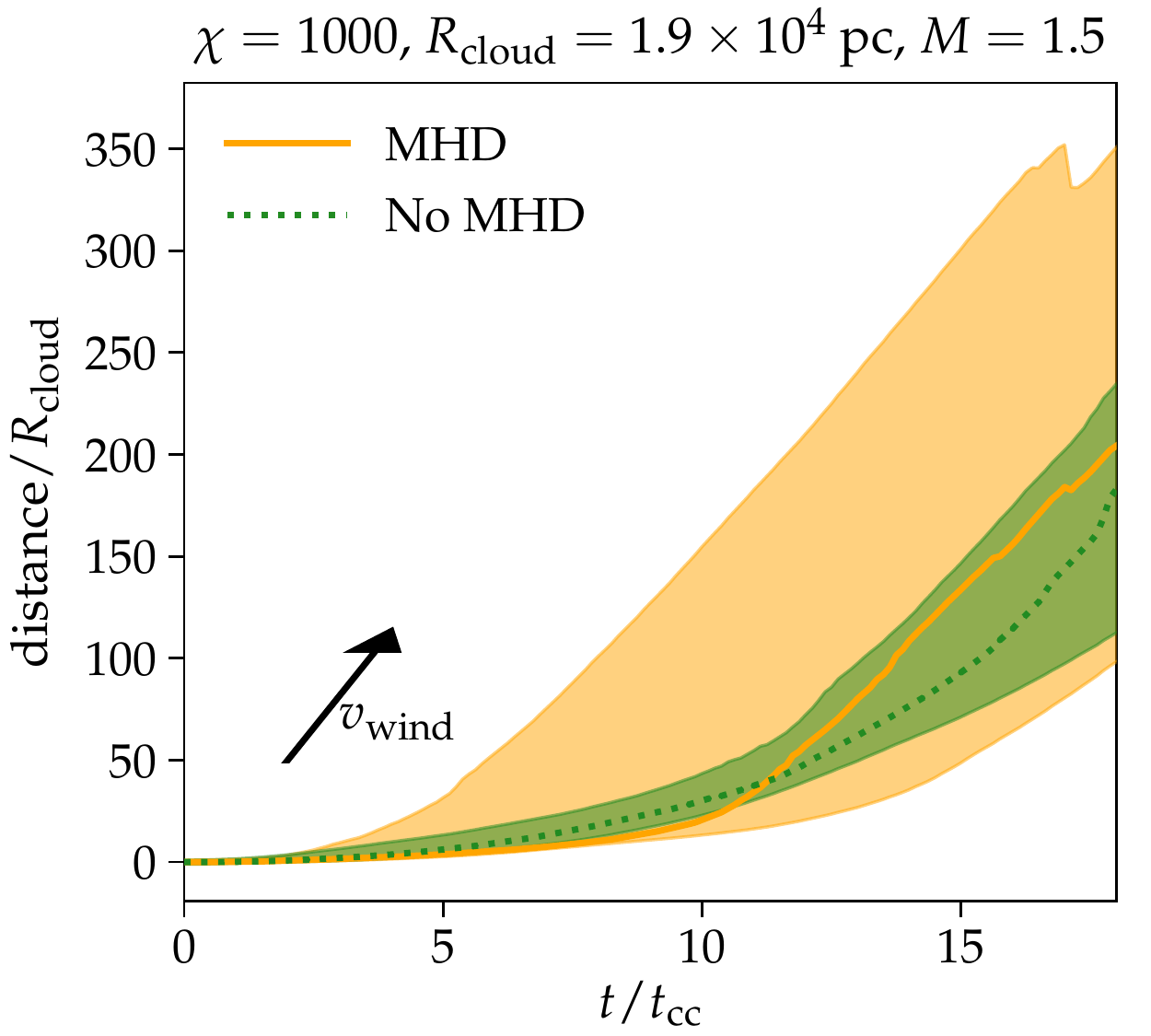}
\caption{An analogue of Fig.~\ref{Fig001_CloudSurvival_Distance_3D} for the simulation with $M=1.5$, $\chi=1000$ and $R_\text{cloud}=1.9 \times 10^4$ pc, where the dense gas mass grows in time. Lines show the median and the contours show 5--95 percentiles of the position of the dense gas (with $n\geq \sqrt{n_\text{cloud}n_\text{wind}}$). The median profile reveals the majority of the dense gas in the MHD simulation to be co-moving with the wind already at $12t_\text{cc}$ (the wind speed is indicated by the arrow). For the non-MHD simulation this occurs later, at $18 t_\text{cc}$. The 5--95 percentile distributions reveal, that the gas is distributed at a larger distance interval in the MHD simulation, compared to the non-MHD version. Overall, the presence of a magnetic field accelerates the dense phase (in the ram-pressure stripped tail) more efficiently through the tension force of the wind magnetic field that is anchored and flux-frozen in the hot wind.}
\label{Fig001_CloudSurvival_Distance_AccretionRegime_3D}
\end{figure}

\subsection{The role of magnetic fields}\label{Sec:MagFieldGrowthCriterion}

\subsubsection{How magnetic fields affect the cloud growth criterion}\label{subSec:RoleOfMHD}

Magnetic fields are not expected to play a major role in deciding whether a cloud is in the growth or destruction regime. The magnetic field strength is for example not explicitly present in our cloud growth criterion in Eq.~\eqref{OurFinalCrit}. In Sect.~\ref{Sec:Results} we demonstrated that a turbulent wind extends the cloud lifetime by a factor of 1.5 to 2. We have accounted for this by including a factor 2 to Eq.~\eqref{DefinitionF}.

To demonstrate that magnetic fields are not altering the cloud growth criterion beyond this expectation, we ran the simulations with $M=1.5$ from Table~\ref{Table:SimulationOverviewB} with MHD disabled. The evolution of these simulations are presented in Appendix~\ref{AppendixMHDsim}. All MHD simulations with clouds in the growth regime are also in this regime in the hydrodynamic simulations without magnetic fields. Neglecting magnetic fields also does not change any of our conclusions regarding the transition from the destruction to the growth regime. The simulation with $R_\text{cloud}=47$ pc and $\chi=100$ is still close to the transition between the growth and destruction regime in the hydrodynamical simulation, but the growth is slower in comparison to the MHD case (both simulations are, however, still in the destruction regime, because of their lack of growth at $12.5 t_\text{cc}$). The comparison between MHD and non-MHD simulations confirms our expectation that the transition radius between the growth and destruction regime is only mildly affected by a magnetic field with a beta factor $\beta\gtrsim 10$. 

\begin{figure*}
\centering
\begin{minipage}{.11\textwidth}
\includegraphics[width=\linewidth]{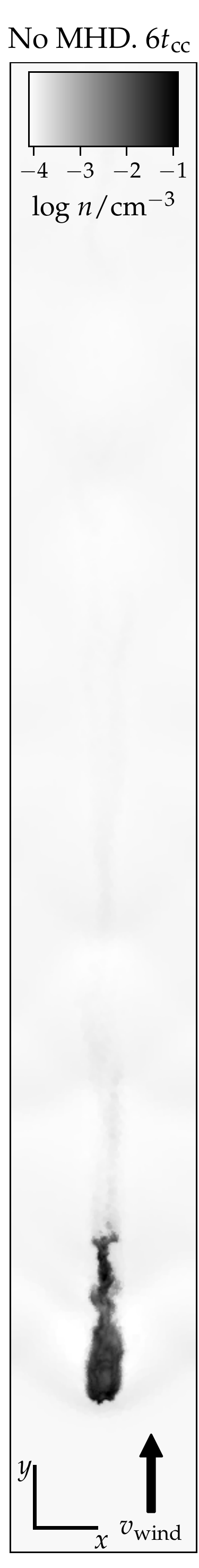}
\end{minipage}
\begin{minipage}{.11\textwidth}
\includegraphics[width=\linewidth]{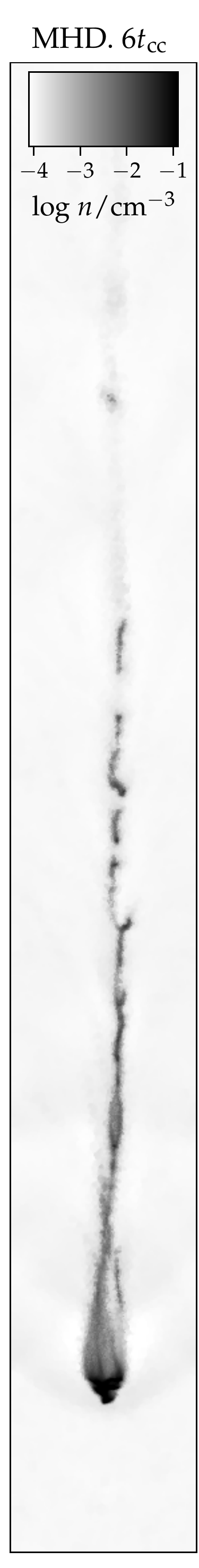}
\end{minipage}
\begin{minipage}{.11\textwidth}
\includegraphics[width=\linewidth]{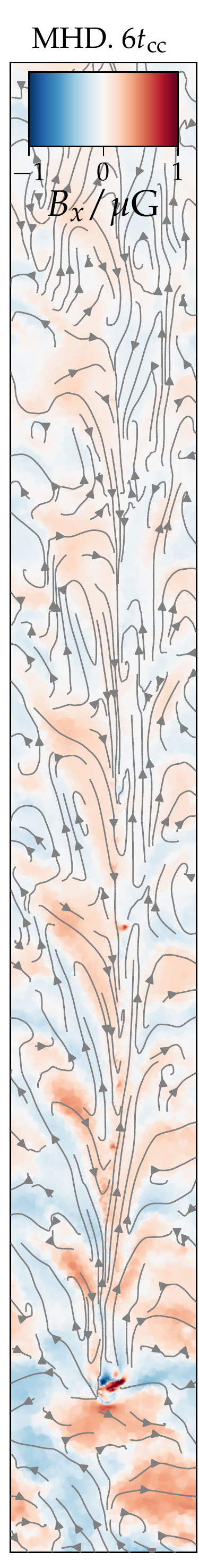}
\end{minipage}
\begin{minipage}{.11\textwidth}
\includegraphics[width=\linewidth]{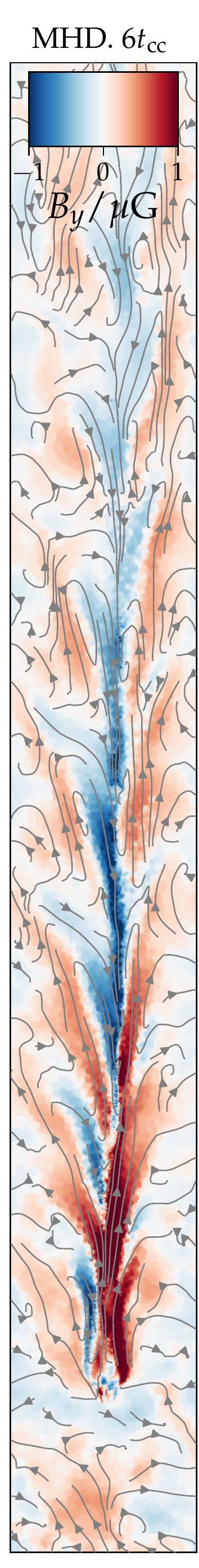}
\end{minipage}
\begin{minipage}{.05\textwidth}
$\phantom{.}$
\end{minipage}
\begin{minipage}{.11\textwidth}
\includegraphics[width=\linewidth]{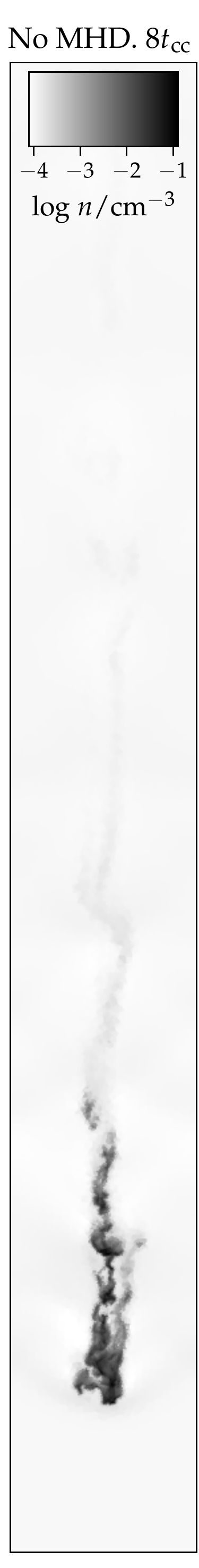}
\end{minipage}
\begin{minipage}{.11\textwidth}
\includegraphics[width=\linewidth]{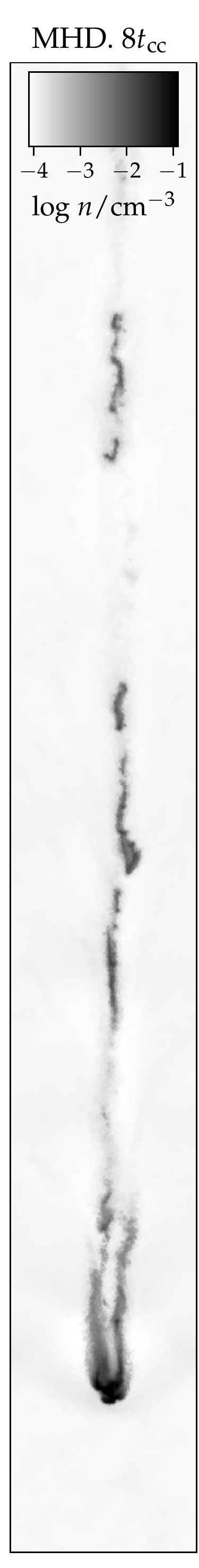}
\end{minipage}
\begin{minipage}{.11\textwidth}
\includegraphics[width=\linewidth]{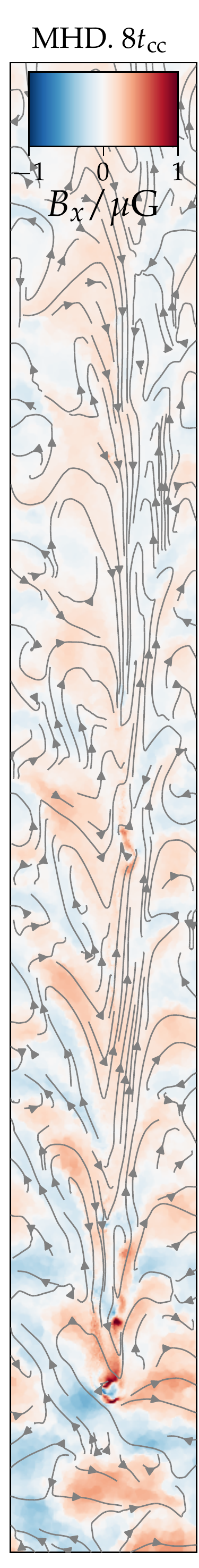}
\end{minipage}
\begin{minipage}{.11\textwidth}
\includegraphics[width=\linewidth]{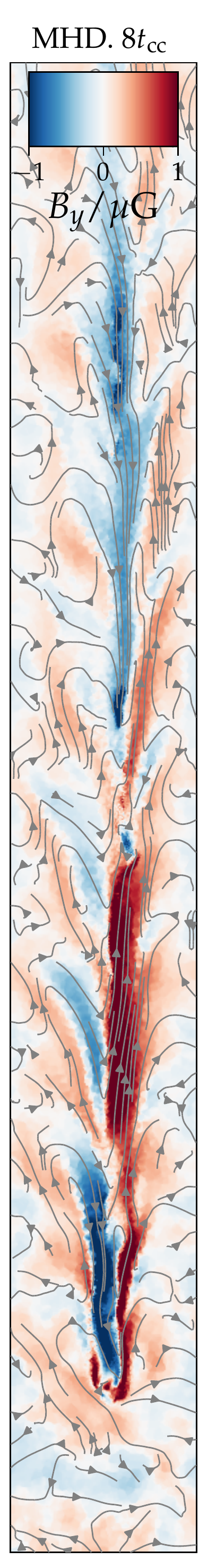}
\end{minipage}
\caption{The gas structure at $6t_\text{cc}$ (first four panels) and $8 t_\text{cc}$ (last four panels) with and without MHD for the same simulations as in Fig.~\ref{Fig001_CloudSurvival_Distance_AccretionRegime_3D}. Each panel shows the average of 50 layers within $z = \pm 0.5 R_\text{cloud}$. We show the number density, and for the simulation with MHD also $B_x$ and $B_y$ (with $\bs{B}$-field stream-lines shown in gray). The stream-lines show the magnetic field in the slice. Magnetic fields facilitate the formation of long, filamentary tails far downstream from the cloud. The dense gas, downstream of the main cloud, has a magnetic field aligned with the cloud's tail.}
\label{FigX880}
\end{figure*}

\subsubsection{How magnetic fields accelerate growing clouds}

While magnetic fields are not able to drastically affect the condition for cloud growth, they are able to significantly change the structure of the cold, ram-pressure stripped tails. In Fig.~\ref{Fig001_CloudSurvival_Distance_AccretionRegime_3D} we show the distance travelled by the dense gas (with $n\geq \sqrt{n_\text{cloud}n_\text{wind}}$) as a function of time for the MHD and non-MHD version of the simulations with $M=1.5$, $\chi=1000$ and $R_\text{cloud}=1.9 \times 10^4$ pc. The cloud becomes co-moving with the wind earlier in the MHD simulation in comparison to the non-MHD simulation (at $12t_\text{cc}$ and $18t_\text{cc}$, respectively). Furthermore, the gas is spread over a larger volume in the MHD case, which is revealed by the 5--95 percentile of the distance travelled. The distribution of the gas is visualised in Fig.~\ref{FigX880}, which shows that more dense gas is transported downstream in the MHD simulations in comparison to the non-MHD analogues at $6t_\text{cc}$ and $8 t_\text{cc}$. Most notably, magnetic fields facilitate the formation of a long tail of dense, cold material in the downstream of the cloud. This cold material is seeded by ram-pressure stripped cloud material that (partially) mixes with the hot wind and effectively causes a net accretion of wind mass to the cloud filamentary tail. Magnetic fields therefore very drastically change the appearance and observability of gas clouds interacting with a hot wind.

For our simulations in the growth regime, we confirm the conclusion from \citet{2015MNRAS.449....2M} that magnetic fields enhance cloud acceleration, especially in the late non-linear stages ($t\gtrsim 8 t_\text{cc}$). However, we find that magnetic fields do not provide the key for the survival of clouds in a wind. This is in agreement with \citealt{2019MNRAS.tmp.2995G} (see their section 5.4). We note that the role of the magnetic field in our simulations is consistent with the recent work of \citet{2020arXiv200207804C}.

\section{Discussion}\label{Sec:Discussion}

The criterion for whether gas clouds are in the destruction or growth regime is potentially important for several applications, which we will now discuss.

\subsection{Jellyfish galaxies}

The tails of jellyfish galaxies form when the dense ISM of a galaxy gets stripped by the ram pressure that the galaxy experiences as it moves through the ICM. For a tail to remain dense and to survive several tens of kpc, as it interacts with the hot gas in the cluster, it has to be in the cloud growth regime. We have shown that if the wind or the cloud are magnetised, this may only mildly modify the growth criterion but it drastically influences the tail morphology because magnetic fields suppress Kelvin Helmholtz instabilities and help accelerating dense clouds downstream from the main gas cloud (i.e., what would be the ISM for a jellyfish galaxy).

The most promising observational method for constraining the magnetic field of the tail of jellyfish galaxies are radio synchrotron observations \citep{2009AJ....137.4436M,2020MNRAS.tmp.2017C}. Using polarization measurements enables us to determine the in-plane magnetic field \citep{1979rpa..book.....R}. Interpretation of such an observation would also require a measurement of the effect of Faraday rotation of the polarization angle from the plasma in between the observer and the region emitting synchrotron radiation \citep[e.g.;][]{1966MNRAS.133...67B,2009A&A...495..697W}. Our simulations predict that if a jellyfish galaxy has a magnetised tail, the magnetic field should be well-aligned with the tail, as it is seen in Fig.~\ref{FigX880}. Such an observation would be of high importance, since it would demonstrate that the magnetic field plays an important role in shaping the distribution of gas in the tail of a jelly-fish galaxies.

A natural extension of this paper would be to simulate a full galaxy in a windtunnel setup to reveal the exact nature of how gas is stripped in clusters. Several papers \citep{2007MNRAS.380.1399R,2012MNRAS.422.1609T,2014ApJ...784...75R,2015A&A...582A..23N,2016A&A...591A..51S,2019A&A...624A..11H} performed such a study, but their physical model did not include magnetic fields, which we believe is a key for shaping the morphology of the stripped gas, and it is certainly necessary to predict synchrotron observables.

A novel method, consisting of performing idealised simulations of galaxies in a cluster, was used to study ram-pressure stripping in \citet{2015MNRAS.449.2312V} and \citet{2017ApJ...841...38V}. The latter paper identified an enhanced magnetic field aligned with the tail of ram-pressure stripped (jellyfish) galaxies. These trends are in perfect agreement with our simulations.

State-of-the-art cosmological MHD simulations (such as Illustris TNG50, \citealt{2019MNRAS.490.3234N,2019MNRAS.490.3196P}) may also shed light on the physics of jellyfish galaxy tails and how they are connected to the structure of magnetic fields; analysing the plasma-$\beta$ parameter, the magnetic field's orientation and producing mock observations of synchrotron emission would potentially provide remarkable insight. As always, an advantage of such simulation is that they are cosmologically self-consistent, but it comes at the cost that they are often hard to interpret in comparison to controlled idealised simulations.

\subsection{CGM in cosmological galaxy simulations}

To resolve the CGM in cosmological simulations, one necessary (but potentially not sufficient) criterion is  a spatial resolution better than (or at least comparable to) the critical cloud size, where clouds transition from the growth to the destruction regime. If a gas cloud is under-resolved so that the cell-size is larger than the criterion revealed by Eq.~\eqref{OurFinalCrit}, it may be growing instead of undergoing destruction, simply because of a lack of resolution.

For the physical conditions of our simulations, the spatial resolution required to resolve the cloud growth criterion of 0.1 cm$^{-3}$ clouds in the CGM is $\gtrsim 100~\rmn{pc}$ for wind temperatures of $T_\rmn{wind}\geq 10^6$~K. This resolution is comparable to what is obtained in recent cosmological simulations specifically targeting a high resolution in the CGM \citep{2019MNRAS.482L..85V,2019MNRAS.483.4040S,2019ApJ...882..156H,2018arXiv181006566P}. Furthermore, to resolve whether a cloud is growing or being destroyed it will likely be necessary to have multiple (i.e., 5--10) cells per cloud radius. While challenging, this is generally promising for our ability to resolve the CGM in cosmological simulations.

It is, however, important to note, that just because a cosmological simulation resolves the critical transition scale for a gas cloud to move from the destruction to the growth regime, it is not guaranteed that the simulation is converged. Numerous papers have shown that clouds with sizes larger than the cooling length are unstable and undergo fragmentation \citep{2018MNRAS.473.5407M,2018arXiv180610688L, 2019MNRAS.482.5401S,2020MNRAS.494L..27G}. \citealt{2019ApJ...882..156H} suggested that this fragmentation could be accounted for by subgrid models describing the sub-resolution distribution of clouds. Thus, a possible approach for future cosmological simulations would need to ensure a sufficient resolution of the CGM, so that the transition radius between the destruction or growth regime of clouds interacting with a warm/hot ambient medium is resolved. Furthermore this approach would need to adopt a subgrid model, which accounts for the unresolved structure on scales in between the resolution and the cooling length.

\subsection{Cold streams}

Streams of cold gas, which have been argued to be important for fuelling high-redshift star formation in galaxies \citep{2005MNRAS.363....2K}, may fragment easily when radiative cooling is included in simulations \citep[see fig.~2 in][]{2019arXiv191005344M}, but fragmentation could be suppressed by adding magnetic fields \citep{2019MNRAS.489.3368B}. A key requirement for the survival of a sequence of clouds formed from a stream is that they are in the growth regime. Compared to a single cloud interacting with a hot wind, a sequence of clouds may shield each other from the instabilities disrupting the dense gas phase \citep[as shown by][]{2019AJ....158..124F}. The criterion dividing the growth and destruction phase may therefore have to be slightly modified to account for the evolution of clumps within a stream. This could be accounted for by adding a fudge parameter in Eq.~\eqref{DefinitionF}, such that the radius of streams in the growth regime would be slightly lowered. We will leave it for future work to assess the survival of fragmented clouds formed from a cold stream.

\section{Conclusion}\label{Sec:Conclusion}

Simulations of cold clouds interacting with a hot wind are extremely useful for understanding processes relevant for a range of astrophysical systems. In this work, we have focused on the effects of radiative cooling and magnetohydrodynamics. For the first time we have performed simulations of the interaction of a turbulent magnetic wind with a cold cloud. Our main findings are:

\begin{itemize}
\item We have examined the transition between the growth and destruction regime of cold--dense clouds interacting with a hot--diffuse wind. We find that a criterion based on the cooling time-scale of the hot wind well captures the transition from one regime to the other (our main result is summarised by Fig.~\ref{MachNumber1.5} and Eq.~\ref{OurFinalCrit}). In the literature, a criterion based on the cooling time-scale of the mixed gas -- with an intermediate temperature of $T_\text{mix}=\sqrt{T_\text{wind}T_\text{cloud}}$ -- has also been proposed, but we find that the rate-limiting step for cooling from the hot wind to the cold cloud temperature is the initial phase when the temperature typically declines by a factor of two to three from the hot wind temperature, and not the subsequent fast cooling from intermediate to low temperatures (see Sec.~\ref{MassGrowthOrigin} for a full discussion).
\item Our criterion for cloud growth resembles the criterion by \citet{2019arXiv190902632L}. The differences are: 1) our simulations have a stronger magnetic field, which accounts for a factor 2 in the factor, $f$ (see Eq.~\ref{DefinitionF}), which appears in our criterion (Eq. \ref{OurFinalCrit}), and 2) we find a strong Mach number dependence of $f\propto M^{-2.5}$ in our criterion -- an explicit Mach number dependence is absent in \citet{2019arXiv190902632L}.
\item The exact criterion used to define whether clouds are growing or experiencing destruction matters. In this paper we have used a more conservative criterion than for example the recent work of \citet{2020arXiv200900525K}. In Appendix~\ref{SharmaGrowthCriterion} we show that our simulations are in good agreement with \citet{2020arXiv200900525K} if we use a less stringent growth criterion. We note however, that our criterion appears to be more relevant for studying the tails in the immediate wake of jellyfish galaxies or the mass loading of galactic winds, which is most easily achieved close to the disk where the accelerating forces as a result of momentum deposition from supernovae, radiation and cosmic rays are strongest.
\item In the simulations with a turbulent magnetic wind the Kelvin--Helmholtz instability is significantly suppressed by magnetic draping in comparison to simulations without magnetic fields or with a uniform magnetic field of the wind. We emphasise the importance of including magnetic fields when simulating astrophysical instabilities.
\item The addition of magnetic fields completely changes the morphology of ram-pressure stripped gas from clumpy density distributions to filamentary long tails. These are long-lived for large clouds in the growth regime due to the increase of the mass of cold gas in the tail. This gas accretion amplifies the draped magnetic field via adiabatic compression and velocity shear and aligns it with the filamentary gaseous tail. We specifically predict the tails of jellyfish galaxies to have ordered and aligned magnetic fields that can be observed by polarized radio synchrotron observations of these objects.
\item We conclude that the cloud growth criterion plays an important role for the survival of fragments in cold accretion streams, for mass loading of galactic winds, for the formation and survival of the jellyfish galaxy tails, and for future subgrid models of the circumgalactic medium in cosmological galaxy formation models.
\end{itemize}

\section*{Acknowledgements}
We thank Max Gronke, Peng Oh, Prateek Sharma, Philipp Girichidis and Thomas Berlok for useful comments and discussions. We also thank the referee for insightful comments and suggestions. We acknowledge support by the European Research Council under ERC-CoG grant CRAGSMAN-646955. This research was supported in part by the National Science Foundation under Grant No. NSF PHY-1748958.

\section*{Data availability}
The simulations and data analysis scripts underlying this article will be shared on reasonable request to the corresponding author. The Arepo code is publicly available.

\footnotesize{
\bibliographystyle{mnras}
\bibliography{ref}

\begin{thebibliography}{}
\makeatletter
\relax
\def\mn@urlcharsother{\let\do\@makeother \do\$\do\&\do\#\do\^\do\_\do\%\do\~}
\def\mn@doi{\begingroup\mn@urlcharsother \@ifnextchar [ {\mn@doi@}
  {\mn@doi@[]}}
\def\mn@doi@[#1]#2{\def\@tempa{#1}\ifx\@tempa\@empty \href
  {http://dx.doi.org/#2} {doi:#2}\else \href {http://dx.doi.org/#2} {#1}\fi
  \endgroup}
\def\mn@eprint#1#2{\mn@eprint@#1:#2::\@nil}
\def\mn@eprint@arXiv#1{\href {http://arxiv.org/abs/#1} {{\tt arXiv:#1}}}
\def\mn@eprint@dblp#1{\href {http://dblp.uni-trier.de/rec/bibtex/#1.xml}
  {dblp:#1}}
\def\mn@eprint@#1:#2:#3:#4\@nil{\def\@tempa {#1}\def\@tempb {#2}\def\@tempc
  {#3}\ifx \@tempc \@empty \let \@tempc \@tempb \let \@tempb \@tempa \fi \ifx
  \@tempb \@empty \def\@tempb {arXiv}\fi \@ifundefined
  {mn@eprint@\@tempb}{\@tempb:\@tempc}{\expandafter \expandafter \csname
  mn@eprint@\@tempb\endcsname \expandafter{\@tempc}}}

\bibitem[\protect\citeauthoryear{{Angl{\'e}s-Alc{\'a}zar},
  {Faucher-Gigu{\`e}re}, {Kere{\v{s}}}, {Hopkins}, {Quataert}  \&
  {Murray}}{{Angl{\'e}s-Alc{\'a}zar} et~al.}{2017}]{2017MNRAS.470.4698A}
{Angl{\'e}s-Alc{\'a}zar} D.,  {Faucher-Gigu{\`e}re} C.-A.,  {Kere{\v{s}}} D.,
  {Hopkins} P.~F.,  {Quataert} E.,   {Murray} N.,  2017, \mn@doi [\mnras]
  {10.1093/mnras/stx1517}, \href
  {https://ui.adsabs.harvard.edu/abs/2017MNRAS.470.4698A} {470, 4698}

\bibitem[\protect\citeauthoryear{{Armillotta}, {Fraternali}, {Werk},
  {Prochaska}  \& {Marinacci}}{{Armillotta} et~al.}{2017}]{2017MNRAS.470..114A}
{Armillotta} L.,  {Fraternali} F.,  {Werk} J.~K.,  {Prochaska} J.~X.,
  {Marinacci} F.,  2017, \mn@doi [\mnras] {10.1093/mnras/stx1239}, \href
  {http://adsabs.harvard.edu/abs/2017MNRAS.470..114A} {470, 114}

\bibitem[\protect\citeauthoryear{{Aung}, {Mandelker}, {Nagai}, {Dekel}  \&
  {Birnboim}}{{Aung} et~al.}{2019}]{2019MNRAS.490..181A}
{Aung} H.,  {Mandelker} N.,  {Nagai} D.,  {Dekel} A.,   {Birnboim} Y.,  2019,
  \mn@doi [\mnras] {10.1093/mnras/stz1964}, \href
  {https://ui.adsabs.harvard.edu/abs/2019MNRAS.490..181A} {490, 181}

\bibitem[\protect\citeauthoryear{{Banda-Barrag{\'a}n}, {Parkin}, {Federrath},
  {Crocker}  \& {Bicknell}}{{Banda-Barrag{\'a}n}
  et~al.}{2016}]{2016MNRAS.455.1309B}
{Banda-Barrag{\'a}n} W.~E.,  {Parkin} E.~R.,  {Federrath} C.,  {Crocker} R.~M.,
    {Bicknell} G.~V.,  2016, \mn@doi [\mnras] {10.1093/mnras/stv2405}, \href
  {https://ui.adsabs.harvard.edu/abs/2016MNRAS.455.1309B} {455, 1309}

\bibitem[\protect\citeauthoryear{{Berlok} \& {Pfrommer}}{{Berlok} \&
  {Pfrommer}}{2019}]{2019MNRAS.489.3368B}
{Berlok} T.,  {Pfrommer} C.,  2019, \mn@doi [\mnras] {10.1093/mnras/stz2347},
  \href {https://ui.adsabs.harvard.edu/abs/2019MNRAS.489.3368B} {489, 3368}

\bibitem[\protect\citeauthoryear{{Bordoloi} et~al.,}{{Bordoloi}
  et~al.}{2014}]{2014ApJ...796..136B}
{Bordoloi} R.,  et~al., 2014, \mn@doi [\apj] {10.1088/0004-637X/796/2/136},
  \href {https://ui.adsabs.harvard.edu/abs/2014ApJ...796..136B} {796, 136}

\bibitem[\protect\citeauthoryear{{Br{\"u}ggen} \& {Scannapieco}}{{Br{\"u}ggen}
  \& {Scannapieco}}{2016}]{2016ApJ...822...31B}
{Br{\"u}ggen} M.,  {Scannapieco} E.,  2016, \mn@doi [\apj]
  {10.3847/0004-637X/822/1/31}, \href
  {http://adsabs.harvard.edu/abs/2016ApJ...822...31B} {822, 31}

\bibitem[\protect\citeauthoryear{{Burn}}{{Burn}}{1966}]{1966MNRAS.133...67B}
{Burn} B.~J.,  1966, \mn@doi [\mnras] {10.1093/mnras/133.1.67}, \href
  {https://ui.adsabs.harvard.edu/abs/1966MNRAS.133...67B} {133, 67}

\bibitem[\protect\citeauthoryear{{Chen} et~al.,}{{Chen}
  et~al.}{2020}]{2020MNRAS.tmp.2017C}
{Chen} H.,  et~al., 2020, \mn@doi [\mnras] {10.1093/mnras/staa1868}, \href
  {https://ui.adsabs.harvard.edu/abs/2020MNRAS.tmp.2017C} {}

\bibitem[\protect\citeauthoryear{{Chevalier} \& {Clegg}}{{Chevalier} \&
  {Clegg}}{1985}]{1985Natur.317...44C}
{Chevalier} R.~A.,  {Clegg} A.~W.,  1985, \mn@doi [\nat] {10.1038/317044a0},
  \href {http://adsabs.harvard.edu/abs/1985Natur.317...44C} {317, 44}

\bibitem[\protect\citeauthoryear{{Corlies}, {Peeples}, {Tumlinson}, {O'Shea},
  {Lehner}, {Howk}, {O'Meara}  \& {Smith}}{{Corlies}
  et~al.}{2020}]{2018arXiv181105060C}
{Corlies} L.,  {Peeples} M.~S.,  {Tumlinson} J.,  {O'Shea} B.~W.,  {Lehner} N.,
   {Howk} J.~C.,  {O'Meara} J.~M.,   {Smith} B.~D.,  2020, \mn@doi [\apj]
  {10.3847/1538-4357/ab9310}, \href
  {https://ui.adsabs.harvard.edu/abs/2020ApJ...896..125C} {896, 125}

\bibitem[\protect\citeauthoryear{{Cottle}, {Scannapieco}, {Br{\"u}ggen},
  {Banda-Barrag{\'a}n}  \& {Federrath}}{{Cottle}
  et~al.}{2020}]{2020arXiv200207804C}
{Cottle} J.,  {Scannapieco} E.,  {Br{\"u}ggen} M.,  {Banda-Barrag{\'a}n} W.,
  {Federrath} C.,  2020, \mn@doi [\apj] {10.3847/1538-4357/ab76d1}, \href
  {https://ui.adsabs.harvard.edu/abs/2020ApJ...892...59C} {892, 59}

\bibitem[\protect\citeauthoryear{{Cramer}, {Kenney}, {Sun}, {Crowl}, {Yagi},
  {J{\'a}chym}, {Roediger}  \& {Waldron}}{{Cramer}
  et~al.}{2019}]{2019ApJ...870...63C}
{Cramer} W.~J.,  {Kenney} J.~D.~P.,  {Sun} M.,  {Crowl} H.,  {Yagi} M.,
  {J{\'a}chym} P.,  {Roediger} E.,   {Waldron} W.,  2019, \mn@doi [\apj]
  {10.3847/1538-4357/aaefff}, \href
  {https://ui.adsabs.harvard.edu/abs/2019ApJ...870...63C} {870, 63}

\bibitem[\protect\citeauthoryear{{Dekel} \& {Birnboim}}{{Dekel} \&
  {Birnboim}}{2006}]{2006MNRAS.368....2D}
{Dekel} A.,  {Birnboim} Y.,  2006, \mn@doi [\mnras]
  {10.1111/j.1365-2966.2006.10145.x}, \href
  {https://ui.adsabs.harvard.edu/abs/2006MNRAS.368....2D} {368, 2}

\bibitem[\protect\citeauthoryear{{Dekel} et~al.,}{{Dekel}
  et~al.}{2009}]{2009Natur.457..451D}
{Dekel} A.,  et~al., 2009, \mn@doi [\nat] {10.1038/nature07648}, \href
  {https://ui.adsabs.harvard.edu/abs/2009Natur.457..451D} {457, 451}

\bibitem[\protect\citeauthoryear{{Dursi}}{{Dursi}}{2007}]{2007ApJ...670..221D}
{Dursi} L.~J.,  2007, \mn@doi [\apj] {10.1086/521997}, \href
  {https://ui.adsabs.harvard.edu/abs/2007ApJ...670..221D} {670, 221}

\bibitem[\protect\citeauthoryear{{Dursi} \& {Pfrommer}}{{Dursi} \&
  {Pfrommer}}{2008}]{2008ApJ...677..993D}
{Dursi} L.~J.,  {Pfrommer} C.,  2008, \mn@doi [\apj] {10.1086/529371}, \href
  {http://adsabs.harvard.edu/abs/2008ApJ...677..993D} {677, 993}

\bibitem[\protect\citeauthoryear{{Ehlert}, {Weinberger}, {Pfrommer}, {Pakmor}
  \& {Springel}}{{Ehlert} et~al.}{2018}]{2018MNRAS.481.2878E}
{Ehlert} K.,  {Weinberger} R.,  {Pfrommer} C.,  {Pakmor} R.,   {Springel} V.,
  2018, \mn@doi [Monthly Notices of the Royal Astronomical Society]
  {10.1093/mnras/sty2397}, \href
  {https://ui.adsabs.harvard.edu/abs/2018MNRAS.481.2878E} {481, 2878}

\bibitem[\protect\citeauthoryear{{Fall} \& {Efstathiou}}{{Fall} \&
  {Efstathiou}}{1980}]{1980MNRAS.193..189F}
{Fall} S.~M.,  {Efstathiou} G.,  1980, \mn@doi [\mnras]
  {10.1093/mnras/193.2.189}, \href
  {https://ui.adsabs.harvard.edu/abs/1980MNRAS.193..189F} {193, 189}

\bibitem[\protect\citeauthoryear{{Faucher-Gigu{\`e}re}, {Lidz}, {Zaldarriaga}
  \& {Hernquist}}{{Faucher-Gigu{\`e}re} et~al.}{2009}]{2009ApJ...703.1416F}
{Faucher-Gigu{\`e}re} C.-A.,  {Lidz} A.,  {Zaldarriaga} M.,   {Hernquist} L.,
  2009, \mn@doi [\apj] {10.1088/0004-637X/703/2/1416}, \href
  {http://adsabs.harvard.edu/abs/2009ApJ...703.1416F} {703, 1416}

\bibitem[\protect\citeauthoryear{{Ferland}, {Korista}, {Verner}, {Ferguson},
  {Kingdon}  \& {Verner}}{{Ferland} et~al.}{1998}]{1998PASP..110..761F}
{Ferland} G.~J.,  {Korista} K.~T.,  {Verner} D.~A.,  {Ferguson} J.~W.,
  {Kingdon} J.~B.,   {Verner} E.~M.,  1998, \mn@doi [\pasp] {10.1086/316190},
  \href {http://adsabs.harvard.edu/abs/1998PASP..110..761F} {110, 761}

\bibitem[\protect\citeauthoryear{{Ferland} et~al.,}{{Ferland}
  et~al.}{2013}]{2013RMxAA..49..137F}
{Ferland} G.~J.,  et~al., 2013, \rmxaa, \href
  {http://adsabs.harvard.edu/abs/2013RMxAA..49..137F} {49, 137}

\bibitem[\protect\citeauthoryear{{Fielding} et~al.,}{{Fielding}
  et~al.}{2020}]{2020arXiv200616316F}
{Fielding} D.~B.,  et~al., 2020, arXiv e-prints, \href
  {https://ui.adsabs.harvard.edu/abs/2020arXiv200616316F} {p. arXiv:2006.16316}

\bibitem[\protect\citeauthoryear{{Forbes} \& {Lin}}{{Forbes} \&
  {Lin}}{2019}]{2019AJ....158..124F}
{Forbes} J.~C.,  {Lin} D. N.~C.,  2019, \mn@doi [\aj]
  {10.3847/1538-3881/ab3230}, \href
  {https://ui.adsabs.harvard.edu/abs/2019AJ....158..124F} {158, 124}

\bibitem[\protect\citeauthoryear{{Genel}, {Vogelsberger}, {Nelson}, {Sijacki},
  {Springel}  \& {Hernquist}}{{Genel} et~al.}{2013}]{2013MNRAS.435.1426G}
{Genel} S.,  {Vogelsberger} M.,  {Nelson} D.,  {Sijacki} D.,  {Springel} V.,
  {Hernquist} L.,  2013, \mn@doi [\mnras] {10.1093/mnras/stt1383}, \href
  {https://ui.adsabs.harvard.edu/abs/2013MNRAS.435.1426G} {435, 1426}

\bibitem[\protect\citeauthoryear{{Genel} et~al.,}{{Genel}
  et~al.}{2014}]{2014MNRAS.445..175G}
{Genel} S.,  et~al., 2014, \mn@doi [\mnras] {10.1093/mnras/stu1654}, \href
  {http://adsabs.harvard.edu/abs/2014MNRAS.445..175G} {445, 175}

\bibitem[\protect\citeauthoryear{{Gregori}, {Miniati}, {Ryu}  \&
  {Jones}}{{Gregori} et~al.}{1999}]{1999ApJ...527L.113G}
{Gregori} G.,  {Miniati} F.,  {Ryu} D.,   {Jones} T.~W.,  1999, \mn@doi [\apjl]
  {10.1086/312402}, \href
  {https://ui.adsabs.harvard.edu/abs/1999ApJ...527L.113G} {527, L113}

\bibitem[\protect\citeauthoryear{{Grimes} et~al.,}{{Grimes}
  et~al.}{2009}]{2009ApJS..181..272G}
{Grimes} J.~P.,  et~al., 2009, \mn@doi [\apjs] {10.1088/0067-0049/181/1/272},
  \href {https://ui.adsabs.harvard.edu/abs/2009ApJS..181..272G} {181, 272}

\bibitem[\protect\citeauthoryear{{Gronke} \& {Oh}}{{Gronke} \&
  {Oh}}{2018}]{2018MNRAS.480L.111G}
{Gronke} M.,  {Oh} S.~P.,  2018, \mn@doi [\mnras] {10.1093/mnrasl/sly131},
  \href {https://ui.adsabs.harvard.edu/abs/2018MNRAS.480L.111G} {480, L111}

\bibitem[\protect\citeauthoryear{{Gronke} \& {Oh}}{{Gronke} \&
  {Oh}}{2019}]{2019MNRAS.tmp.2995G}
{Gronke} M.,  {Oh} S.~P.,  2019, \mn@doi [\mnras] {10.1093/mnras/stz3332},
  \href {https://ui.adsabs.harvard.edu/abs/2019MNRAS.tmp.2995G} {p.~2995}

\bibitem[\protect\citeauthoryear{{Gronke} \& {Oh}}{{Gronke} \&
  {Oh}}{2020}]{2020MNRAS.494L..27G}
{Gronke} M.,  {Oh} S.~P.,  2020, \mn@doi [\mnras] {10.1093/mnrasl/slaa033},
  \href {https://ui.adsabs.harvard.edu/abs/2020MNRAS.494L..27G} {494, L27}

\bibitem[\protect\citeauthoryear{{Hausammann}, {Revaz}  \&
  {Jablonka}}{{Hausammann} et~al.}{2019}]{2019A&A...624A..11H}
{Hausammann} L.,  {Revaz} Y.,   {Jablonka} P.,  2019, \mn@doi [\aap]
  {10.1051/0004-6361/201834871}, \href
  {https://ui.adsabs.harvard.edu/abs/2019A&A...624A..11H} {624, A11}

\bibitem[\protect\citeauthoryear{{Huang}, {Katz}, {Scannapieco}, {Cottle},
  {Dav{\'e}}, {Weinberg}, {Peeples}  \& {Br{\"u}ggen}}{{Huang}
  et~al.}{2020}]{2020MNRAS.tmp.2098H}
{Huang} S.,  {Katz} N.,  {Scannapieco} E.,  {Cottle} J.,  {Dav{\'e}} R.,
  {Weinberg} D.~H.,  {Peeples} M.~S.,   {Br{\"u}ggen} M.,  2020, \mn@doi
  [\mnras] {10.1093/mnras/staa1978}, \href
  {https://ui.adsabs.harvard.edu/abs/2020MNRAS.tmp.2098H} {}

\bibitem[\protect\citeauthoryear{{Hummels} et~al.,}{{Hummels}
  et~al.}{2019}]{2019ApJ...882..156H}
{Hummels} C.~B.,  et~al., 2019, \mn@doi [\apj] {10.3847/1538-4357/ab378f},
  \href {https://ui.adsabs.harvard.edu/abs/2019ApJ...882..156H} {882, 156}

\bibitem[\protect\citeauthoryear{{Jones}, {Ryu}  \& {Tregillis}}{{Jones}
  et~al.}{1996}]{1996ApJ...473..365J}
{Jones} T.~W.,  {Ryu} D.,   {Tregillis} I.~L.,  1996, \mn@doi [\apj]
  {10.1086/178151}, \href
  {https://ui.adsabs.harvard.edu/abs/1996ApJ...473..365J} {473, 365}

\bibitem[\protect\citeauthoryear{{Kanjilal}, {Dutta}  \& {Sharma}}{{Kanjilal}
  et~al.}{2020}]{2020arXiv200900525K}
{Kanjilal} V.,  {Dutta} A.,   {Sharma} P.,  2020, arXiv e-prints, \href
  {https://ui.adsabs.harvard.edu/abs/2020arXiv200900525K} {p. arXiv:2009.00525}

\bibitem[\protect\citeauthoryear{{Katz}, {Weinberg}  \& {Hernquist}}{{Katz}
  et~al.}{1996}]{1996ApJS..105...19K}
{Katz} N.,  {Weinberg} D.~H.,   {Hernquist} L.,  1996, \mn@doi [\apjs]
  {10.1086/192305}, \href {http://adsabs.harvard.edu/abs/1996ApJS..105...19K}
  {105, 19}

\bibitem[\protect\citeauthoryear{{Kere{\v{s}}}, {Katz}, {Weinberg}  \&
  {Dav{\'e}}}{{Kere{\v{s}}} et~al.}{2005}]{2005MNRAS.363....2K}
{Kere{\v{s}}} D.,  {Katz} N.,  {Weinberg} D.~H.,   {Dav{\'e}} R.,  2005,
  \mn@doi [\mnras] {10.1111/j.1365-2966.2005.09451.x}, \href
  {https://ui.adsabs.harvard.edu/abs/2005MNRAS.363....2K} {363, 2}

\bibitem[\protect\citeauthoryear{{Leroy} et~al.,}{{Leroy}
  et~al.}{2015}]{2015ApJ...814...83L}
{Leroy} A.~K.,  et~al., 2015, \mn@doi [\apj] {10.1088/0004-637X/814/2/83},
  \href {https://ui.adsabs.harvard.edu/abs/2015ApJ...814...83L} {814, 83}

\bibitem[\protect\citeauthoryear{{Li}, {Hopkins}, {Squire}  \& {Hummels}}{{Li}
  et~al.}{2019}]{2019arXiv190902632L}
{Li} Z.,  {Hopkins} P.~F.,  {Squire} J.,   {Hummels} C.,  2019, \mn@doi
  [\mnras] {10.1093/mnras/stz3567}, \href
  {https://ui.adsabs.harvard.edu/abs/2019MNRAS.tmp.3180L} {p.~3180}

\bibitem[\protect\citeauthoryear{{Liang} \& {Remming}}{{Liang} \&
  {Remming}}{2020}]{2018arXiv180610688L}
{Liang} C.~J.,  {Remming} I.,  2020, \mn@doi [\mnras] {10.1093/mnras/stz3403},
  \href {https://ui.adsabs.harvard.edu/abs/2020MNRAS.491.5056L} {491, 5056}

\bibitem[\protect\citeauthoryear{{Mac Low}, {McKee}, {Klein}, {Stone}  \&
  {Norman}}{{Mac Low} et~al.}{1994}]{1994ApJ...433..757M}
{Mac Low} M.-M.,  {McKee} C.~F.,  {Klein} R.~I.,  {Stone} J.~M.,   {Norman}
  M.~L.,  1994, \mn@doi [\apj] {10.1086/174685}, \href
  {https://ui.adsabs.harvard.edu/abs/1994ApJ...433..757M} {433, 757}

\bibitem[\protect\citeauthoryear{{Mandelker}, {Padnos}, {Dekel}, {Birnboim},
  {Burkert}, {Krumholz}  \& {Steinberg}}{{Mandelker}
  et~al.}{2016}]{2016MNRAS.463.3921M}
{Mandelker} N.,  {Padnos} D.,  {Dekel} A.,  {Birnboim} Y.,  {Burkert} A.,
  {Krumholz} M.~R.,   {Steinberg} E.,  2016, \mn@doi [\mnras]
  {10.1093/mnras/stw2267}, \href
  {https://ui.adsabs.harvard.edu/abs/2016MNRAS.463.3921M} {463, 3921}

\bibitem[\protect\citeauthoryear{{Mandelker}, {Nagai}, {Aung}, {Dekel},
  {Padnos}  \& {Birnboim}}{{Mandelker} et~al.}{2019}]{2019MNRAS.484.1100M}
{Mandelker} N.,  {Nagai} D.,  {Aung} H.,  {Dekel} A.,  {Padnos} D.,
  {Birnboim} Y.,  2019, \mn@doi [\mnras] {10.1093/mnras/stz012}, \href
  {https://ui.adsabs.harvard.edu/abs/2019MNRAS.484.1100M} {484, 1100}

\bibitem[\protect\citeauthoryear{{Mandelker}, {Nagai}, {Aung}, {Dekel},
  {Birnboim}  \& {van den Bosch}}{{Mandelker}
  et~al.}{2020}]{2019arXiv191005344M}
{Mandelker} N.,  {Nagai} D.,  {Aung} H.,  {Dekel} A.,  {Birnboim} Y.,   {van
  den Bosch} F.~C.,  2020, \mn@doi [\mnras] {10.1093/mnras/staa812}, \href
  {https://ui.adsabs.harvard.edu/abs/2020MNRAS.tmp.1082M} {}

\bibitem[\protect\citeauthoryear{{Marinacci} et~al.,}{{Marinacci}
  et~al.}{2018}]{2018MNRAS.480.5113M}
{Marinacci} F.,  et~al., 2018, \mn@doi [\mnras] {10.1093/mnras/sty2206}, \href
  {https://ui.adsabs.harvard.edu/abs/2018MNRAS.480.5113M} {480, 5113}

\bibitem[\protect\citeauthoryear{{McCourt}, {O'Leary}, {Madigan}  \&
  {Quataert}}{{McCourt} et~al.}{2015}]{2015MNRAS.449....2M}
{McCourt} M.,  {O'Leary} R.~M.,  {Madigan} A.-M.,   {Quataert} E.,  2015,
  \mn@doi [\mnras] {10.1093/mnras/stv355}, \href
  {http://adsabs.harvard.edu/abs/2015MNRAS.449....2M} {449, 2}

\bibitem[\protect\citeauthoryear{{McCourt}, {Oh}, {O'Leary}  \&
  {Madigan}}{{McCourt} et~al.}{2018}]{2018MNRAS.473.5407M}
{McCourt} M.,  {Oh} S.~P.,  {O'Leary} R.,   {Madigan} A.-M.,  2018, \mn@doi
  [\mnras] {10.1093/mnras/stx2687}, \href
  {http://adsabs.harvard.edu/abs/2018MNRAS.473.5407M} {473, 5407}

\bibitem[\protect\citeauthoryear{{Miller}, {Hornschemeier}  \&
  {Mobasher}}{{Miller} et~al.}{2009}]{2009AJ....137.4436M}
{Miller} N.~A.,  {Hornschemeier} A.~E.,   {Mobasher} B.,  2009, \mn@doi [\aj]
  {10.1088/0004-6256/137/5/4436}, \href
  {https://ui.adsabs.harvard.edu/abs/2009AJ....137.4436M} {137, 4436}

\bibitem[\protect\citeauthoryear{{Miniati}, {Ryu}, {Ferrara}  \&
  {Jones}}{{Miniati} et~al.}{1999}]{1999ApJ...510..726M}
{Miniati} F.,  {Ryu} D.,  {Ferrara} A.,   {Jones} T.~W.,  1999, \mn@doi [\apj]
  {10.1086/306599}, \href
  {https://ui.adsabs.harvard.edu/abs/1999ApJ...510..726M} {510, 726}

\bibitem[\protect\citeauthoryear{{Naiman} et~al.,}{{Naiman}
  et~al.}{2018}]{2018MNRAS.477.1206N}
{Naiman} J.~P.,  et~al., 2018, \mn@doi [\mnras] {10.1093/mnras/sty618}, \href
  {https://ui.adsabs.harvard.edu/abs/2018MNRAS.477.1206N} {477, 1206}

\bibitem[\protect\citeauthoryear{{Nelson}, {Vogelsberger}, {Genel}, {Sijacki},
  {Kere{\v{s}}}, {Springel}  \& {Hernquist}}{{Nelson}
  et~al.}{2013}]{2013MNRAS.429.3353N}
{Nelson} D.,  {Vogelsberger} M.,  {Genel} S.,  {Sijacki} D.,  {Kere{\v{s}}} D.,
   {Springel} V.,   {Hernquist} L.,  2013, \mn@doi [\mnras]
  {10.1093/mnras/sts595}, \href
  {https://ui.adsabs.harvard.edu/abs/2013MNRAS.429.3353N} {429, 3353}

\bibitem[\protect\citeauthoryear{{Nelson} et~al.,}{{Nelson}
  et~al.}{2019}]{2019MNRAS.490.3234N}
{Nelson} D.,  et~al., 2019, \mn@doi [\mnras] {10.1093/mnras/stz2306}, \href
  {https://ui.adsabs.harvard.edu/abs/2019MNRAS.490.3234N} {490, 3234}

\bibitem[\protect\citeauthoryear{{Nichols}, {Revaz}  \& {Jablonka}}{{Nichols}
  et~al.}{2015}]{2015A&A...582A..23N}
{Nichols} M.,  {Revaz} Y.,   {Jablonka} P.,  2015, \mn@doi [\aap]
  {10.1051/0004-6361/201526113}, \href
  {https://ui.adsabs.harvard.edu/abs/2015A&A...582A..23N} {582, A23}

\bibitem[\protect\citeauthoryear{{Oppenheimer}, {Schaye}, {Crain}, {Werk}  \&
  {Richings}}{{Oppenheimer} et~al.}{2018}]{2018MNRAS.481..835O}
{Oppenheimer} B.~D.,  {Schaye} J.,  {Crain} R.~A.,  {Werk} J.~K.,   {Richings}
  A.~J.,  2018, \mn@doi [\mnras] {10.1093/mnras/sty2281}, \href
  {https://ui.adsabs.harvard.edu/abs/2018MNRAS.481..835O} {481, 835}

\bibitem[\protect\citeauthoryear{{Padnos}, {Mandelker}, {Birnboim}, {Dekel},
  {Krumholz}  \& {Steinberg}}{{Padnos} et~al.}{2018}]{2018MNRAS.477.3293P}
{Padnos} D.,  {Mandelker} N.,  {Birnboim} Y.,  {Dekel} A.,  {Krumholz} M.~R.,
  {Steinberg} E.,  2018, \mn@doi [\mnras] {10.1093/mnras/sty789}, \href
  {https://ui.adsabs.harvard.edu/abs/2018MNRAS.477.3293P} {477, 3293}

\bibitem[\protect\citeauthoryear{{Pakmor}, {Springel}, {Bauer}, {Mocz},
  {Munoz}, {Ohlmann}, {Schaal}  \& {Zhu}}{{Pakmor}
  et~al.}{2016}]{2016MNRAS.455.1134P}
{Pakmor} R.,  {Springel} V.,  {Bauer} A.,  {Mocz} P.,  {Munoz} D.~J.,
  {Ohlmann} S.~T.,  {Schaal} K.,   {Zhu} C.,  2016, \mn@doi [\mnras]
  {10.1093/mnras/stv2380}, \href
  {http://adsabs.harvard.edu/abs/2016MNRAS.455.1134P} {455, 1134}

\bibitem[\protect\citeauthoryear{{Peeples}, {Werk}, {Tumlinson}, {Oppenheimer},
  {Prochaska}, {Katz}  \& {Weinberg}}{{Peeples}
  et~al.}{2014}]{2014ApJ...786...54P}
{Peeples} M.~S.,  {Werk} J.~K.,  {Tumlinson} J.,  {Oppenheimer} B.~D.,
  {Prochaska} J.~X.,  {Katz} N.,   {Weinberg} D.~H.,  2014, \mn@doi [\apj]
  {10.1088/0004-637X/786/1/54}, \href
  {https://ui.adsabs.harvard.edu/abs/2014ApJ...786...54P} {786, 54}

\bibitem[\protect\citeauthoryear{{Peeples} et~al.,}{{Peeples}
  et~al.}{2019}]{2018arXiv181006566P}
{Peeples} M.~S.,  et~al., 2019, \mn@doi [\apj] {10.3847/1538-4357/ab0654},
  \href {https://ui.adsabs.harvard.edu/abs/2019ApJ...873..129P} {873, 129}

\bibitem[\protect\citeauthoryear{{Pfrommer} \& {Dursi}}{{Pfrommer} \&
  {Dursi}}{2010}]{2010NatPh...6..520P}
{Pfrommer} C.,  {Dursi} L.~J.,  2010, \mn@doi [Nature Physics]
  {10.1038/nphys1657}, \href
  {https://ui.adsabs.harvard.edu/abs/2010NatPh...6..520P} {6, 520}

\bibitem[\protect\citeauthoryear{{Pillepich} et~al.,}{{Pillepich}
  et~al.}{2018}]{2018MNRAS.475..648P}
{Pillepich} A.,  et~al., 2018, \mn@doi [\mnras] {10.1093/mnras/stx3112}, \href
  {http://adsabs.harvard.edu/abs/2018MNRAS.475..648P} {475, 648}

\bibitem[\protect\citeauthoryear{{Pillepich} et~al.,}{{Pillepich}
  et~al.}{2019}]{2019MNRAS.490.3196P}
{Pillepich} A.,  et~al., 2019, \mn@doi [\mnras] {10.1093/mnras/stz2338}, \href
  {https://ui.adsabs.harvard.edu/abs/2019MNRAS.490.3196P} {490, 3196}

\bibitem[\protect\citeauthoryear{{Rees} \& {Ostriker}}{{Rees} \&
  {Ostriker}}{1977}]{1977MNRAS.179..541R}
{Rees} M.~J.,  {Ostriker} J.~P.,  1977, \mn@doi [\mnras]
  {10.1093/mnras/179.4.541}, \href
  {https://ui.adsabs.harvard.edu/abs/1977MNRAS.179..541R} {179, 541}

\bibitem[\protect\citeauthoryear{{Richter} et~al.,}{{Richter}
  et~al.}{2017}]{2017A&A...607A..48R}
{Richter} P.,  et~al., 2017, \mn@doi [\aap] {10.1051/0004-6361/201630081},
  \href {https://ui.adsabs.harvard.edu/abs/2017A&A...607A..48R} {607, A48}

\bibitem[\protect\citeauthoryear{{Roediger} \& {Br{\"u}ggen}}{{Roediger} \&
  {Br{\"u}ggen}}{2007}]{2007MNRAS.380.1399R}
{Roediger} E.,  {Br{\"u}ggen} M.,  2007, \mn@doi [\mnras]
  {10.1111/j.1365-2966.2007.12241.x}, \href
  {https://ui.adsabs.harvard.edu/abs/2007MNRAS.380.1399R} {380, 1399}

\bibitem[\protect\citeauthoryear{{Rupke} \& {Veilleux}}{{Rupke} \&
  {Veilleux}}{2013}]{2013ApJ...768...75R}
{Rupke} D.~S.~N.,  {Veilleux} S.,  2013, \mn@doi [\apj]
  {10.1088/0004-637X/768/1/75}, \href
  {http://adsabs.harvard.edu/abs/2013ApJ...768...75R} {768, 75}

\bibitem[\protect\citeauthoryear{{Ruszkowski}, {En{\ss}lin}, {Br{\"u}ggen},
  {Heinz}  \& {Pfrommer}}{{Ruszkowski} et~al.}{2007}]{2007MNRAS.378..662R}
{Ruszkowski} M.,  {En{\ss}lin} T.~A.,  {Br{\"u}ggen} M.,  {Heinz} S.,
  {Pfrommer} C.,  2007, \mn@doi [\mnras] {10.1111/j.1365-2966.2007.11801.x},
  \href {https://ui.adsabs.harvard.edu/abs/2007MNRAS.378..662R} {378, 662}

\bibitem[\protect\citeauthoryear{{Ruszkowski}, {Br{\"u}ggen}, {Lee}  \&
  {Shin}}{{Ruszkowski} et~al.}{2014}]{2014ApJ...784...75R}
{Ruszkowski} M.,  {Br{\"u}ggen} M.,  {Lee} D.,   {Shin} M.~S.,  2014, \mn@doi
  [\apj] {10.1088/0004-637X/784/1/75}, \href
  {https://ui.adsabs.harvard.edu/abs/2014ApJ...784...75R} {784, 75}

\bibitem[\protect\citeauthoryear{{Rybicki} \& {Lightman}}{{Rybicki} \&
  {Lightman}}{1979}]{1979rpa..book.....R}
{Rybicki} G.~B.,  {Lightman} A.~P.,  1979, {Radiative processes in
  astrophysics}.
Wiley

\bibitem[\protect\citeauthoryear{{Scannapieco} \& {Br{\"u}ggen}}{{Scannapieco}
  \& {Br{\"u}ggen}}{2015}]{2015ApJ...805..158S}
{Scannapieco} E.,  {Br{\"u}ggen} M.,  2015, \mn@doi [\apj]
  {10.1088/0004-637X/805/2/158}, \href
  {http://adsabs.harvard.edu/abs/2015ApJ...805..158S} {805, 158}

\bibitem[\protect\citeauthoryear{{Schneider} \& {Robertson}}{{Schneider} \&
  {Robertson}}{2017}]{2017ApJ...834..144S}
{Schneider} E.~E.,  {Robertson} B.~E.,  2017, \mn@doi [\apj]
  {10.3847/1538-4357/834/2/144}, \href
  {http://adsabs.harvard.edu/abs/2017ApJ...834..144S} {834, 144}

\bibitem[\protect\citeauthoryear{{Schneider} \& {Robertson}}{{Schneider} \&
  {Robertson}}{2018}]{2018ApJ...860..135S}
{Schneider} E.~E.,  {Robertson} B.~E.,  2018, \mn@doi [\apj]
  {10.3847/1538-4357/aac329}, \href
  {https://ui.adsabs.harvard.edu/abs/2018ApJ...860..135S} {860, 135}

\bibitem[\protect\citeauthoryear{{Sparre}, {Pfrommer}  \&
  {Vogelsberger}}{{Sparre} et~al.}{2019}]{2019MNRAS.482.5401S}
{Sparre} M.,  {Pfrommer} C.,   {Vogelsberger} M.,  2019, \mn@doi [\mnras]
  {10.1093/mnras/sty3063}, \href
  {https://ui.adsabs.harvard.edu/abs/2019MNRAS.482.5401S} {482, 5401}

\bibitem[\protect\citeauthoryear{{Springel}}{{Springel}}{2010}]{2010MNRAS.401..791S}
{Springel} V.,  2010, \mn@doi [\mnras] {10.1111/j.1365-2966.2009.15715.x},
  \href {http://adsabs.harvard.edu/abs/2010MNRAS.401..791S} {401, 791}

\bibitem[\protect\citeauthoryear{{Springel} et~al.,}{{Springel}
  et~al.}{2018}]{2018MNRAS.475..676S}
{Springel} V.,  et~al., 2018, \mn@doi [\mnras] {10.1093/mnras/stx3304}, \href
  {http://adsabs.harvard.edu/abs/2018MNRAS.475..676S} {475, 676}

\bibitem[\protect\citeauthoryear{{Steinhauser}, {Schindler}  \&
  {Springel}}{{Steinhauser} et~al.}{2016}]{2016A&A...591A..51S}
{Steinhauser} D.,  {Schindler} S.,   {Springel} V.,  2016, \mn@doi [\aap]
  {10.1051/0004-6361/201527705}, \href
  {https://ui.adsabs.harvard.edu/abs/2016A&A...591A..51S} {591, A51}

\bibitem[\protect\citeauthoryear{{Strickland} \& {Heckman}}{{Strickland} \&
  {Heckman}}{2009}]{2009ApJ...697.2030S}
{Strickland} D.~K.,  {Heckman} T.~M.,  2009, \mn@doi [\apj]
  {10.1088/0004-637X/697/2/2030}, \href
  {http://cdsads.u-strasbg.fr/abs/2009ApJ...697.2030S} {697, 2030}

\bibitem[\protect\citeauthoryear{{Suresh}, {Nelson}, {Genel}, {Rubin}  \&
  {Hernquist}}{{Suresh} et~al.}{2019}]{2019MNRAS.483.4040S}
{Suresh} J.,  {Nelson} D.,  {Genel} S.,  {Rubin} K. H.~R.,   {Hernquist} L.,
  2019, \mn@doi [\mnras] {10.1093/mnras/sty3402}, \href
  {https://ui.adsabs.harvard.edu/abs/2019MNRAS.483.4040S} {483, 4040}

\bibitem[\protect\citeauthoryear{{Tonnesen} \& {Bryan}}{{Tonnesen} \&
  {Bryan}}{2012}]{2012MNRAS.422.1609T}
{Tonnesen} S.,  {Bryan} G.~L.,  2012, \mn@doi [\mnras]
  {10.1111/j.1365-2966.2012.20737.x}, \href
  {https://ui.adsabs.harvard.edu/abs/2012MNRAS.422.1609T} {422, 1609}

\bibitem[\protect\citeauthoryear{{Tumlinson}, {Peeples}  \& {Werk}}{{Tumlinson}
  et~al.}{2017}]{2017ARA&A..55..389T}
{Tumlinson} J.,  {Peeples} M.~S.,   {Werk} J.~K.,  2017, \mn@doi [\araa]
  {10.1146/annurev-astro-091916-055240}, \href
  {http://adsabs.harvard.edu/abs/2017ARA%26A..55..389T} {55, 389}

\bibitem[\protect\citeauthoryear{{Veilleux}, {Maiolino}, {Bolatto}  \&
  {Aalto}}{{Veilleux} et~al.}{2020}]{2020A&ARv..28....2V}
{Veilleux} S.,  {Maiolino} R.,  {Bolatto} A.~D.,   {Aalto} S.,  2020, \mn@doi
  [\aapr] {10.1007/s00159-019-0121-9}, \href
  {https://ui.adsabs.harvard.edu/abs/2020A&ARv..28....2V} {28, 2}

\bibitem[\protect\citeauthoryear{{Vijayaraghavan} \& {Ricker}}{{Vijayaraghavan}
  \& {Ricker}}{2015}]{2015MNRAS.449.2312V}
{Vijayaraghavan} R.,  {Ricker} P.~M.,  2015, \mn@doi [\mnras]
  {10.1093/mnras/stv476}, \href
  {https://ui.adsabs.harvard.edu/abs/2015MNRAS.449.2312V} {449, 2312}

\bibitem[\protect\citeauthoryear{{Vijayaraghavan} \& {Ricker}}{{Vijayaraghavan}
  \& {Ricker}}{2017}]{2017ApJ...841...38V}
{Vijayaraghavan} R.,  {Ricker} P.~M.,  2017, \mn@doi [\apj]
  {10.3847/1538-4357/aa6eac}, \href
  {https://ui.adsabs.harvard.edu/abs/2017ApJ...841...38V} {841, 38}

\bibitem[\protect\citeauthoryear{{Vogelsberger}, {Genel}, {Sijacki}, {Torrey},
  {Springel}  \& {Hernquist}}{{Vogelsberger}
  et~al.}{2013}]{2013MNRAS.436.3031V}
{Vogelsberger} M.,  {Genel} S.,  {Sijacki} D.,  {Torrey} P.,  {Springel} V.,
  {Hernquist} L.,  2013, \mn@doi [\mnras] {10.1093/mnras/stt1789}, \href
  {http://adsabs.harvard.edu/abs/2013MNRAS.436.3031V} {436, 3031}

\bibitem[\protect\citeauthoryear{{Vogelsberger} et~al.,}{{Vogelsberger}
  et~al.}{2014}]{2014Natur.509..177V}
{Vogelsberger} M.,  et~al., 2014, \mn@doi [\nat] {10.1038/nature13316}, \href
  {http://adsabs.harvard.edu/abs/2014Natur.509..177V} {509, 177}

\bibitem[\protect\citeauthoryear{{Waelkens}, {Jaffe}, {Reinecke}, {Kitaura}  \&
  {En{\ss}lin}}{{Waelkens} et~al.}{2009}]{2009A&A...495..697W}
{Waelkens} A.,  {Jaffe} T.,  {Reinecke} M.,  {Kitaura} F.~S.,   {En{\ss}lin}
  T.~A.,  2009, \mn@doi [\aap] {10.1051/0004-6361:200810564}, \href
  {https://ui.adsabs.harvard.edu/abs/2009A&A...495..697W} {495, 697}

\bibitem[\protect\citeauthoryear{{Werk} et~al.,}{{Werk}
  et~al.}{2014}]{2014ApJ...792....8W}
{Werk} J.~K.,  et~al., 2014, \mn@doi [\apj] {10.1088/0004-637X/792/1/8}, \href
  {http://adsabs.harvard.edu/abs/2014ApJ...792....8W} {792, 8}

\bibitem[\protect\citeauthoryear{{White} \& {Frenk}}{{White} \&
  {Frenk}}{1991}]{1991ApJ...379...52W}
{White} S. D.~M.,  {Frenk} C.~S.,  1991, \mn@doi [\apj] {10.1086/170483}, \href
  {https://ui.adsabs.harvard.edu/abs/1991ApJ...379...52W} {379, 52}

\bibitem[\protect\citeauthoryear{{White} \& {Rees}}{{White} \&
  {Rees}}{1978}]{1978MNRAS.183..341W}
{White} S.~D.~M.,  {Rees} M.~J.,  1978, \mn@doi [\mnras]
  {10.1093/mnras/183.3.341}, \href
  {https://ui.adsabs.harvard.edu/abs/1978MNRAS.183..341W} {183, 341}

\bibitem[\protect\citeauthoryear{{Wisotzki} et~al.,}{{Wisotzki}
  et~al.}{2018}]{2018Natur.562..229W}
{Wisotzki} L.,  et~al., 2018, \mn@doi [\nat] {10.1038/s41586-018-0564-6}, \href
  {https://ui.adsabs.harvard.edu/#abs/2018Natur.562..229W} {562, 229}

\bibitem[\protect\citeauthoryear{{Yu}, {Owen}, {Wu}  \& {Ferreras}}{{Yu}
  et~al.}{2020}]{2020arXiv200104384Y}
{Yu} B.~P.~B.,  {Owen} E.~R.,  {Wu} K.,   {Ferreras} I.,  2020, \mn@doi
  [\mnras] {10.1093/mnras/staa021}, \href
  {https://ui.adsabs.harvard.edu/abs/2020MNRAS.492.3179Y} {492, 3179}

\bibitem[\protect\citeauthoryear{{Yun} et~al.,}{{Yun}
  et~al.}{2019}]{2019MNRAS.483.1042Y}
{Yun} K.,  et~al., 2019, \mn@doi [\mnras] {10.1093/mnras/sty3156}, \href
  {https://ui.adsabs.harvard.edu/abs/2019MNRAS.483.1042Y} {483, 1042}

\bibitem[\protect\citeauthoryear{{van de Voort}, {Springel}, {Mandelker}, {van
  den Bosch}  \& {Pakmor}}{{van de Voort} et~al.}{2019}]{2019MNRAS.482L..85V}
{van de Voort} F.,  {Springel} V.,  {Mandelker} N.,  {van den Bosch} F.~C.,
  {Pakmor} R.,  2019, \mn@doi [\mnras] {10.1093/mnrasl/sly190}, \href
  {https://ui.adsabs.harvard.edu/\#abs/2019MNRAS.482L..85V} {482, L85}

\bibitem[\protect\citeauthoryear{{van de Voort}, {Bieri}, {Pakmor},
  {G{\'o}mez}, {Grand}  \& {Marinacci}}{{van de Voort}
  et~al.}{2020}]{2020arXiv200807537V}
{van de Voort} F.,  {Bieri} R.,  {Pakmor} R.,  {G{\'o}mez} F.~A.,  {Grand} R.
  J.~J.,   {Marinacci} F.,  2020, arXiv e-prints, \href
  {https://ui.adsabs.harvard.edu/abs/2020arXiv200807537V} {p. arXiv:2008.07537}

\makeatother
\end{thebibliography}
}


\appendix

\section{Dense gas mass and the growth criterion}

\begin{figure*}
\centering
\begin{minipage}{.48\textwidth}
\includegraphics[width=\linewidth]{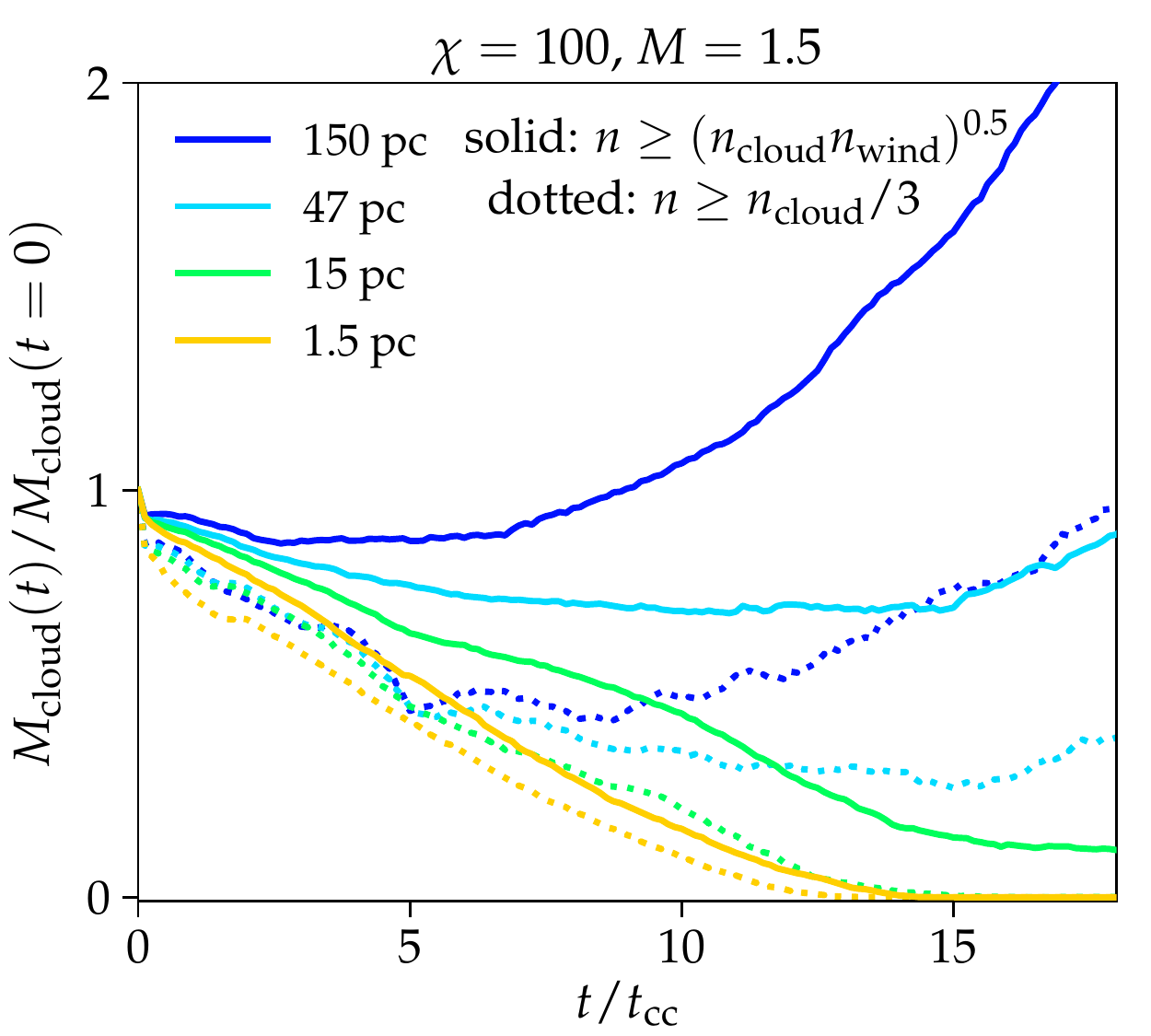}
\end{minipage}
\begin{minipage}{.48\textwidth}
\centering
\includegraphics[width=\linewidth]{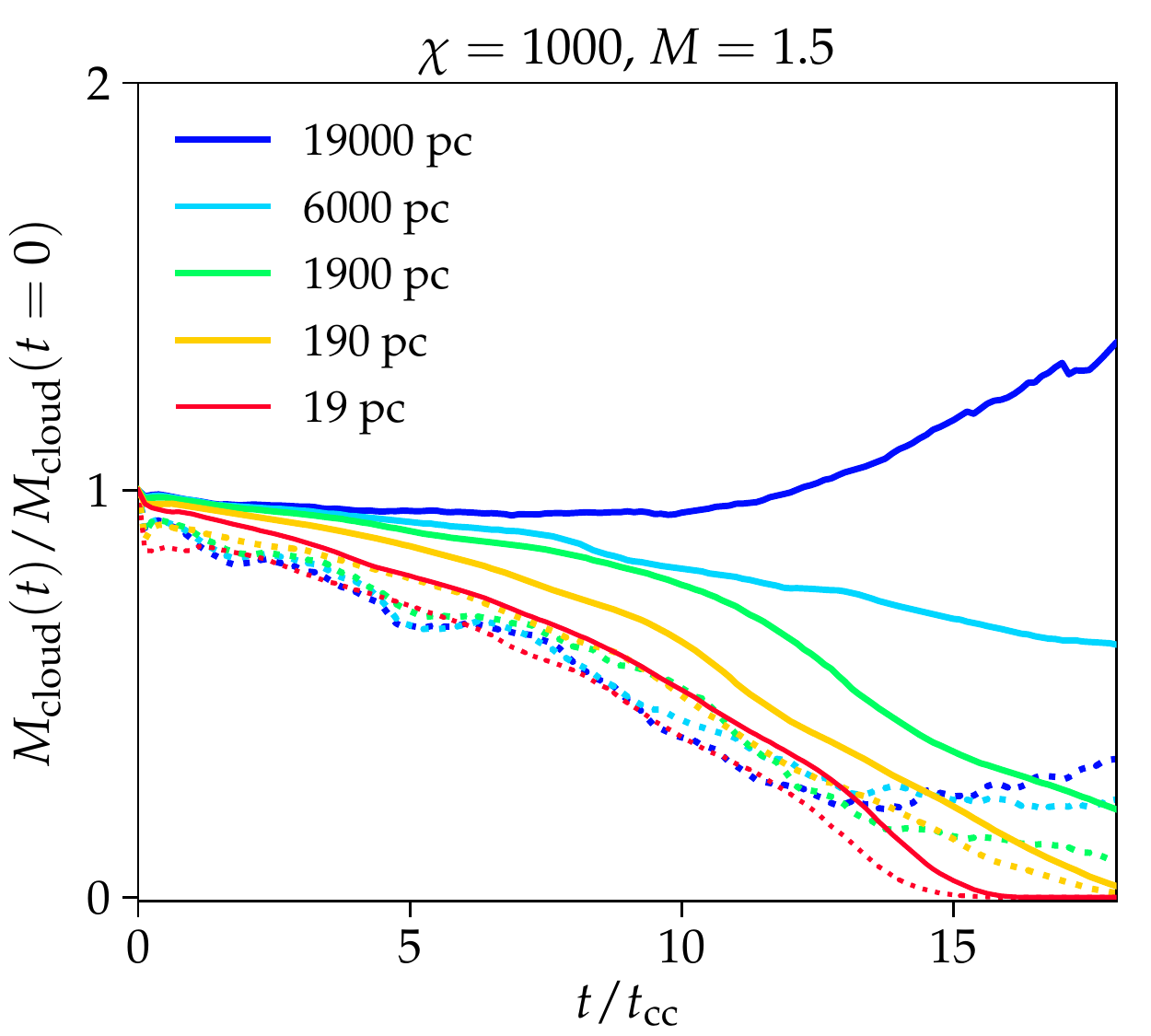}
\end{minipage}
\caption{We demonstrate the effect of using two different density thresholds for defining the mass in dense phase. Using the geometric mean (solid lines) yields a smoother and more monotonic evolution compared to using a third of the initial cloud density (dotted lines).}
\label{FigX122_CloudGrowth_chi1000_Also0.333rhocrit_3D}
\end{figure*}

\begin{figure}
\centering
\includegraphics[width=\linewidth]{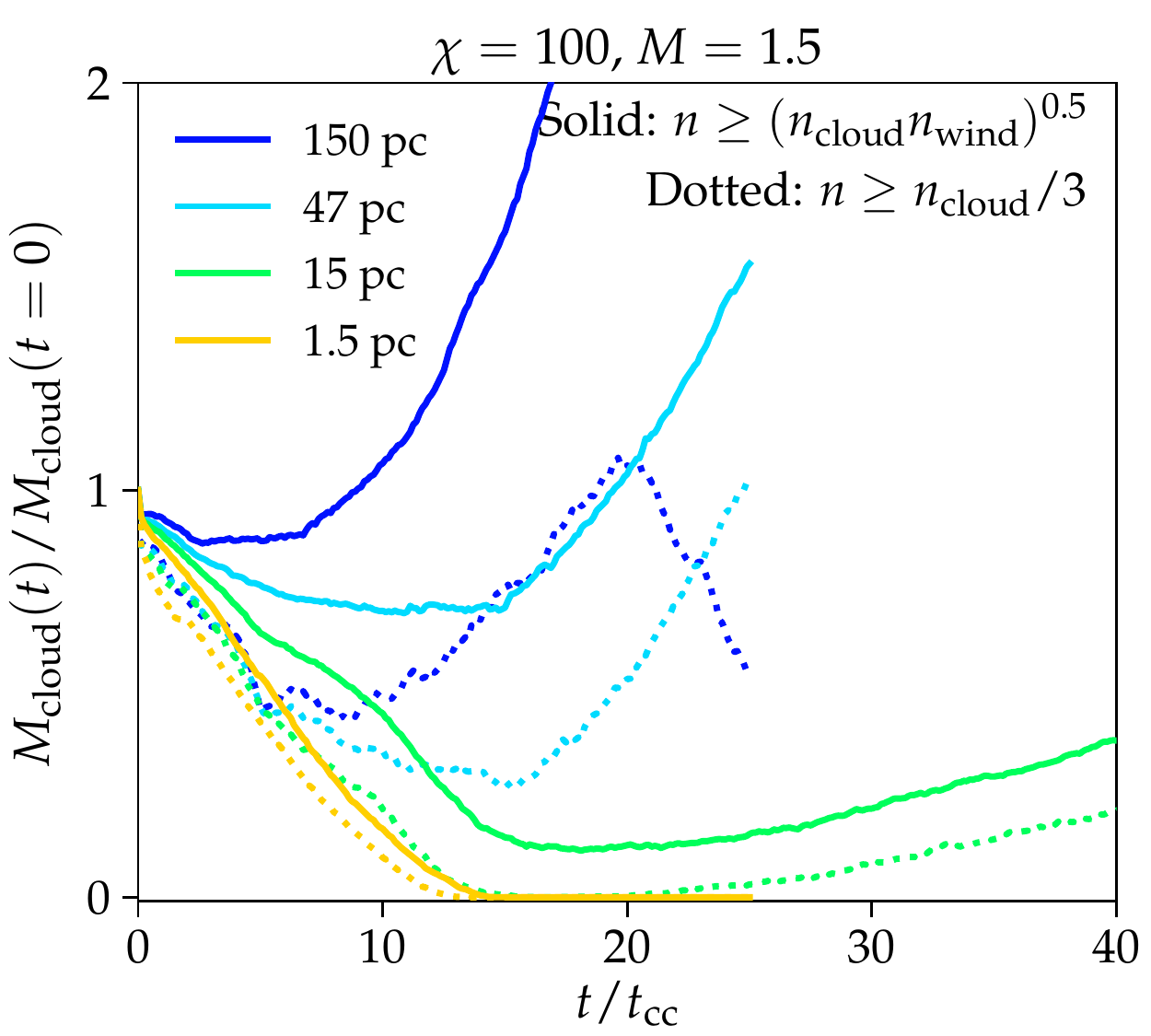}
\caption{We test whether our simulations with $M=1.5$ and $\chi=100$ grow in mass at later times and contrast this to our growth criterion, that quantifies the change in mass accretion rate at 12.5~$t_\text{cc}$. If we had defined growth based on an arbitrary late time, our simulations with a radius of 15 and 47 pc would also be in the growth regime. It is not surprising that a more relaxed growth criterion yields a lower transition radius from the destruction to the growth regime.}
\label{FigX121_CloudGrowth_Chi100_ToPrateek_3DA}
\end{figure}

\subsection{Defining the dense gas}\label{MassDefinition}
In Fig.~\ref{FigX122_CloudGrowth_chi1000_Also0.333rhocrit_3D} we explore the evolution of the dense gas in simulations with $M=1.5$. We either define the dense gas phase as $n\geq n_\text{cloud}/3$ or $n\geq \sqrt{n_\text{cloud}n_\text{wind}}$.

In the main paper we classify a simulation to be in the growth or destruction regime based on whether we see an increasing mass in the dense phase at $12.5t_\text{cc}$. For $\chi=100$ the clouds with $R_\text{cloud}\leq 47$ pc are undergoing destruction at this time, independent of which density threshold is used to define the dense gas phase. For $\chi=1000$ the situation is different. Here the simulation with 19000 pc is in the destruction regime (at $12.5t_\text{cc}$) if we use a threshold of $n\geq n_\text{cloud}/3$, but in the growth regime for $n\geq \sqrt{n_\text{cloud}n_\text{wind}}$. If we instead had defined the regime based on the behaviour at $15t_\text{cc}$ either density threshold would yield a cloud in the growth regime. Hence, if we are close to the transition radius between the growth and the destruction regime, the density threshold used to define the dense gas may change the regime of a cloud, but in most cases the regime is independent of the density threshold.

\subsection{The growth criterion}\label{SharmaGrowthCriterion}

In Sec.~\ref{GrowthCriterion} we define clouds to be growing if the dense gas mass increases at 12.5 $t_\text{cc}$ for $M=0.5$ and $M=1.5$. For $M=4.5$ we define growth at a later time, because a longer life-time is expected for clouds with high Mach numbers (as shown e.g., by \citealt{2015ApJ...805..158S}).

\citet{2020arXiv200900525K} recently presented simulations similar to ours. They analyse simulations with $M=1.5$ and $\chi=100$, and find a transition radius of $\lesssim 7.16$ pc, which is much smaller than our finding in Fig.~\ref{MachNumber1.5} (we report a transition radius of $\lesssim 150$ pc). They define (i) the dense gas phase with a density $n\geq n_\text{cloud}/3$ and (ii) clouds to grow if the dense gas mass increases at any point throughout the simulations. To test whether our different dense gas criteria and growth definitions cause this discrepancy in transition radius, we further analyse our simulations with $M=1.5$ and $\chi=100$ in Fig.~\ref{FigX121_CloudGrowth_Chi100_ToPrateek_3DA}.

We see that the simulation with a radius of 150 pc is in the growth regime at early time. However, at $t\geq 20t_\text{cc}$ the mass in dense gas with $n\geq n_\text{cloud}/3$ declines, because runaway cooling occurs in the wind. The simulations $R_\text{cloud}=15$ pc and $47$ pc experience growth at late times, so they would also be in the growth regime according to \citet{2020arXiv200900525K}, but not according to our stricter criterion requiring growth at $12.5 t_\text{cc}$. We have run the simulation with $R_\text{cloud}=15$ pc for a longer time, to show that the growth occurring at $t\lesssim 25 t_\text{cc}$ is persistent. The simulation $R_\text{cloud}=1.5$ pc gets completely dissolved and is in the destruction regime according to our criterion and that suggested by \citet{2020arXiv200900525K}.

In summary, if we weaken our growth criterion, and define clouds to be growing if they experienced growth at any time during a simulation, we would find a lower transition radius of $\lesssim 15 $ pc, which is in good agreement with \citet{2020arXiv200900525K}. We note that the dense gas mass definition does not alter whether simulations are growing or dissolving in these simulations.

\section{Simulations with $\chi=333$}\label{AppendixChi333}

\begin{figure}
\centering
\includegraphics[width=\linewidth]{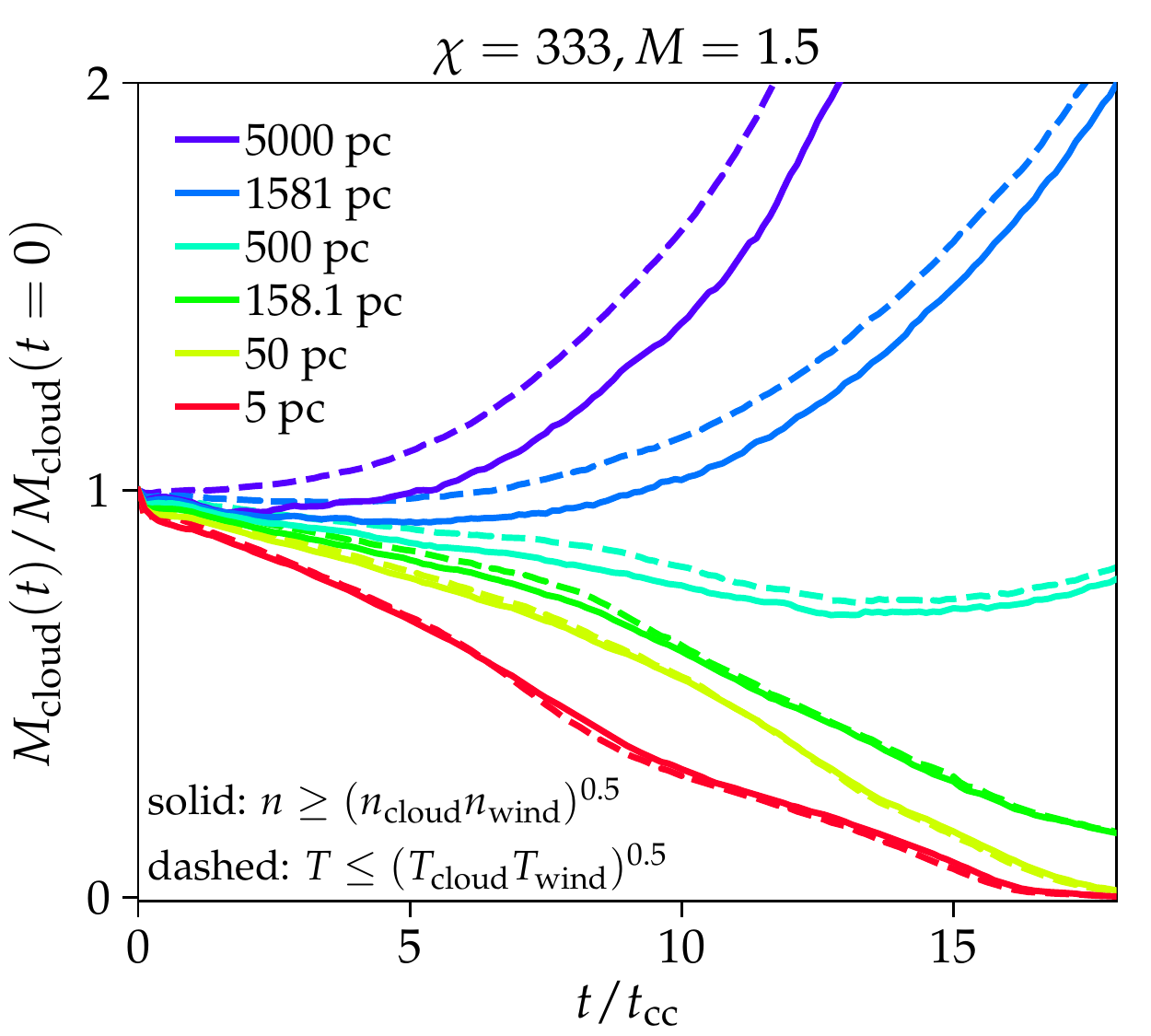}
\caption{For $\chi=333$ we show the evolution of the dense phase (solid lines) and the cold phase (dashed lines). The dense and cold phases have a comparable evolution, so we are not in the runaway cooling regime in any of these simulations. In Sect.~\ref{AppendixChi333} we classify each cloud as being either in the destruction or growth regime.}\label{FigX121_CloudGrowth_Chi333_3D}
\end{figure}

The simulations with $\chi=333$ are shown in Fig.~\ref{FigX121_CloudGrowth_Chi333_3D}. They all have $M=1.5$. The clouds with $R_\text{cloud}=1581$ pc and $5000$ pc have an increasing mass in the dense phase at $12.5 t_\text{cc}$, so they are in the growth regime. The cloud with $R_\text{cloud}=500$ pc is on the edge between the growth and the destruction regime, as it experiences a decline in dense mass for $t\leq 12.5t_\text{cc}$ and afterwards it starts growing. We mark this simulation as being in the destruction regime, due to its decay at $12.5t_\text{cc}$. The clouds with $R_\text{cloud}\leq 158.1$ pc are clearly in the destruction regime.

\section{Numerical convergence}\label{AppendixConvergence}

\begin{figure*}
\centering
\begin{minipage}{.48\textwidth}
\includegraphics[width=\linewidth]{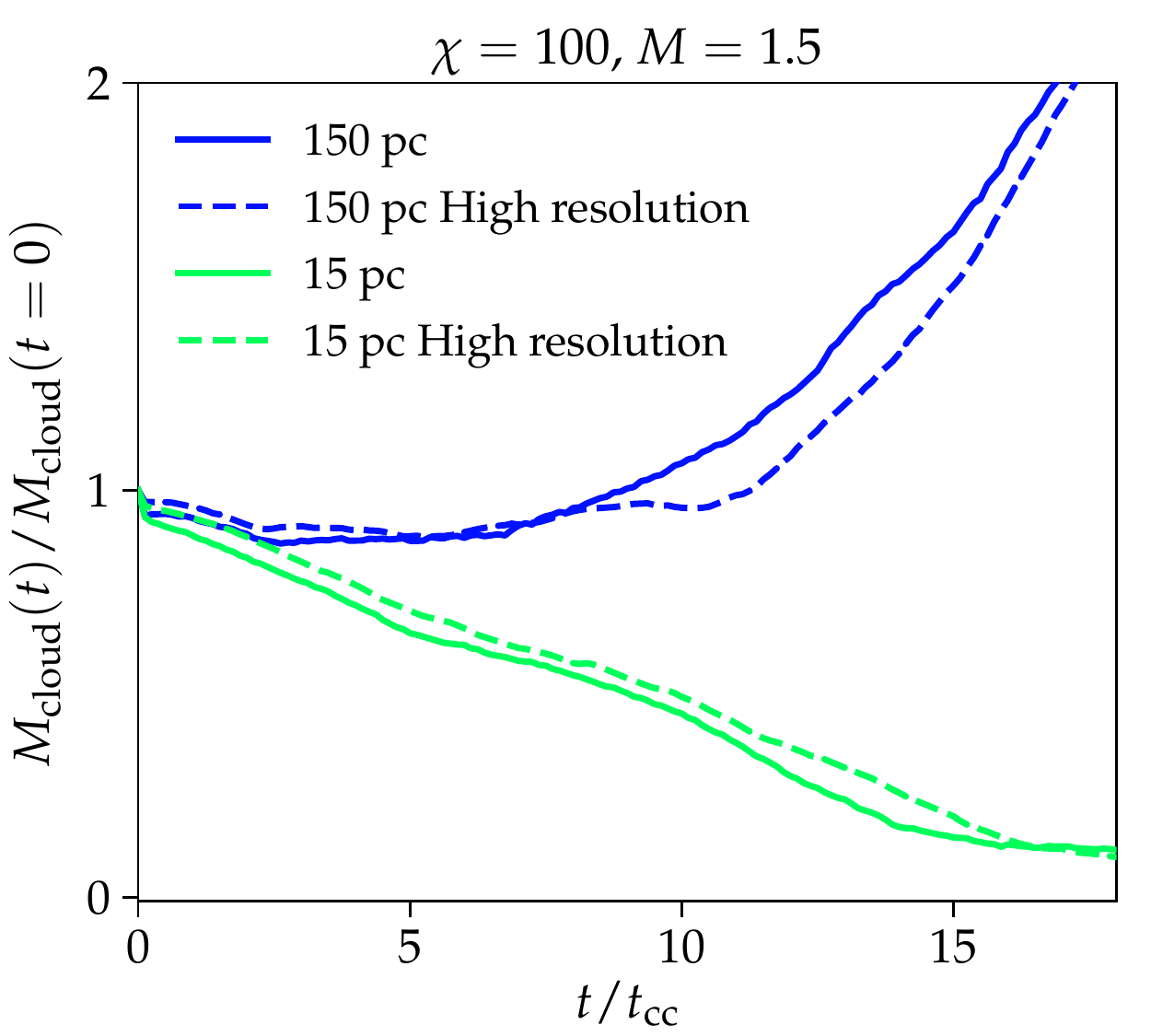}
\end{minipage}
\begin{minipage}{.48\textwidth}
\centering
\includegraphics[width=\linewidth]{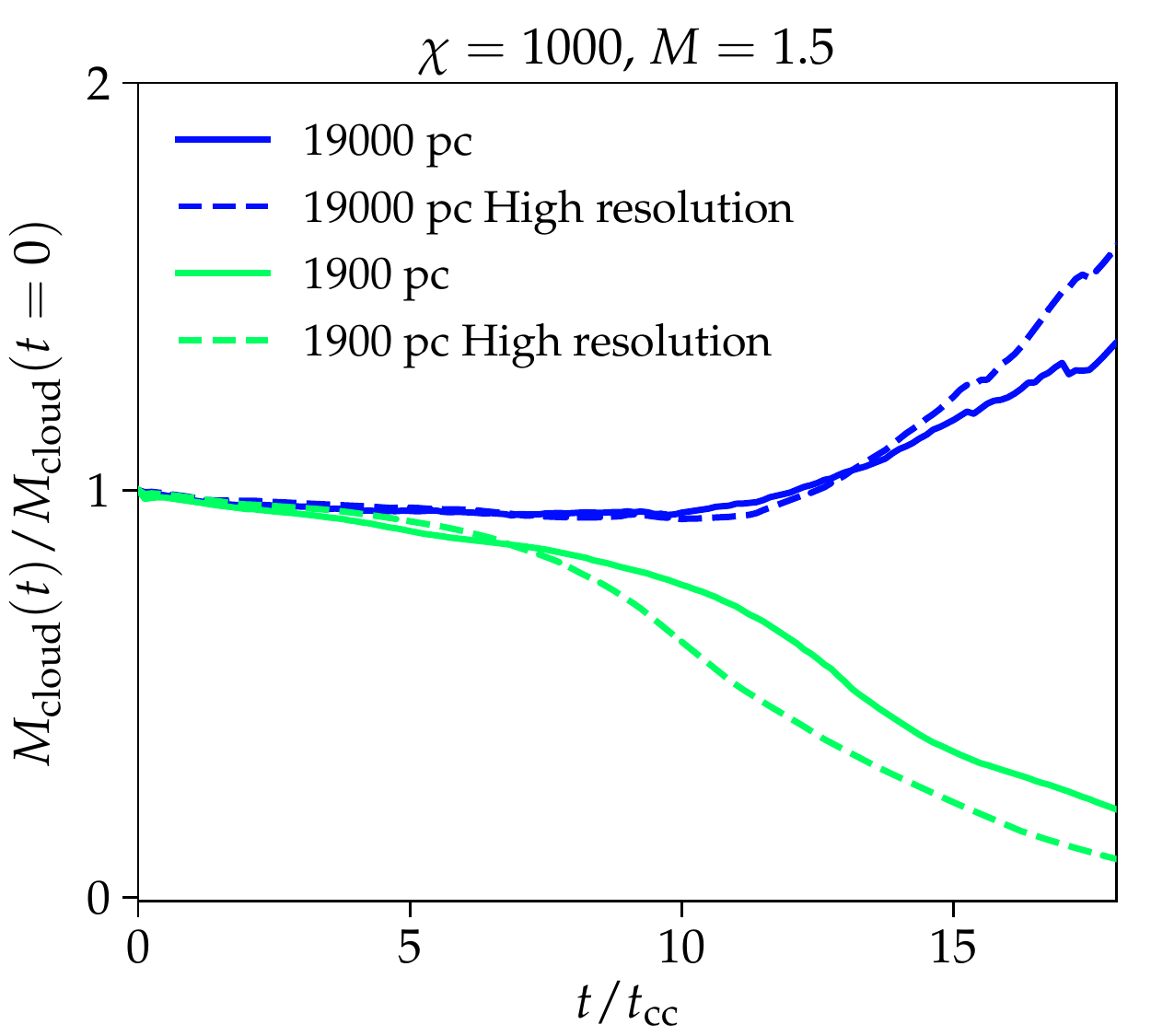}
\end{minipage}
\caption{Convergence tests demonstrating that the transition between the destruction and growth regime is unchanged at one resolution level higher at both $\chi=100$ and $1000$, compared to what is presented in the main paper.}
\label{FigX126_Convergence_Chi1000_3D}\label{FigX125_Convergence_Chi100_3D}
\end{figure*}

The simulations presented in Sect.~\ref{GrowthRegime} (Table~\ref{Table:SimulationOverviewB}) were run at a resolution of 7 and 15 cells per cloud radii for $\chi=100$ and $\chi=1000$. This resolution is lower than typically used in modern cloud crushing simulations. The adapted resolution is a compromise between the requirement of very large simulation boxes to ensure that no dense gas leaves the simulated domain and the resolution needed to obtain a converged solution.

Here we perform a convergence test for the $M=1.5$ simulations. For $\chi=100$ we select the two simulations with a radius of 15 and 150 pc. These parameters have been chosen because in-between these two cloud sizes, we expect the transition between the growth and destruction regime. In high-resolution simulations, with a two times better spatial resolution, we confirm in Fig.~\ref{FigX126_Convergence_Chi1000_3D} (left panel) that neither of these simulations changes regime, when moving to a higher resolution. For $\chi=1000$ we perform a similar test for a 1900 pc and a 19000 pc cloud. Here we also see that increasing the spatial resolution does not change the regime of a cloud (right panel of Fig.~\ref{FigX126_Convergence_Chi1000_3D}). We conclude that the determination of the growth regime from Sect.~\ref{GrowthRegime} is converged.

\section{The transition between the growth and destruction regime without MHD}\label{AppendixMHDsim}

\begin{figure*}
\centering
\begin{minipage}{.48\textwidth}
\includegraphics[width=\linewidth]{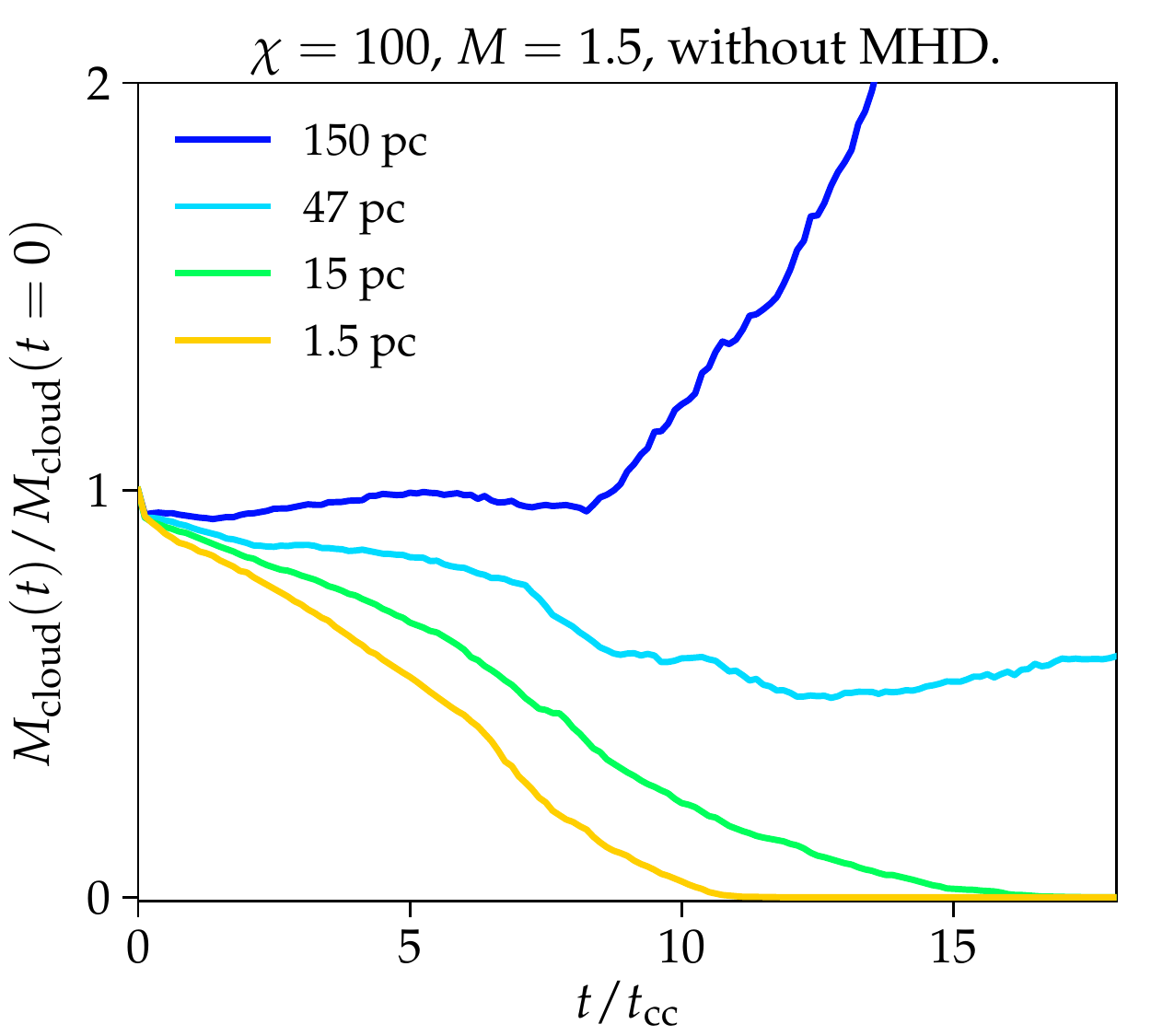}
\end{minipage}
\begin{minipage}{.48\textwidth}
\centering
\includegraphics[width=\linewidth]{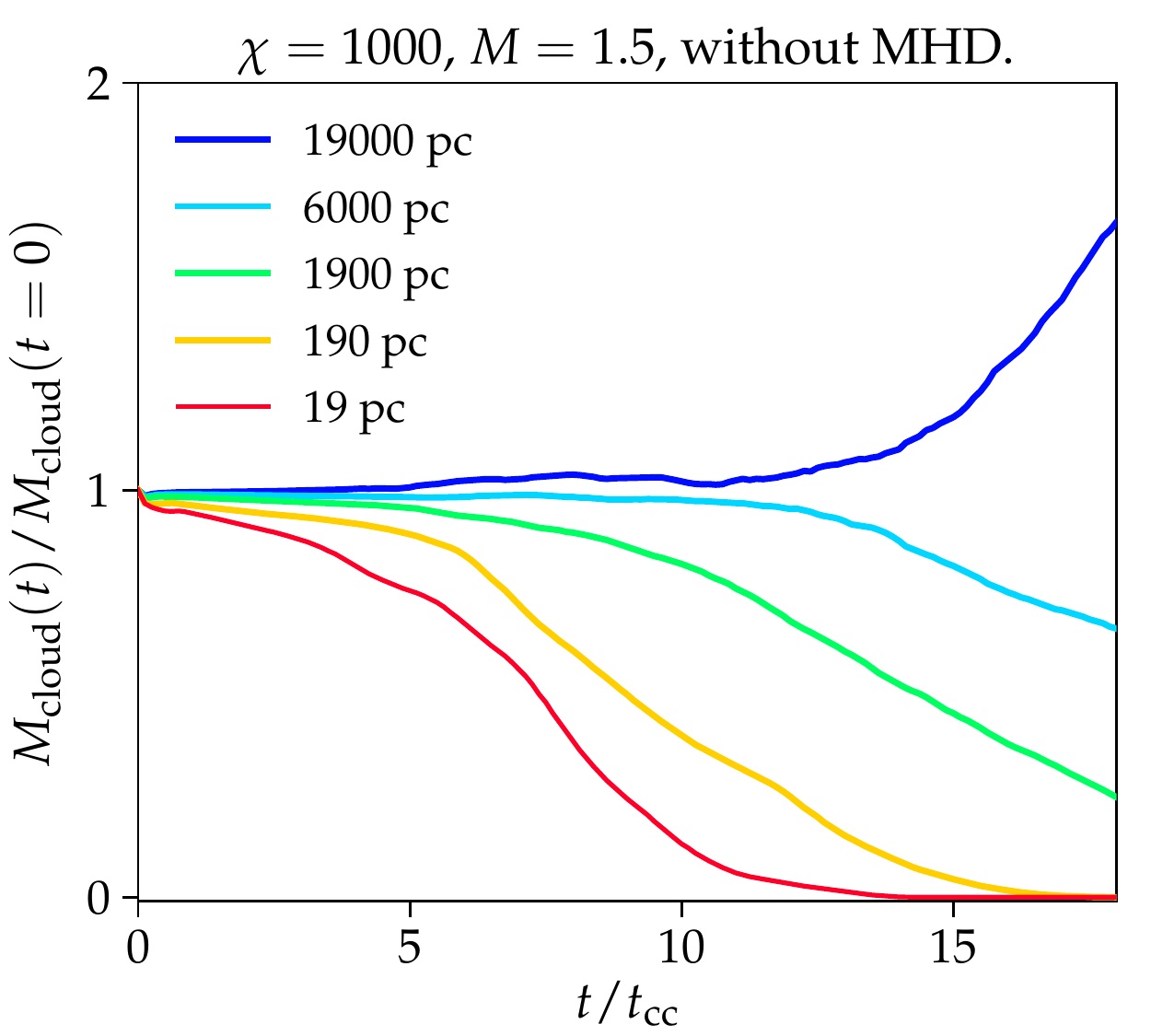}
\end{minipage}
\caption{Same as Fig.~\ref{MachNumber1.5}, but with MHD disabled. The radius, where clouds transition from the destruction to the growing regime, is comparable to that obtained in our MHD simulations. For the $\chi=100$ simulations the simulations with $R_\text{cloud}= 47$ pc is growing after $t=13t_\text{cc}$, and the most notable difference is that the growth occurs faster and more steadily in the MHD version in Fig.~\ref{MachNumber1.5} (this simulation is undergoing destruction according to our criterion, see text for details).}
\label{FigX221}\label{FigX222}
\end{figure*}

In Sect.~\ref{GrowthCriterion} and Fig.~\ref{MachNumber1.5} we determined the radius for which a cloud transitions from the destruction to the growth regime. This was done for both of our setups with $\chi=100$ and $\chi=1000$, both with a turbulent magnetic field in the wind and a tangled magnetic field in the cloud. The magnetic field strength corresponded to $\beta=10$.

To test the role of magnetic fields we have rerun the $M=1.5$ simulations without MHD. The evolution of the simulations is shown in Fig.~\ref{FigX221}. This figure can be compared directly to the MHD-simulations with $M=1.5$ in Fig.~\ref{MachNumber1.5}. By comparing the two figures we see that magnetic fields do alter the detailed qualitative evolution of the mass survival fraction, but the simulations, which were in the growth (destruction) regime in the MHD simulations are also in the growth (destruction) regime without MHD.

The cloud with 47 pc and $\chi=100$ is close to the transition radius between the growth and destruction regime in both the hydrodynamical and MHD simulation (we have marked both of them to be in the destruction regime, because of their lack of growth at $12.5 t_\text{cc}$). It is noticeable that the mass of the dense gas is lower near the end of the hydrodynamical simulation in comparison to the MHD simulation. This is indeed an indication that inclusion of a $\beta=10$ magnetic field does mildly affect the cloud growth criterion. We discuss the growth criterion further in Sect.~\ref{Sec:MagFieldGrowthCriterion}.

\bsp	
\label{lastpage}
\end{document}